\documentclass[journal]{IEEEtran}
\usepackage{amssymb,amsmath,epsfig,psfrag,epstopdf,enumerate}
\usepackage{multirow}

\newcommand*\ud[1]{%
  \underaccent{\dot}{#1}}

\usepackage[english]{babel}
\usepackage{color}
\usepackage{graphicx}
\usepackage{epstopdf}

\usepackage[normalem]{ulem}
\usepackage{varioref}
\usepackage{balance}
\usepackage{accents}

\newtheorem{corollary}{Corollary}[section]
\newtheorem{definition}{Definition}[section]
\newtheorem{lemma}{Lemma}[section]
\newtheorem{proposition}{Proposition}[section]
\newtheorem{remark}{Remark}[section]
\newtheorem{theorem}{Theorem}[section]
\newtheorem{example}{Example}[section]

\begin{document}


\title{The Matrix Exponential Distribution -- \\
A Tool for Wireless System Performance Analysis}

\author{
\IEEEauthorblockN{Peter Larsson, \emph{Student
Member, IEEE},  Lars K. Rasmussen, \emph{Senior
Member, IEEE},\\ Mikael Skoglund, \emph{Senior
Member, IEEE}}%
\thanks{The authors are with the ACCESS Linnaeus Center and the School of
Electrical Engineering at  KTH Royal Institute of  Technology, SE-100 44
Stockholm, Sweden.}
}
\maketitle

\begin{abstract} 
In \cite{LarssonRasmSkog16a}, we introduced a new, matrix algebraic, performance analysis framework for wireless systems with fading channels based on the matrix exponential distribution. The main idea was to use the compact, powerful, and easy-to-use, matrix exponential (ME)-distribution for i) modeling the unprocessed channel signal to noise ratio (SNR), ii) exploiting the closure property of the ME-distribution for SNR processing operations to give the effective channel random variable (r.v.) on ME-distribution form, and then to iii) express the performance measure in a closed-form based on ME-distribution matrix/vector parameters only. In this work, we aim to more clearly present, formalize, refine and develop this unified bottom-up analysis framework, show its versatility to handle important communication cases, performance evaluation levels, and performance metrics. The bivariate ME-distribution is introduced here as yet another useful ME-tool, e.g. to account for dependency among two r.v.s.
We propose that the ME-distribution may, in addition to fading, also characterize the pdf of discrete-time signal r.v.s,  thus extending the ME-distribution matrix form to new generalized 1D/2D-Gaussian-, and Rayleigh-, distribution-like matrix forms.
Our findings here, strengthen the observation from \cite{LarssonRasmSkog16a, LarssonRasmSkog16b}, and indicates that the ME-distribution can be a promising tool for wireless system modeling and performance analysis.
\end{abstract}


\section{Introduction}
\label{sec:Sec6d1}

\IEEEPARstart{A}{nalytical} performance studies of wireless communication systems with fading channels play an important role in our understanding of novel, as well as well-established, communication schemes.
The performance analysis may regard low level physical (Phy) functions (e.g. involving combining, modulation, detection) \cite{Wilson96, ProakisMano96}, or high level Phy functions (e.g. involving retransmissions, multinode cooperation) \cite{CaireTuni01, LanemanTseWorn04, HunterNosra06}, for which relevant performance metrics are studied. For such analysis, analytical fading channel models, like Rayleigh-, Nakagami-$m$-, Ricean-, Hoyt-, Log-normal-distributed fading, \cite{ProakisMano96, Rappaport01, Shankar12}, are commonly encountered. 

In \cite{LarssonRasmSkog16a}, extending the work  \cite{LarssonRasmSkog14b} on a Laplace transform (LT)-oriented throughput analysis, we introduced a new performance analysis framework for wireless communication systems with fading channels. This analysis framework modeled fading channels with the matrix exponential distribution (ME-distribution), and gave analytical performance expressions in the parameters of the ME-distribution. Specifically, we focused on a two-node network for automatic repeat request (ARQ) \cite{CaireTuni01, LarssonRasmSkog14b, BetteshSham06, ShenLiuFitz08}, and Hybrid-ARQ (HARQ) systems \cite{CaireTuni01, LarssonRasmSkog14b, Mandelbaum74, Sindhu77, Chase85, Benelli85, WuJind10, SzczecinskiKDR13}. In this context, we examined aspects of diversity signal processing, such as maximum ratio combining (MRC) and selection diversity combining (SDC) \cite{Wilson96, ProakisMano96, Rappaport01, TseVisw04}, multiple antenna communication schemes, such as orthogonal space-time block coding (OSTBC) \cite{TarokhSeshCald98} and Alamouti's TX diversity \cite{AlamoutiTaro97P, Alamouti98}. Performance were evaluated with respect to (wrt) throughput, outage probability, mean-number of transmissions, and packet loss rate. ME-distribution tools, such as the closure of the convolution and integration, were introduced for system modeling and performance analysis. In \cite{LarssonRasmSkog16b}, we extended the ME-distribution approach to the effective capacity performance measure for (H)ARQ systems.
The insight, and motivation, to introduce the ME-distribution approach was due to recognizing the opportunity to give a unified matrix-algebraic bottom-up performance evaluation framework, for a wide class of fading channels, which gives very compact performance expressions. Specifically, the unprocessed wireless channel signal-to-noise ratio (SNR) r.v. is compactly expressed as a ME-distribution, where (possible) subsequent processing steps, due to  communication system modeling, then gives an effective SNR (or mutual information (MI)) r.v. on a ME-distribution form. Then, the final performance evaluation step can express a performance metric of interest with the ME-distribution parameters for the unprocessed wireless channel SNR, or alternatively, for the effective channel.

\subsection{Related Works}
\label{sec:Sec6d1d2}
The list of works on performance evaluation of wireless systems with fading channels is exceedingly long, and a complete account can not be given. Some notable reference works, however, are  \cite{Wilson96, ProakisMano96, Rappaport01, Shankar12, TseVisw04, TarokhSeshCald98, Alamouti98, SimonAlouini05, Molisch05, BiglieriProaSham98}, and references therein. Some significant performance studies on communication cases with fading channels that consider higher Phy functions are, e.g.,  \cite{CaireTuni01, LanemanTseWorn04, HunterNosra06, WuJind10, WuNegi03}. There are also some important works focusing on lower Phy functions, analyzing outage probability and diversity combining for general fading channels, e.g., \cite{KoAlouSimo00, JabiSzczBenj12}.

All works in those areas have, as far as we know, used less versatile, and more specialized, SNR fading channel models (such as Rayleigh and Nakagami-$m$ fading) compared to the proposed ME-distribution model in \cite{LarssonRasmSkog16a}, and developed further here. Other works do, as far as we know, also not consistently use (or allow for) expression  on the same form, such as the proposed ME-distribution form, from modeling the unprocessed wireless channel, via possible signal processing steps, as input to higher layer performance evaluation. Notably, many works do not aim for a complete bottom-up system analysis, but stop at evaluating, e.g., symbol-error-rate (SER), bit-error-rate (BER), or outage probability. Thus, in contrast to \cite{LarssonRasmSkog16a}, many prior performance analysis works are executed on a per-case basis, for specialized channels, and for specialized communication systems only. On the other hand, the ME-distribution has been explored in a wide range of fields, such as economics (risk and ruin probabilities), control theory (linear ordinary differential equations), queuing theory, see e.g. \cite{BeanFackTayl08, AsmussenBlad96}, and references therein. Despite this, the ME-distribution has, apart from \cite{LarssonRasmSkog16a, LarssonRasmSkog16b}, not been considered for wireless channel SNR (or effective channel) modeling, nor for performance evaluation for wireless systems with fading channels. Only two other works, \cite{MehdiLiefReec95,McMillan95}, which like \cite{LarssonRasmSkog16a, LarssonRasmSkog16b}, consider wireless systems, have used the ME-distribution. Yet, then only for queuing, but not for fading channel, analysis. However, some shortcomings of \cite{LarssonRasmSkog16a} is that it focused primarily on (H)ARQ system analysis, and may not have presented the ME-distribution approach in a sufficiently well-structured and detailed manner in order to appreciate the full value. Also, while \cite{LarssonRasmSkog16a} used the ME-distribution approach to analyze (H)ARQ, the studied cases are unnecessarily limited, and the derived performance expressions can be simplified further. The wider application of the ME-distribution approach, apart from analyzing single-/multiple-antenna and (H)ARQ systems is recognized, but not discussed in great detail. In \cite{ LarssonRasmSkog16b}, the ME-distribution approach was applied to yet another performance measure, the effective capacity, but the focus remained on (H)ARQ system analysis, not the method itself. Thus, the aim of this work is to emphasize and focus on the ME-distribution approach as a promising tool for performance analysis of wireless communication systems with fading channels.
When it comes to generalizing the ME-distribution form to 1D/2D-Gaussian-, and Rayleigh-, like forms, which is treated at the end of this paper, we are not aware of any such works at all.
The contributions are given next.

\subsection{Contributions}
\label{sec:Sec6d1d3}
This work offers several different contributions, at different levels. First, a number of results, expressions, and useful ME-properties that have, to our knowledge, not been reported in the literature are presented. Specifically, closed-form performance expressions, expressed in ME-distribution matrix parameters, are given for the following cases:  i) Rate adaptive transmission, Theorem~\ref{thm:Thm6d4} and \ref{thm:Thm6d5}, ii) Network coded bidirectional relaying (NCBR), Theorem~\ref{thm:Thm6d11}, iii) ARQ with identical independent distributed (iid) ME-distributed signal and interferers, Theorem~\ref{thm:Thm6d12} and with Sylvester's equation \ref{thm:Thm6d13}, iv) Differential binary phase shift keying (PSK) and frequency shift keying (FSK) with non-coherent detection, Theorem~\ref{thm:Thm6d14}, v) Binary PSK and FSK with coherent detection, Theorem~\ref{thm:Thm6d15}, vi) Coded transmission with independent fading, Theorem~\ref{thm:Thm6d16}. We also give closed-form throughput expressions for vii) ARQ, Theorem~\ref{thm:Thm6d8}, viii) Truncated-HARQ, Theorem~\ref{thm:Thm6d9}, ix) Persistent-HARQ, Theorem~\ref{thm:Thm6d10}, for ME-distributed effective channels which are on more general, but simpler, forms than in \cite{LarssonRasmSkog16a}, particularly for Truncated-HARQ. Corollary~\ref{cr:Crl6d4} handles, in contrast to \cite{LarssonRasmSkog16a}, also $N$-fold diversity.
Some additional new results are; x) The integral expressions in Theorem~\ref{thm:Thm6d1}, Corollary~\ref{cr:Crl6d6} and Lemma~\ref{lm:Lm6d4}, xi) The expression for the maximum of two ME-distributed r.v.s in Theorem~\ref{thm:Thm6d2}, xii) The integral expression in Theorem~\ref{thm:Thm6d3}.

Second, compared to \cite{LarssonRasmSkog16a}, and in addition to the above, the ME-distribution performance analysis framework is more clearly, and better, motivated, defined, and explored. It is made clear that the framework is generally applicable to various communication problems and performance measures. A more extensive background on well-known ME-distribution properties is given, and some new ones are introduced. We generalize the (H)ARQ cases in \cite{LarssonRasmSkog16a}, and  refine and simplify the performance expressions. 

Third, summarizing on a higher level, we consider, analyze and give closed-form performance expressions for many new communication cases never treated with the ME-distribution before, introduce the bi- (multivariate) ME-distribution for wireless system performance analysis, extend the use of the ME-distribution (density) to model discrete-time r.v. signals, and generalize the ME-distribution to Rayleigh-, univariate Gaussian, and bivariate Gaussian-like probability densities.

\subsection{Outline}
\label{sec:Sec6d1d4}
In Section \ref{sec:Sec6d2}, we review the ME-function and the ME-distribution. We motivate why  the ME-distribution is introduced for wireless fading channel modeling and system performance analysis in Section \ref{sec:Sec6d3}. In Section \ref{sec:Sec6d4}, we then structure the performance analysis framework, introduce the unprocessed ME-distributed SNR channel, and give some useful ME-properties. Performance evaluation, wrt relevant performance measures, for various communication cases, such as multiple antenna systems, rate-adaptive systems, modulation schemes, and (H)ARQ w/wo interference, takes place in Section \ref{sec:Sec6d5}. In Section \ref{sec:Sec6d6}, the notion of ME-distributed discrete-time signals is proposed, and in Section \ref{sec:Sec6d7}, generalizations of the ME-distribution is considered. The paper is summarized and concluded in Section \ref{sec:Sec6d8}.

\section{Preliminaries}
\label{sec:Sec6d2}
We start by introducing the notion and by reviewing some basic properties of the ME-function and the ME-distribution.

\subsection{Notation} 
\label{sec:Sec6d2d1}
We let $x(\cdot)$, $x$, $\mathbf{x}$, $\mathbf{X}$ denote polynomials, scalars, vectors, and matrices. The Kronecker-product, Kronecker-sum, convolution, $k$-fold convolution, and the matrix transpose, are indicated by $\oplus$, $\otimes$, $*$, $(\cdot)^{k\circledast}$, and $(\cdot)^\textrm{T}$, respectively. The expectation and the probability of a r.v. uses the notation $\mathbb{E}\{\cdot\}$ and $\mathbb{P}\{\cdot\}$. Special constants are the standard basis unit vector $\mathbf{e}_t$ (with a one at the $t$th position), the identity matrix $\mathbf{I}$, the shift matrix $\mathbf{S}$ (with all ones on the super-diagonal, otherwise all zero entries). The pdf, cdf, and the Laplace transform of a pdf of a r.v. $T$ are written as $f_T(t)$, $F_T(t)$ and $F(s)$. Effective channel parameters are indicated with a tilde, e.g. as $\mathbf{\tilde x}$.

\subsection{Matrix Exponential} 
\label{sec:Sec6d2d2}
Consider the square complex valued matrix $\mathbf{X}\in \mathbb{C}^{d\times d}$, where $d\in \mathbb{N}^+$ and $t$ is a scalar. Then, the matrix exponential can be defined as
\begin{align}
    \mathrm{e}^{t\mathbf{X}}
    \triangleq\sum_{k=0}^\infty\frac{(t\mathbf{X})^k}{k!},
    \label{eq:Eq6d1}
\end{align}
where $\mathbf{X}^0\triangleq\mathbf{I}$. The ME-function is, e.g., also possible to write as the limit
\begin{align}
    \mathrm{e}^{t\mathbf{X}}
    =\lim_{k\rightarrow \infty}\left(1+\frac{t\mathbf{X}}{k}\right)^k.
    \label{eq:Eq6d2}
\end{align}
Using the right-hand-side (RHS) of \eqref{eq:Eq6d1}, it is seen that the derivative of the ME-function is
\begin{align}
    \frac{\mathrm{d}}{\mathrm{d}t}\mathrm{e}^{t\mathbf{X}}
    =\mathbf{X}\mathrm{e}^{t\mathbf{X}}.
    \label{eq:Eq6d3}
\end{align}
Further, using the RHS of \eqref{eq:Eq6d1}, it is noted that $\mathbf{X}\mathrm{e}^{t\mathbf{X}}=\mathrm{e}^{t\mathbf{X}}\mathbf{X}$ commute. The integral of the ME-function, which is a scalar integral in $t$ with a matrix parameter $\mathbf{X}$, can be expressed as
\begin{align}
    \int_a^b\mathrm{e}^{t\mathbf{X}}\,\mathrm{d}t
    =\mathbf{X}^{-1}\left(\mathrm{e}^{t\mathbf{X}}-\mathbf{I}\right)|_a^b,
    \label{eq:Eq6d4}
\end{align}
given that $\mathbf{X}$ is non-singular. The integral \eqref{eq:Eq6d4} is easily proven by using the RHS of \eqref{eq:Eq6d1}. Note also that $\mathbf{X}^{-1}\mathrm{e}^{t\mathbf{X}}=\mathrm{e}^{t\mathbf{X}}\mathbf{X}^{-1}$.

A good overview of \emph{Nineteen dubious ways to compute the exponential of a matrix} is found in \cite{MolerLoan03}. The ME-function is also surveyed in \cite{Higham08}. More practically, the ME-function is, e.g., implemented in \textsc{Matlab}, Mathematica, and Maple. Note that in \textsc{Matlab}, the matrix- and scalar-exponential commands differ, and are $\mathrm{expm}(\mathbf{X})$, $\mathrm{exp}(x)$, respectively.

\subsection{Matrix Exponential Distribution} 
\label{sec:Sec6d2d3}
Next, we review the ME-distribution and several well-known properties which are often presented in the literature, \cite{BeanFackTayl08,AsmussenBlad96,Fackrell03,AsmussenOcin04,RuizCastro13}.

The cdf of a ME-distributed r.v. $T$ is commonly written as\footnote{Sometimes, the ME-distribution class includes a point-mass at zero. However, continuous real-world wireless channels, as considered here, do not have such property. Therefore, any point-mass at zero is omitted in the following.}
\begin{align}
    F_T(t)
    =1+\mathbf{x}\mathrm{e}^{t\mathbf{Y}}\mathbf{Y}^{-1}\mathbf{z}, t\geq0,
    \label{eq:Eq6d5}
\end{align}
where in general $\mathbf{x}\in \mathbb{C}^{1\times d}$, $\mathbf{Y}\in \mathbb{C}^{d\times d}$, and $\mathbf{z}\in \mathbb{C}^{d\times 1}$. The only requirement on $\mathbf{x}$, $\mathbf{Y}$, and $\mathbf{z}$ are that $F_T(t)$ corresponds to a cdf, i.e. non-decreasing, right-continuous, $\lim_{t\rightarrow 0}F_T(t)=0$, and $\lim_{t\rightarrow \infty}F_T(t)=1$.
From \eqref{eq:Eq6d5}, the pdf is found to be
\begin{align}
    f_T(t)
    =\mathbf{x}\mathrm{e}^{t\mathbf{Y}}\mathbf{z}, t\geq0,
    \label{eq:Eq6d6}
\end{align}
which is known to correspond to a sum of exponential-polynomial-trigonometric terms
\cite{AsmussenOcin04}. The moments, easily derived via partial integration, are
\begin{align}
    \mathbb{E}\{T^k\}=(-1)^{k+1}k!\mathbf{x}\mathbf{Y}^{-(k+1)}\mathbf{z}.
    \label{eq:Eq6d7}
\end{align}
Note that the form of \eqref{eq:Eq6d5} and \eqref{eq:Eq6d6} differ from the form of the scalar exponential distribution wrt negative signs.\footnote{The scalar exponential distribution is generally defined to have cdf on the form $F_T(t)=1-\mathrm{e}^{-y t}$, and pdf $f_T(t)=y\mathrm{e}^{-y t}$. The analogous ME-cdf form would be $F_T(t)=1-\mathbf{x}\mathrm{e}^{-t\mathbf{Y}}\mathbf{z}$, with ME-pdf $f_T(t)=\mathbf{x}\mathbf{Y}\mathrm{e}^{-t\mathbf{Y}}\mathbf{z}$. The pdf would, however, have a more complicated LT than \eqref{eq:Eq6d8}}
The Laplace-Stieltje's transform (LST) of \eqref{eq:Eq6d5}, corresponds to the Laplace transform (LT) of \eqref{eq:Eq6d6}, which is
\begin{align}
    F(s)
    =\mathbf{x}(s\mathbf{I}-\mathbf{Y})^{-1}\mathbf{z}.
    \label{eq:Eq6d8}
\end{align}
Eq. \eqref{eq:Eq6d8} is also known (as discussed below) to correspond to a ratio of two polynomials expressed in the Laplace variable $s$. It has been shown in \cite{AsmussenBlad96} that the class of rational LSTs is equivalent to the class of ME-distributions. It may be noted that phase-type distributions, introduced by Neuts and treated in detail in \cite{Neuts81}, have the same form as the ME-distribution, but phase-type distributions have certain parameter constraints and allows for a probabilistic interpretation \cite{Fackrell03}. Neuts \cite{Neuts81} states that the phase-type distribution is dense on $[0,\infty)$, and  Ruiz-Castro in \cite{RuizCastro13} extends this statement to the wider class of ME-distributions. However, Neuts \cite{Neuts81} also points out that \textit{the value of this theorem as an approximation theorem is largely illusory. No general approximation results are, in fact, known.}

As discussed after \eqref{eq:Eq6d5}, the ME-distribution allows for a flexible parameter choice of $\mathbf{x}$, $\mathbf{Y}$, and $\mathbf{z}$. A convenient and often occurring real-valued companion matrix-based parametrization, originally given in \cite{AsmussenBlad96}, has been treated in, e.g., \cite{AsmussenOcin04} and \cite[Theorem~2.1]{BeanFackTayl08}. With our notation, this translates to
\begin{align}
    \mathbf{Y}
    \triangleq
    \begin{bmatrix}
    0 & 1 & 0 & \cdots & 0 & 0 \\
    0 & 0 & 1 & \ddots & 0 & 0 \\
    0 & 0 & 0 & \ddots & 0 & 0 \\
    \vdots & \ddots & \ddots & \ddots & \ddots & \ddots \\
    0 & 0 & 0 & \cdots & 0 & 1 \\
    -y_1 & -y_2 & -y_3 & \cdots & -y_{d-1} & -y_{d} \\
    \end{bmatrix},
    \label{eq:Eq6d9}
\end{align}
or equivalently $\mathbf{Y}=\mathbf{S}-\mathbf{z}\mathbf{y}$, where $\mathbf{S}$ is a shift matrix of appropriate dimension, and
\begin{align}
    \mathbf{x}
    &\triangleq[x_1 \ x_2 \ \ldots \ x_{d-1} \ x_{d}]\in\mathbb{R}^{1 \times d}, \label{eq:Eq6d10}\\
    \mathbf{y}
    &\triangleq[y_1 \ y_2 \ \ldots \ y_{d-1} \ y_{d}]\in\mathbb{R}^{1 \times d}, \label{eq:Eq6d11}\\
    \mathbf{z}
    &\triangleq[0 \ 0 \ \ldots \ 0 \ 1]^\textrm{T}\in\mathbb{R}^{d \times 1},
    \label{eq:Eq6d12}
\end{align}
with the corresponding parametrization of the rational LT,
\begin{align}
    F(s)
    &=\frac{x(s)}{y(s)},
    \label{eq:Eq6d13}\\
    x(s)
    &\triangleq x_{d}s^{d-1} + x_{d-1}s^{d-2}+ \ldots +x_{2}s^{1}+x_{1},
    \label{eq:Eq6d14}\\
    y(s)
    &\triangleq s^d+y_{d}s^{d-1} + y_{d-1}s^{d-2}+ \ldots +y_{2}s^{1}+y_{1}.
    \label{eq:Eq6d15}
\end{align}

Using the final-, and initial-, value theorem, it can be shown, as e.g. in \cite{LarssonRasmSkog16a}, that necessary, but not sufficient, conditions for $F(s)$ to correspond to a pdf $f_Z(z)$, without a point mass at zero, are $\deg(x(s))<\deg(y(s))$, and $x_1=y_1$.

Two works that inspired us in \cite{LarssonRasmSkog16a} to consider the closure of convolutions for the ME-distribution class are \cite{RuizCastro13} and \cite{Neuts81}. In fact, it is well-known that the class of ME-distributions is closed under many different operations, such as the convolution, maximum and minimum of two r.v.s. An excellent overview of various closure properties for the ME-distribution class is found in \cite{RuizCastro13}. For phase-type distributions, which have the same form as ME-distributions, closure properties are given in \cite[Section~2.2]{Neuts81}. We review those three cases, convolution, maximum and minimum below, and refer the interested reader to the literature for further details.\footnote{Convolution (e.g. for MRC, truncated-HARQ, and SDC) and the maximum operator (e.g. for SDC) were used for signal processing and performance analysis in \cite{LarssonRasmSkog16a}, which motivates reviewing those properties here.}

\begin{proposition}
\label{pr:Pr6d1}
(Convolution of two ME-distributed r.v.s. \cite[Proposition 3.1]{RuizCastro13})
Let the r.v.s. $T_j, j=\{1,2\}$ have pdfs $f_{T}^{(j)}(t)=\mathbf{x}_j\mathrm{e}^{t\mathbf{Y}_j}\mathbf{z}_j$. Then, $T=T_1+T_2$ has the pdf
\begin{align}
    f_T(t)
    &=\mathbf{x} \mathrm{e}^{t\mathbf{ Y}}\mathbf{z},
    \label{eq:Eq6d16}
\end{align}
where
\begin{align}
    \mathbf{x}
    &=\begin{bmatrix}
        \mathbf{x}_1 & \mathbf{0}
    \end{bmatrix},
    \label{eq:Eq6d17}\\
    \mathbf{Y}
    &=\begin{bmatrix}
        \mathbf{Y}_1 &\mathbf{z}_1\mathbf{x}_2\\
        \mathbf{0} & \mathbf{Y}_2
    \end{bmatrix}
    \label{eq:Eq6d18},\\
    \mathbf{z}
    &=\begin{bmatrix}
        \mathbf{0} \\
        \mathbf{z}_2
    \end{bmatrix}
    \label{eq:Eq6d19}.
\end{align}
In the above, and henceforth, vectors/matrices indicated as $\mathbf{0}$ are for notational convenience, with appropriate dimensions given by the problem.
\end{proposition}
\begin{IEEEproof}
The proof is reviewed here for completeness.
\begin{align}
&\mathbf{x} _1\mathrm{e}^{t\mathbf{Y}_1}\mathbf{z_1}*\mathbf{x} _2\mathrm{e}^{t\mathbf{Y}_2}\mathbf{z_2}\notag\\
&=\mathcal{L}^{-1}_t\left \{ \mathbf{x} _1(s\mathbf{I}-\mathbf{Y}_1)^{-1}\mathbf{z_1} \cdot  \mathbf{x} _2(s\mathbf{I}-\mathbf{Y}_2)^{-1}\mathbf{z_2}\right \}\notag\\
&=
\mathcal{L}^{-1}_t\left \{
\begin{bmatrix}
\mathbf{x}_1 & \mathbf{0}\\
\end{bmatrix}
\begin{bmatrix}
\mathbf{Y}_1-s\mathbf{I} & \mathbf{z}_1\mathbf{x}_2  \\
0 & \mathbf{Y}_2-s\mathbf{I}\\
\end{bmatrix}^{-1}
\begin{bmatrix}
\mathbf{0} \\ \mathbf{z}_2\\
\end{bmatrix}
\right \}\notag\\
&=\begin{bmatrix}
\mathbf{x}_1 & \mathbf{0}\\
\end{bmatrix}
\mathrm{e}^{t\begin{bmatrix} \mathbf{Y}_1 &\mathbf{z}_1\mathbf{x}_2 \\ \mathbf{0}  & \mathbf{Y}_2\\ \end{bmatrix}}
\begin{bmatrix}
\mathbf{0} \\ \mathbf{z}_2\\
\end{bmatrix}.\notag
\end{align}
Proof by \cite[Proposition 3.1]{RuizCastro13}.
\end{IEEEproof}

\begin{proposition}
\label{pr:Pr6d2}
(Maximum of two ME-distributed r.v.s.  \cite[Proposition~3.5]{RuizCastro13},  \cite{BeanFackTayl08}).
Consider the ME-distributions $F_{T}^{(j)}(t)=1+\mathbf{x}_j\mathrm{e}^{t\mathbf{Y}_j}\mathbf{Y}_j^{-1}\mathbf{z}_j, t\geq0,j\in\{1,2\}$. Then, $T=\max\{T_1,T_2\}$ has the ME-distribution
\begin{align}
    F_{T}(t)=1+\mathbf{x}\mathrm{e}^{t\mathbf{Y}}\mathbf{Y}^{-1}\mathbf{z},
    \label{eq:Eq6d20}
\end{align}
where
\begin{align}
    \mathbf{x}
    &=
    \begin{bmatrix}
        \mathbf{x}_1  \! \! \otimes \mathbf{x}_2 & \mathbf{x}_1 & \mathbf{x}_2
    \end{bmatrix},
    \label{eq:Eq6d21}\\
    \mathbf{Y}
    &=
    \begin{bmatrix}
        \mathbf{Y}_1 \! \! \oplus \mathbf{Y}_2 & \mathbf{0} & \mathbf{0}\\
        \mathbf{0} & \mathbf{Y}_1 & \mathbf{0}\\
        \mathbf{0} & \mathbf{0} & \mathbf{Y}_2\\
    \end{bmatrix},
    \label{eq:Eq6d22}\\
    \mathbf{z}
    &=
    \begin{bmatrix}
    (\mathbf{Y}_1^{-1}\oplus\mathbf{Y}_2^{-1})(\mathbf{z}_1 \! \! \otimes \mathbf{z}_2)\\
    \mathbf{z}_1\\
    \mathbf{z}_2
    \end{bmatrix}.
    \label{eq:Eq6d23}
\end{align}
\end{proposition}
\begin{IEEEproof}
Proof by \cite[Proposition~3.5]{RuizCastro13}.
\end{IEEEproof}

\begin{proposition}
\label{pr:Pr6d3}
(Minimum of two ME-distributed r.v.s \cite[Proposition~3.6]{RuizCastro13}, \cite{BeanFackTayl08}).
Consider the ME-distributions $F_{T}^{(j)}(t)=1+\mathbf{x}_j\mathrm{e}^{t\mathbf{Y}_j}\mathbf{Y}_j^{-1}\mathbf{z}_j, t\geq0,j\in\{1,2\}$. Then, $T=\min\{T_1,T_2\}$ has the ME-distribution
\begin{align}
    F_{T}(t)
    =1+\mathbf{x}\mathrm{e}^{t\mathbf{Y}}\mathbf{Y}^{-1}\mathbf{z},
    \label{eq:Eq6d24}
\end{align}
where
\begin{align}
    \mathbf{x}
    &=\mathbf{x}_1 \!\! \otimes \mathbf{x}_2,
    \label{eq:Eq6d25}\\
    \mathbf{Y}
    &=\mathbf{Y}_1 \! \! \oplus \mathbf{Y}_2,
    \label{eq:Eq6d26}\\
    \mathbf{z}
    &=-(\mathbf{Y}_1^{-1} \! \! \oplus \mathbf{Y}_2^{-1})(\mathbf{z}_1 \otimes \mathbf{z}_2).
    \label{eq:Eq6d27}
\end{align}
\end{proposition}
\begin{IEEEproof}
Proof by \cite[Proposition 3.6]{RuizCastro13}.
\end{IEEEproof}

Good surveys of the class of ME-distribution, its use, applications and properties, are e.g. found in, \cite{BeanFackTayl08,AsmussenBlad96,AsmussenOcin04,RuizCastro13}.
Finally note also that the ME-distribution has the same analytical form as the phase type-distribution \cite{Neuts81}, but the parameters in $\mathbf{x}, \mathbf{Y}$, and $\mathbf{z}$ are, in contrast to the phase type-distribution, not restricted to have a probabilistic interpretation. Hence, much of the known properties in the literature of phase type-distributions carry over to the ME-distribution class.

\begin{table*}[t]
\normalsize
\begin{center}
  \begin{tabular}{|c|c|c|c|c|}
    \hline
    \begin{tabular}{c}
	\textbf{Bivariate pdf}\\
$f_{h}(h)=$
    \end{tabular}
	&
    \begin{tabular}{c}
	    \textbf{Amplitude pdf}\\
$f_{|h|}(|h|)=$
    \end{tabular}
    &
    \begin{tabular}{c}
	     \textbf{SNR pdf}\\
$f_{G}(g)=$
    \end{tabular}
    &
    \begin{tabular}{c}
	     \textbf{LT of SNR pdf} \\
$F(s)=$
    \end{tabular}
    &
    \begin{tabular}{c}
	       \textbf{SNR cdf} \\
$F_G(g)=$
    \end{tabular}
    \\
    \hline
    \begin{tabular}{c}
    Bivariate Gaussian distr. \\ 
$\frac{1}{\pi \Omega}\mathrm{e}^{-(h_\textrm{r}^2+h_\textrm{i}^2)/\Omega}$
    \end{tabular}
    &
    \begin{tabular}{c}
    Rayleigh distr. \\ $\frac{2|h|}{\Omega}\mathrm{e}^{-|h|^2/\Omega}$
    \end{tabular}
    &
    \begin{tabular}{c}
    Exponentially distr. \\ 
	$\frac{1}{S}\mathrm{e}^{-g/S}$
    \end{tabular}
    &
    $\frac{1}{1+sS}$
    &
    $1-\mathrm{e}^{-g/S}$ \\
    \hline
    \begin{tabular}{c}
       --
    \end{tabular}
    &
    \begin{tabular}{c}
        Nakagami-$m$ distr. \\
        $\frac{2m^m|h|^{2m-1}}{\Gamma(m)\Omega^m}\mathrm{e}^{-m|h|^2/\Omega}$
    \end{tabular}
    &
    \begin{tabular}{c}
        Gamma distr. \\
        $\frac{m^mg^{m-1}}{\Gamma(m)S^m}\mathrm{e}^{-mg/S}$
    \end{tabular}
    &
    $\left(\frac{1}{1+sS/m}\right)^{m}$
    &
    $\frac{1}{\Gamma(m)}\gamma(m,mg/S)$ \\
    \hline
    \begin{tabular}{c}
        (Unnamed distr.) \\
        $\frac{1}{\pi}\mathbf{p}_{|h|}\mathrm{e}^{(h_\textrm{r}^2+h_\textrm{i}^2)\mathbf{Q}_{|h|}}\mathbf{r}$
    \end{tabular}
    &
    \begin{tabular}{c}
        (Unnamed distr.) \\
        $2|h|\mathbf{p}_{|h|}\mathrm{e}^{|h|^2\mathbf{Q}_{|h|}}\mathbf{r}$
    \end{tabular}
    &
    \begin{tabular}{c}
        ME-distr. \\
        $\mathbf{p}\mathrm{e}^{g\mathbf{Q}}\mathbf{r}$
    \end{tabular}
    &
    $\frac{p(s)}{q(s)}$
    &
    \begin{tabular}{c}
        $1+\mathbf{p}\mathrm{e}^{g\mathbf{Q}}\mathbf{Q}^{-1}\mathbf{r}$ \\
    \end{tabular}\\
    \hline
  \end{tabular}
\caption{Comparison of pdfs (and cdfs) for unprocessed fading wireless channels SNRs. The following notion is used: The instantaneous SNR is $g\triangleq |h|^2P/\sigma^2$, where $|h|$ is the channel amplitude gain, $P$ is the received power, $\sigma^2$ is the receiver noise power. The mean SNR is $S\triangleq \mathbb{E}\{g\}$. The complex amplitude gain is $h\triangleq h_\textrm{r}+ih_\textrm{i}$, and $\Omega\triangleq\mathbb{E}\{|h|\}$.}
\label{tab:Tab6d1}
\end{center}
\end{table*}

\section{ME-distribution in Wireless Communications}
\label{sec:Sec6d3}
In this section, we present some observations\footnote{Some observations where given already in \cite{LarssonRasmSkog16a}, but here, we give a more structured and detailed treatment.} that motivate us to consider the ME-distribution as a basis for performance analysis of wireless communication systems with fading.

\subsection{Unprocessed Wireless Channel SNR}
\label{sec:Sec6d3d1}
We start with the following example.
\begin{example}
\label{ex:Ex6d1}
The Nakagami-$m$ channel is a relatively versatile channel model, with Rayleigh fading as a special case when the nakagami-$m$ parameter $m^\textrm{N}=1$, and no fading when $m^\textrm{N}\rightarrow \infty$. The Laplace transform for the (gamma-distributed) SNR pdf is $F(s)={1}/{(1+sS/m^\textrm{N})^{m^\textrm{N}}}$,
where  $S$ denotes the mean SNR of the channel, and $m^\textrm{N}\geq \frac{1}{2}$. For the special case $m^\textrm{N}\in \mathbb{N}^+$, $F(s)$ is on a rational LT form, and thus belongs to the ME-distribution class.\footnote{This is more clearly seen when writing the LT
as $F(s)={(1/\tilde {S})^{m^\textrm{N}}}/(1/\tilde {S}+s)^{m^\textrm{N}} \!$, and then on a rational polynomial form as in \eqref{eq:Eq6d13}.}
\end{example}
To show the versatility of the ME-distribution, going beyond the simplicity of Ex. \ref{ex:Ex6d1}, consider the following due to \cite{AsmussenBlad96}.
\begin{example}
\label{ex:Ex6d2}
The pdf $f_T(t)=(1+7^{-2})\left(1-\cos(7t)\right)\mathrm{e}^{-t}$ has the rational LT $F(s)=50/(s^3+3s^2+52s+50)$ and is ME-distributed. With the oscillatory decaying nature for the pdf, it is noted that the ME-distribution-form can also capture relatively complex behaviors already with low degree LTs.
\end{example}

\begin{example}
\label{ex:Ex6d3}
In \cite[Sec. IV.F.4]{LarssonRasmSkog16a}, we also proposed modeling the unprocessed channel SNR pdf with a ME-density, $f_G(g)=\mathbf{p}\mathrm{e}^{g \mathbf{Q}}\mathbf{r}$, with a corresponding rational LT $F(s)=p(s)/q(s)$ where the $p(s)$ and $q(s)$ are polynomials. The motivation is that this general LT form have the potential to model (exactly, or approximately) the statistical characteristics of many different fading channel SNRs. This is so since the ME-distribution is dense on $(0,\infty]$, \cite{RuizCastro13, Neuts81}.
\end{example}

Not all SNR pdfs of well-known wireless channel models are in the ME-distribution class, and therefore do not have a rational LT. Examples of non-rational functions are, e.g., $1/\sqrt{1+s}$ and $\mathrm{e}^{-s}$, which require polynomials of infinite degrees. Other examples are Rician and log-normal fading, which do not have ME-distributed SNRs, and thus also no rational LTs. Hence, using the ME-distribution as an approximation to model wireless channel SNR fading is an interesting option.
The idea of approximating given pdfs, having non-rational LTs, with ME-pdfs, having rational LTs, has been studied extensively in the literature. An excellent overview of state-of-the-art techniques, and review of related works, for approximating a pdf with phase type- or ME-distributions is given in \cite{Fackrell03}. A detailed review is outside the scope of this work, but the main principles are generally built on norm minimization, either wrt a pdf or its LT. Not only unprocessed wireless channel SNR pdfs can be approximated with pdfs on ME-distribution-form, but also, if desired, the SNR pdfs after SNR processing.
More explicitly, we showed in \cite{LarssonRasmSkog16a} that the mean number of transmissions (and hence the throughput) of persistent-IR operating in a Rayleigh fading channel could be arbitrarily well approximated with a ME-distribution-form using a truncated continued fraction form. Moreover, in \cite{LarssonRasmSkog16a} we mentioned, the possibility of using a continuous least squares approximation in the pdf-domain. However, this is hard to solve explicitly. An alternative idea, also proposed in  \cite{LarssonRasmSkog16a},  was to approximate pdfs, with non-rational LTs, by using a rational Pad\'{e} approximation in the LT domain. A new idea, briefly mentioned in \cite{LarssonRasmSkog16b}, is to consider fitting a ME-distribution, or -density, directly to measured fading channel gains. In this way, the channel model would be formulated directly as a ME-distribution. We leave this interesting idea for future research.

In Tab. \ref{tab:Tab6d1}, we illustrate mathematical expressions for the proposed ME-distributed wireless channel SNR alongside with Gamma-distributed fading, and exponentially distributed fading SNR. The pdfs, the LTs of the pdfs, and the cdfs, are shown from the middle to the rightmost columns.
In the amplitude domain, the familiar Rayleigh and Nakagami-$m$ pdfs are shown. Through variable substitution, we also introduce the corresponding (hitherto unnamed) amplitude pdf for the ME-distribution case (second left column). It is well-known that the Rayleigh distribution can be derived from the bivariate Gaussian distribution. In an analogous manner, using the same variable substitutions, we generalize the ME-distribution to a (hitherto unnamed) bivariate pdf expressed in ME-distribution matrix-, and vector-, parameters (left column). Note that this generalized bivariate-pdf degenerates to the bivariate Gaussian pdf for the scalar case. The ME-distribution generalization are treated further in Section \ref{sec:Sec6d7}.

\subsection{Processing and Effective Channel SNR}
\label{sec:Sec6d3d2}

Below, we illustrate the connection between the ME-distribution and signal processing in wireless communication with four motivating examples. The first two, receiver-MRC and -SDC, illustrate diversity-based signal processing facilitated by antenna hardware capability only. The other two, OSTBC and spatially-multiplexed zero-forcing MIMO (ZF-MIMO), shows signal processing at both transmitter- and receiver-side which also involves special signal-design and -processing. 

\begin{example}
\label{ex:Ex6d4}
(MRC)
Consider a receiver with $N_\textrm{rx}$ antennas, exponentially distributed SNRs $Z_n, n\in\{1,2,\ldots N_\textrm{rx}\}$, each with mean SNR $S_n$. The LT of the MRC SNR, $Z=\sum_{n=1}^{N_\textrm{rx}}Z_n$, is then $F(s)={1}/{\prod_{n=1}^{N_\textrm{rx}}(1+sS_n)}$,
which is on a rational form, and hence correspond to a ME-distribution.
\end{example}

In SDC, the signal with the greatest SNR is selected. The following example on SDC is considered in \cite{LarssonRasmSkog16a}.
\begin{example}
\label{ex:Ex6d5}
(SDC)
Consider $N$-fold SDC, with effective SNR $Z=\max(Z_1,Z_2,\ldots Z_N)$, where the SNRs $Z_n,n\in\{1,2,\ldots N\}$ are iid exponentially distributed, with mean SNR $S$, and cdfs $F_Z(z)=1-\mathrm{e}^{-zS^{-1}}$. For this case, the pdf is $f_Z(z)=\frac{\mathrm{d}}{\mathrm{d}z}F_Z(z)^{N}=Nf_Z(z)F_Z(z)^{N-1}=NS^{-1} \mathrm{e}^{-zS^{-1}}(1-\mathrm{e}^{-zS^{-1}})^{N-1}$.
The Laplace transform of $f_Z(z)$ can be written as
$F(s)={N!}/{\prod_{n=1}^N(n+sS)}$ (the proof is given in Appendix), which is on rational form and also represents a special case of a ME-distribution. Interestingly, note that the product-form of $F(s)$ can be interpreted as $N$ convolutions, corresponding to a  summation of $N$ iid exponentially distributed r.v.s with SNRs $S/n, \ n\in\{1,2,\ldots N\}$.
\end{example}

\begin{example}
\label{ex:Ex6d6}
(OSTBC+MRC)
We now consider a multi-antenna OSTBC+MRC channel, with $N_\textrm{tx}$ ($N_\textrm{rx}$) transmit (receive) antennas, diversity order $N=N_\textrm{rx}N_\textrm{tx}$, and OSTBC code rate $R_\textrm{stc}$. We have previously discussed this channel in \cite{LarssonRasmSkog14b} and found that the LT is $F(s)={1}/{(1+sS/R_\textrm{stc}N_\textrm{tx})^N}$. 
Hence, as this is on a rational polynomial form, the effective SNR for OSTBC+MRC in Rayleigh fading channel is also a special case of a ME-distribution.
\end{example}

Consider, e.g., the mapping $Z=\ln(1+G)$, where $G$ is a ME-distributed SNR r.v., and $Z$ is the mutual information (MI) for a Gaussian distributed signal in additive white Gaussian noise (AWGN). With this mapping, the effective channel MI r.v. has a non-rational LT and is not ME-distributed. For such cases, a ME-distribution can approximate the distribution of the MI r.v., \cite{LarssonRasmSkog16a}.

\section{Performance Analysis Framework, ME-distr. wireless channel SNR model, Tools}
\label{sec:Sec6d4}
Encouraged by the observations reviewed in Section \ref{sec:Sec6d3}, we now formalize the overall performance analysis framework, generalize the ME-distributed wireless channel SNR model, and consider some new mathematical tools.

\subsection{Performance Analysis Framework}
\label{sec:Sec6d4d1}
\begin{figure}[t]
 \centering
 \vspace{+.1 cm}
 \includegraphics[width=13cm]{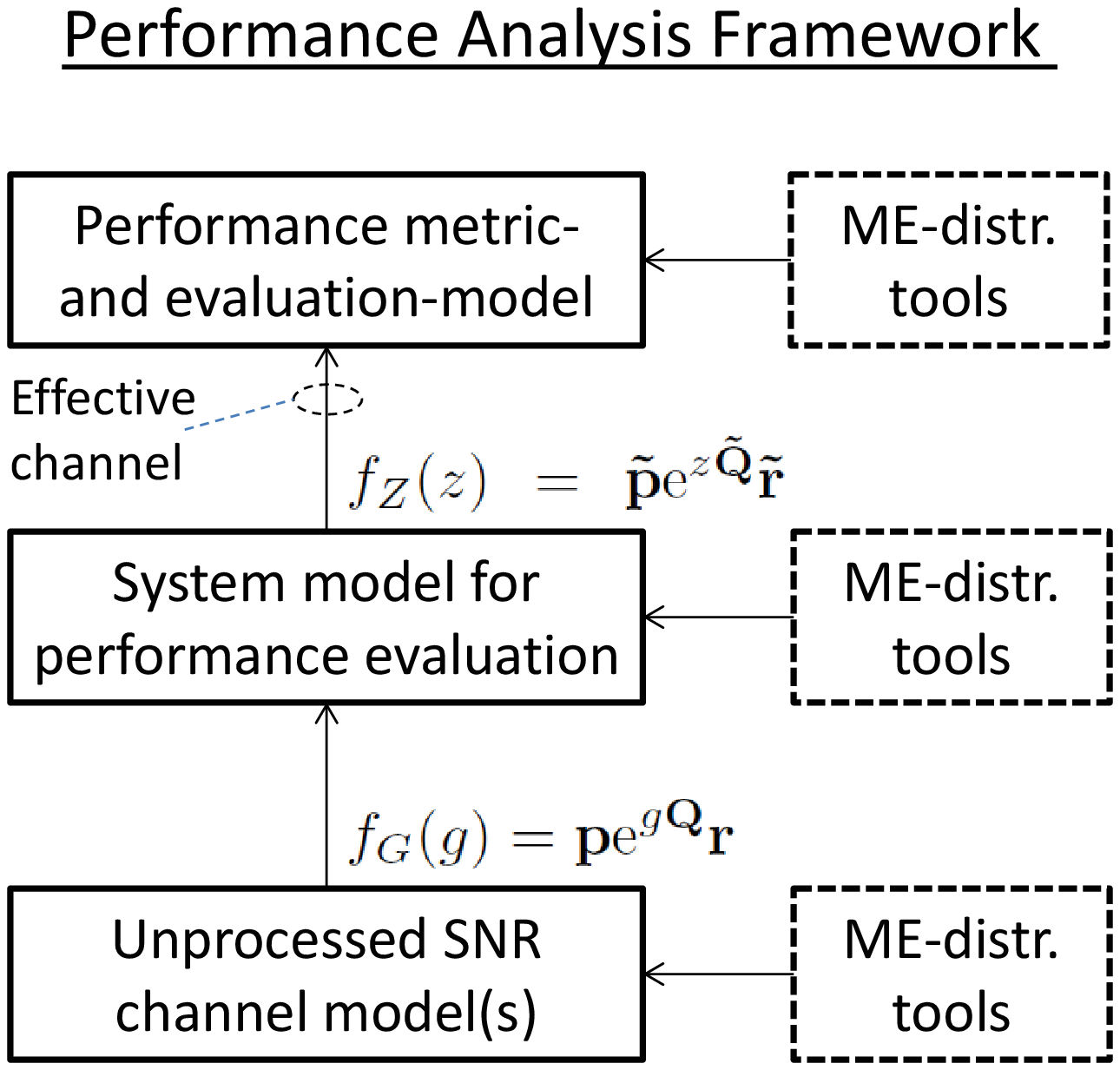}
 \vspace{-3.2cm}
 \caption{Performance analysis framework.}
 \label{fig:PerfFrame}
 \vspace{-0.3cm}
\end{figure}

\begin{figure}[t]
 \centering
 \vspace{+.1 cm}
 \includegraphics[width=9cm]{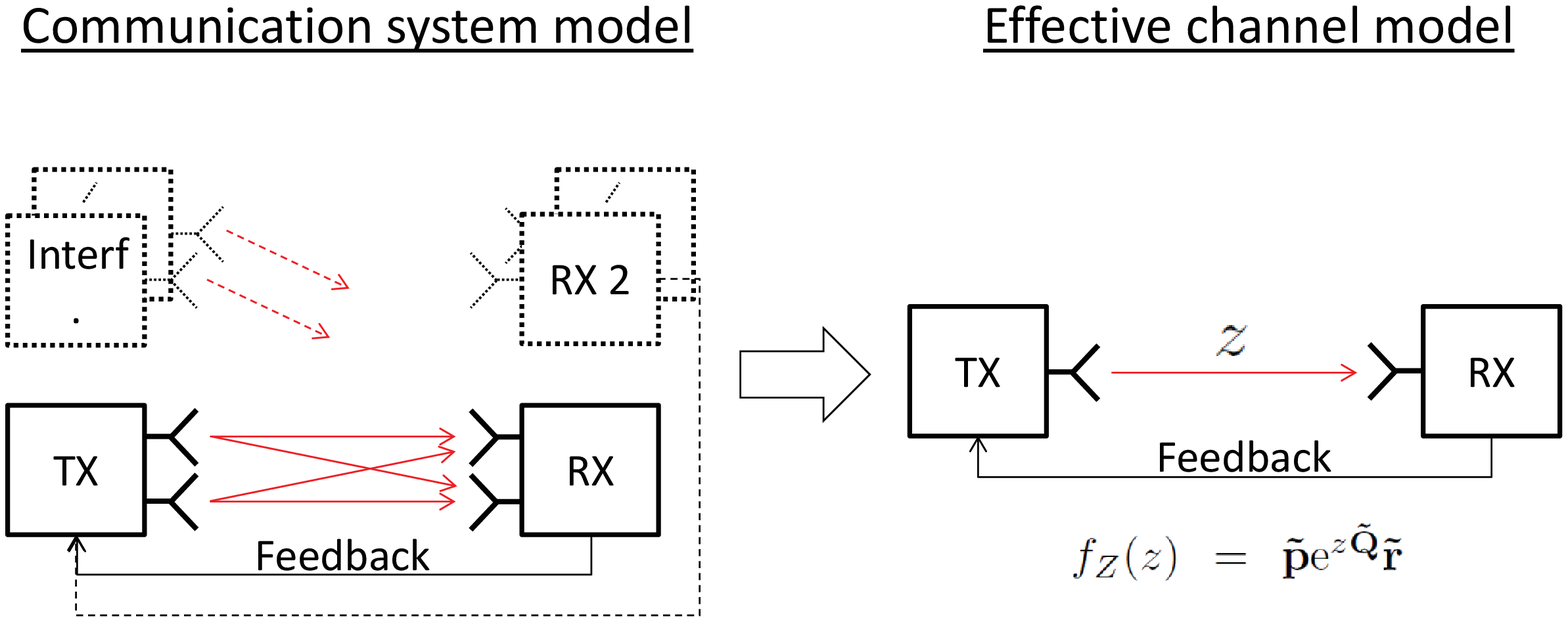}
 \vspace{-3.2cm}
 \caption{Equivalent effective channel model.}
 \label{fig:SystemModel}
 \vspace{-0.0cm}
\end{figure}

The performance analysis framework\footnote{This framework was used in \cite{LarssonRasmSkog16a}, but not explicitly formalized, not sufficiently organized, and not generalized as to the extent here. With some guidance to various sections in \cite{LarssonRasmSkog16a}, we hope to make this clearer.
A full bottom-up perspective was, e.g., considered in \cite[Sec. IV.F.2-4]{LarssonRasmSkog16a}. There, the unprocessed wireless channel SNR r.v. passed through the system level (diversity) signal processing, and performance metrics, such as throughput, mean number of transmissions, and loss rate probability, where determined and expressed directly in the unprocessed wireless channel SNR parameters, $\mathbf{p}, \mathbf{Q}$ (or equivalently as $\mathbf{p}, \mathbf{q}$).
The top-level performance analysis view, taking only the effective channel r.v. (albeit then assuming $Z=G$) as input, was studied in \cite[Sec. IV.D]{LarssonRasmSkog16a}. The middle level, discussing the effective channel SNR due to various signal processing schemes, communication schemes, hardware configurations ($N_\textrm{rx}$-MRC, $N_\textrm{rx}\times N_\textrm{tx}$ OSTBC-MRC, $2\times 1$ Alamouti-TX diversity, $N_\textrm{rx}$-SDC), were addressed in [Sec IV.D]. The bottom level, with unprocessed fading channel SNR, was  discussed in \cite[Sec. IV.D]{LarssonRasmSkog16a} for Nakagami-$m$ and Rayleigh fading as instances of ME-distributed channels, whereas the fully general ME-distributed channel, with $F(s)=p(s)/q(s)$ were handled in \cite[Sec. IV.F.2-4]{LarssonRasmSkog16a}. In \cite{LarssonRasmSkog16b}, we used the ME-distribution framework for yet another performance measure, the  logarithmic moment-generating-function (log-mgf) (the effective capacity) for (H)ARQ systems.}  is shown in Fig.~\ref{fig:PerfFrame}. At the bottom level, the unprocessed SNR channel r.v. $G$ is modeled with pdf $f_G(g)=\mathbf{ p}\mathrm{e}^{g \mathbf{Q}}\mathbf{r}$. At the middle level, a performance analysis system model, accounting for various processing steps in the communication system model of interest, translates the unprocessed channel SNR r.v. $G$ into an effective channel r.v. $Z$ with pdf $f_Z(z)=\mathbf{\tilde p}\mathrm{e}^{z \mathbf{\tilde Q}}\mathbf{\tilde r}$. At the top level, a performance expression, for some metric of choice, is derived and expressed in the unprocessed channel SNR parameters, $\mathbf{p}, \mathbf{Q}$ (and $\mathbf{r}$). Alternatively, if only the performance evaluation step is of interest, the performance metric may be directly evaluated and expressed in the effective channel parameters, $\mathbf{\tilde p}, \mathbf{\tilde Q}$ (and $\mathbf{\tilde r}$). This system modeling abstraction is reflected in Fig.~\ref{fig:SystemModel}, where a more complex communication system model (on the left-hand side (LHS)) has a corresponding effective channel model (on the RHS). The studied communication system may, e.g., range from physical layer performance evaluation of a SISO-system to link- or network-layer performance evaluation of a multi-node MIMO-system. The only requirement is that the effective channel is, or can be approximated as, ME-distributed. If no particular processing of the received signal(s) takes place, we simply set $f_Z(z)=f_G(g)$. Various mathematical tools, suitable for the ME-distribution, can be used at the different performance analysis levels\footnote{Various ME-distribution tools were introduced and used in \cite{LarssonRasmSkog16a}, such as the closure of the convolution (for ME-distributions with different parameters, MRC, multiple transmissions) \cite[Sec. IV.F1]{LarssonRasmSkog16a}, k-fold convolution \cite[Sec. IV.F1]{LarssonRasmSkog16a}, integration (for outage probability) \cite[Sec. IV.F5]{LarssonRasmSkog16a}, derivation (for performance optimization)  \cite[Sec. V]{LarssonRasmSkog16a}, the notion of ME-distribution approximations \cite[Sec. IV.D]{LarssonRasmSkog16a}, and implicitly also the  closure of the maximum operation (for SDC)  \cite[Sec. IV.E]{LarssonRasmSkog16a}.}, e.g. to reflect the communication system operation at the middle level, and enable computation of the performance metric at the top level. It should be emphasized that the analysis framework is exemplified here for the univariate ME-distribution case, but may be generalized to a multivariate ME-distribution case, as e.g. in Section \ref{sec:Sec6d5d8}. In Tab. \ref{tab:Tab6d2}, we indicate that the analysis approach is applicable on many different levels of system complexity, and with different performance metrics. Next, we consider the bottom-level, the unprocessed ME-distributed SNR channel model, introduce some new tools, and subsequently work upwards.

\subsection{Unprocessed ME-distributed Wireless Channel SNR}
\label{sec:Sec6d4d2}
In \cite{LarssonRasmSkog16a}, we assumed that the LT of the wireless channel SNR pdf was on the form $F(s)=p(s)/q(s)$, i.e. a ratio of a numerator polynomial and a denominator polynomial. This form agrees with the companion-form, \eqref{eq:Eq6d9}-\eqref{eq:Eq6d15}, introduced in \cite{AsmussenBlad96}. However, it is sometimes convenient to express the LT of the fading channel SNR in a product form, such as for Nakagami-$m$ channel fading \eqref{eq:Eq6d28}.\footnote{In Section \ref{sec:Sec6d3}, it was also seen that Rayleigh fading channel with OSTBC, MRC and SDC, are readily expressed in a product-form. In Section \ref{sec:Sec6d5d8}, we will see that sum-interference SNR has a rational LT on a product-form.}
To handle such cases, but also the polynomial case, we consider a more general product-polynomial-form in the following corollary.
\begin{corollary}
\label{cr:Crl6d1}
(Unprocessed wireless channel SNR pdf with rational LT on polynomial-product-form).
Let the LT of a unprocessed ME-distributed wireless channel SNR pdf $f_G(g)$ have the form
\begin{align}
    F(s)
    &=\frac{\prod_{j=1}^J \ud{p}_j(s)}{\prod_{j=1}^J \ud{q}_j(s)},
    \label{eq:Eq6d28}\\
    \ud{p}_j(s)
    &\triangleq \ud{p}_{d_j,j}s^{d_j-1} + \ud{p}_{d_j-1,j}s^{d_j-2}+ \ldots +\ud{p}_{1,j},
    \label{eq:Eq6d29}\\
    \ud{q}_j(s)
    &\triangleq s^{d_j}+\ud{q}_{d_j,k}s^{d_j-1} + \ud{q}_{d_j-1,j}s^{d_j-2}+ \ldots +\ud{q}_{1,j}.
    \label{eq:Eq6d30}
\end{align}
Then, the ME-distribution pdf of the unprocessed SNR is on the form
\begin{align}
    f_G(g)
    =\mathbf{p}\mathrm{e}^{g\mathbf{Q}}\mathbf{r}, z\geq0,
    \label{eq:Eq6d31}
\end{align}
where
\begin{align}
    \mathbf{Q}
    &\triangleq
    \begin{bmatrix}
    \mathbf{\ud{Q}}_1 & \mathbf{\ud{P}}_2 & \mathbf{0} & \cdots & \mathbf{0} & \mathbf{0} \\
    \mathbf{0} & \mathbf{\ud{Q}}_2 & \mathbf{\ud{P}}_3 & \cdots & \mathbf{0} & \mathbf{0} \\
    \mathbf{0} & \mathbf{0} & \mathbf{\ud{Q}}_3 & \ddots & \mathbf{0} & \mathbf{0} \\
    \vdots & \vdots & \vdots & \ddots & \vdots & \vdots \\
    \mathbf{0} & \mathbf{0} & \mathbf{0} & \cdots & \mathbf{\ud{Q}}_{J-1} & \mathbf{\ud{P}}_J \\
    \mathbf{0} & \mathbf{0} & \mathbf{0}  & \cdots & \mathbf{0} & \mathbf{\ud{Q}}_J \\
    \end{bmatrix} \in \mathbb{R}^{d \times d},\\
    \mathbf{p}
    &\triangleq [\mathbf{p}_1 \ \mathbf{0}\ \cdots \ \mathbf{0}] \in \mathbb{R}^{1 \times d},\\
    \mathbf{r}
    &\triangleq\mathbf{e}_d \in \mathbb{R}^{d\times 1},\\
    d
    &\triangleq\sum_{j=1}^{J} \ud{d}_j,\\
    \mathbf{\ud{P}}_j
    &\triangleq\mathbf{\ud{r}}_{j-1}\mathbf{\ud{p}}_j  \in \mathbb{R}^{\ud{d}_{j-1}, \times \ud{d}_j},\ j\in\{2,3,\ldots J\},\\
    \mathbf{\ud{Q}}_j
   &\triangleq\mathbf{S}_{d_j}-\mathbf{\ud{r}}_{j}\mathbf{\ud{q}}_j \in \mathbb{R}^{\ud{d}_j \times \ud{d}_j},\ j\in\{1,2,\ldots J\},\\
    \mathbf{\ud{p}}_{j}
    &\in \mathbb{R}^{ 1\times \ud{d}_j},\ j\in\{1,2,\ldots J\},\\
    \mathbf{\ud{q}}_{j}
    &\in \mathbb{R}^{ 1\times \ud{d}_j},\ j\in\{1,2,\ldots J\},\\
    \mathbf{\ud{r}}_{j}
    &\triangleq\mathbf{e}_{\ud{d}_j}\in \mathbb{R}^{ \ud{d}_j \times 1},\ j\in\{1,2,\ldots J\},
\end{align}
and the matrices $\mathbf{0}$ are of appropriate dimensions.
\end{corollary}
\begin{IEEEproof}
Eq. \eqref{eq:Eq6d28} is a product of rational LTs. This corresponds to a rational LT \eqref{eq:Eq6d13}, with parameters \eqref{eq:Eq6d9}-\eqref{eq:Eq6d12}, and convolution operations as in Proposition \ref{pr:Pr6d1}.
\end{IEEEproof}
Thus, Corollary~\ref{cr:Crl6d1} gives a more flexible form than \eqref{eq:Eq6d9}-\eqref{eq:Eq6d15}, and can be adapted to different channel models, purposes, and scenarios as needed.
\begin{table}[t]
\normalsize
\begin{center}
\begin{tabular}{|p{4cm}|p{4cm}|}
    \hline
\textbf{Performance eval. levels}
&
\textbf{Performance metrics} \\
    \hline
Rate adaptive transmissions
&
Ergodic capacity, effective capacity\\
    \hline
ARQ / HARQ
&
Throughput, effective capacity \\
    \hline
Channel coding
&
Outage probability, outage capacity, PEP\\
    \hline
Modulation and detection
&
SER, BER, diversity gain, PEP \\
\hline
System modeling and processing: Combining, communication schemes, interference etc.
&
(Effective SNR and mutual information r.v.)\\
\hline
Discrete-time r.v. signals
&
Mutual information, entropy\\
\hline
  \end{tabular}
\vspace{-0.0cm}
\caption{Examples of performance evaluation levels and performance metrics where the ME-distribution approach may be used.}
\label{tab:Tab6d2}
\end{center}
\vspace{-1.0cm}
\end{table}
\begin{remark}
Note that the unprocessed SNR r.v. $G$ has mean $\mathbb{E}\{G\}=S$. Often, it is convenient to consider the r.v. $G_\textrm{um}$ with unit mean (um) SNR $\mathbb{E}\{G_\textrm{um}\}=1$, ME-distribution parameters $(\mathbf{p}_\textrm{um},\mathbf{Q}_\textrm{um},\mathbf{r}_\textrm{um})$, and work with $g=S g_\textrm{um}$. Thus, by simple variable substitution, $\mathbf{p}\mathrm{e}^{z \mathbf{Q}}\mathbf{r}=S^{-1}\mathbf{p}_\textrm{um}\mathrm{e}^{zS^{-1} \mathbf{Q}_\textrm{um}}\mathbf{r}_\textrm{um}$, which implies $\mathbf{Q}=S^{-1}\mathbf{Q}_\textrm{um}$, $\mathbf{p}=S^{-1}\mathbf{p}_\textrm{um}$, and $\mathbf{r}=\mathbf{r}_\textrm{um}$. The correspondence in the LT-domain is $F(s)=p(s)/q(s)=p_\textrm{um}(sS)/q_\textrm{um}(sS)$.
\end{remark}

\subsection{New ME-distribution Properties}
\label{sec:Sec6d4d3}

In the following, we develop a number of new closed-form expressions that are useful in the analysis. 

\subsubsection{New Expression for the Integral of the ME-density} 
\label{sec:Sec6d4d3d1}
In \eqref{eq:Eq6d4}, we showed the standard approach for integrating ME-functions. The cdf in \eqref{eq:Eq6d5} is an example where this integration approach is used, giving a somewhat messy expression. Moreover, the integration approach in \eqref{eq:Eq6d4} also requires that $\mathbf{Y}$ is non-singular. To handle singular matrices, which arises in practical analysis, we would like to put the integral expression on a more compact, easy-to-manipulate, and tidy form. This is the role of the next theorem.

\begin{theorem}
\label{thm:Thm6d1}
(Integration of ME-function on ME-pdf form).
The integral of $f(t)=\mathbf{x}\mathrm{e}^{t \mathbf{Y}}\mathbf{z}$, with intervals $(0,b)$ can be expressed as
\begin{align}
    \int_0^b \mathbf{x}\mathrm{e}^{t \mathbf{Y}}\mathbf{z} \,\mathrm{d}t
    =\mathbf{E}_{1,d^\textrm{I}},
    \label{eq:Eq6d41}
\end{align}
where
\begin{align}
    \mathbf{E}&\triangleq\mathrm{e}^{b \mathbf{Y}^\textrm{I}},
    \label{eq:Eq6d42}\\
    d^\textrm{I}&= d+1,
    \label{eq:Eq6d43}\\
    \mathbf{Y}^\textrm{I}
    &=
    \begin{bmatrix}
    0 & \mathbf{x}\\
    \mathbf{0} & \mathbf{Y}
    \end{bmatrix}.
    \label{eq:Eq6d44}
\end{align}
\end{theorem}
\begin{IEEEproof}
Integration corresponds to convolution with a step function that has LT $1/s$. Using Proposition \ref{pr:Pr6d1} gives
\begin{align}
    \int_0^b   \mathfrak{L}_t^{-1}  \left\{\frac{x(s)}{y(s)}\right\} \! \mathrm{d}t
    =\mathfrak{L}_b^{-1}  \left\{\frac{1}{s}\frac{x(s)}{y(s)}\right\}=\mathbf{e}_1^\textrm{T}\mathrm{e}^{b \mathbf{Y}^\textrm{I}}   \mathbf{e}_{d^\textrm{I}}
    =\mathbf{E}_{1,d^\textrm{I}}.\notag
\end{align}
\end{IEEEproof}

\begin{example}
\label{ex:Ex6d7}
If $f_Z(z)=\mathbf{p}\mathrm{e}^{z \mathbf{Q}}\mathbf{r}$ is the pdf of a ME-distributed channel SNR r.v. Then, the cdf can be expressed as $F_Z(z)=\mathbf{E}_{1,d^\textrm{I}}$, where $ \mathbf{E}\triangleq\mathrm{e}^{\Theta \mathbf{Q}^\textrm{I}}$,
$d^\textrm{I}= d+1$, $\mathbf{Q}^\textrm{I}=
\begin{bmatrix} 0 & \mathbf{p}\\
\mathbf{0} & \mathbf{Q}
\end{bmatrix}$.
\end{example}

\begin{remark}
\label{rm:Rm6d1}
In \cite[(25),(46)]{LarssonRasmSkog16a}, we introduced the \textit{Singular Matrix Integration by Matrix Augmentation} idea, but on a different form than Theorem~\ref{thm:Thm6d1}. The form in Theorem~\ref{thm:Thm6d1}, as will be seen, allows for even simpler expressions and analysis.
\end{remark}

\subsubsection{New Expression(s) for the Maximum (and Minimum) of ME-distributed r.v.s} \label{sec:Sec6d4d3d2}
In \cite[(35), (36)]{LarssonRasmSkog16a}, and the derivation in Ex.~\ref{ex:Ex6d5}, we considered $N$-branch SDC, i.e.~selecting the signal with maximum SNR, for exponentially distributed r.v.s and showed that the pdf of the max SNR is a ME-distribution. More generally, the characterization of the maximum (and the minimum) of ME-distributed r.v.s is well-known and have been considered in e.g. \cite{BeanFackTayl08}, \cite{RuizCastro13}, as well as reviewed in Proposition \ref{pr:Pr6d2} (and \ref{pr:Pr6d3}). The expressions (and the derivations) in Propositions \ref{pr:Pr6d2} and \ref{pr:Pr6d3} are somewhat inconvenient, and may discourage practical use. Using the integration idea in Theorem~\ref{thm:Thm6d1}, we introduce a simpler, more tractable, expression (and derivation) for the maximum operation in the theorem below.

\begin{theorem}
\label{thm:Thm6d2}
(Maximum of two ME-distributed r.v.s).
Let $T_j,j\in\{1,2\}$ be ME-distributed r.v.s with pdf $f^{(j)}_T(t)=\mathbf{x}_j\mathrm{e}^{t \mathbf{Y}_j}\mathbf{z}_j$, and degree $d_j$. Then, the CDF of the ME-distribution r.v. $T=\max(T_1,T_2)$ can be expressed as
\begin{align}
    F_T^\textrm{max}(t)
    &=\mathbf{E}_{1,d^\textrm{I}_1+d^\textrm{I}_2},
    \label{eq:Eq6d45}
\end{align}
where
\begin{align}
    \mathbf{E}
    &\triangleq \mathrm{e}^{t \left(\mathbf{Y}_1^\textrm{I} \! \oplus  \mathbf{Y}_2^\textrm{I}\right)},
    \label{eq:Eq6d46}\\
    \mathbf{Y}_j^\textrm{I}
    &=
    \begin{bmatrix}
    0 & \mathbf{x}_j\\
    \mathbf{0} & \mathbf{Y}_j
    \end{bmatrix}.
    \label{eq:Eq6d47}
\end{align}
\end{theorem}
\begin{IEEEproof}
\begin{align}
    F_T^\textrm{max}(t)
    &=F^{(1)}_T(t)F^{(2)}_T(t)\notag\\
    &=\left(\int_0^\textrm{T}\mathbf{x}_1\mathrm{e}^{u \mathbf{Y}_1}\mathbf{z}_1 \, \mathrm{d}u\right)
    \left(\int_0^\textrm{T}\mathbf{x}_2\mathrm{e}^{u \mathbf{Y}_2}\mathbf{z}_2 \, \mathrm{d}u\right)\notag\\
    &\overset{(a)}{=}\left((\mathbf{e}_1^{(1)})^\textrm{t}\mathrm{e}^{t \mathbf{Y}_1^\textrm{I}}\mathbf{e}_{d_1^\textrm{I}}^{(1)}\right)
    \left((\mathbf{e}_1^{(2)})^\textrm{t}\mathrm{e}^{t \mathbf{Y}_2^\textrm{I}}\mathbf{e}_{d_2^\textrm{I}}^{(2)}\right)\notag\\
    &\overset{(b)}{=}\left(\mathbf{e}_1^{(1)} \!\! \otimes \mathbf{e}_1^{(2)}\right)^\textrm{T}
    \mathrm{e}^{t \left(\mathbf{Y_1^\textrm{I}}   \oplus  \mathbf{Y_2^\textrm{I}}\right)}
    \left(\mathbf{e}_{d_1^\textrm{I}}^{(1)} \!\!  \otimes \mathbf{e}_{d_2^\textrm{I}}^{(2)}\right)\notag\\
    &=\mathbf{E}_{1,d^\textrm{I}_1+d^\textrm{I}_2}, \ \mathbf{E} \triangleq \mathrm{e}^{t \left(\mathbf{Y}_1^\textrm{I}   \oplus  \mathbf{Y}_2^\textrm{I}\right)},\notag
\end{align}
where we used the integration idea in step (a), and the Kronecker product identities $(\mathbf{X}_1  \otimes \mathbf{Y}_1)(\mathbf{X}_2  \otimes \mathbf{Y}_2)=(\mathbf{X}_1\mathbf{X}_2) \otimes (\mathbf{Y}_1\mathbf{Y}_2)$, $\mathbf{X}  \oplus \mathbf{Y}=\mathbf{X}  \otimes \mathbf{I}_n+\mathbf{I}_m  \otimes \mathbf{Y}$, and $\mathrm{e}^{\mathbf{X}  \oplus \mathbf{Y}}=\mathrm{e}^{\mathbf{X}} \!  \otimes \mathrm{e}^{\mathbf{Y}}$,  with rearrangement in step (b).
\end{IEEEproof}

\begin{remark}
\label{rm:Rm6d2}
(Maximum of two ME-distributed r.v.s - Alternative expression).
It is also seen in Theorem~\ref{thm:Thm6d2}, that from step (a), the cdf can directly be written
\begin{align}
    F_T^\textrm{max}(t)
    &=\mathbf{E}^{(1)}_{1,d_1^\textrm{I}}\mathbf{E}^{(2)}_{1,d_2^\textrm{I}},
    \label{eq:Eq6d48}
\end{align}
where
\begin{align}
    \mathbf{E}^{(j)}
    &\triangleq \mathrm{e}^{t \mathbf{Y}_j^\textrm{I}},
    \label{eq:Eq6d49}\\
    \mathbf{Y}_j^\textrm{I}
    &=
    \begin{bmatrix}
    0 & \mathbf{x}_j\\
    \mathbf{0} & \mathbf{Y}_j
    \end{bmatrix}.
    \label{eq:Eq6d50}
\end{align}
The extension to more than two r.v.s is straightforward.
\end{remark}

\begin{remark}
\label{rm:Rm6d3}
(Minimum of two ME-distributed r.v.s - Alternative expression)
The minimum of two r.v.s, i.e. $T=\min\{T_1,T_2\}$, can be derived analogously to Theorem~\ref{thm:Thm6d2} and Remark~\ref{rm:Rm6d2}, based on $F_T^\textrm{min}(t)
=1-(1-F^{(1)}_T(t))(1-F^{(2)}_T(t))$. With $\mathbf{E}^{(j)}_{1,d_j^\textrm{I}}$, as in Remark~\ref{rm:Rm6d2}, the cdf is simply
\begin{align}
    F_T^\textrm{min}(t)
    &=1-\left(1-\mathbf{E}^{(1)}_{1,d_1^\textrm{I}}\right) \left(1-\mathbf{E}^{(2)}_{1,d_2^\textrm{I}}\right),
\end{align}
which is easily extended to more than two r.v.s.
\end{remark}
Note that while the new expressions for the maximum and the minimum of two ME-distributed r.v.s are convenient and easy to use, they do not, in contrast to Propositions \ref{pr:Pr6d2} and \ref{pr:Pr6d3}, illustrate closure properties.

\subsubsection{Integral of the Product of the ME-density and a Function}
\label{sec:Sec6d4d3d3}

In our analysis, for finding closed-form performance expressions, it is often of interest to determine integrals involving the product of a function and the ME-density.  In the next theorem, we illustrate a useful technique, using a wisely selected integral representation of the function, swapping the order of integrations, and then determining the integrals. This method is an extension, to ME-distributed r.v., of the well-known performance analysis approach by Simon and Alouini, \cite{SimonAlouini05}. Whereas \cite{SimonAlouini05} considered only Rayleigh, Ricean, and Nakagami-$m$ fading, we are able to handle the much wider class of ME-distributions and express all results in closed-forms expressed only in ME-matrix/vector-parameter.

\begin{theorem}
\label{thm:Thm6d3}
(Integral of ME-density- and Function-product).
Let $g(t)$ be a function for which an integral representation  $g(t)=\int_{a_u}^{b_u}g_1(u)\mathrm{e}^{-t g_2(u)} \, \mathrm{d}u$ exist. Then, the expectation of $g(t)$ is
\begin{align}
 &  \mathbb{E}\{g(t)\}
=\int_0^\infty g(t) \mathbf{x}\mathrm{e}^{t\mathbf{Y}}\mathbf{z} \, \mathrm{d}t\notag\\
&=\int_{a_u}^{b_u} g_1(u) \mathbf{x}\left(g_2(u)\mathbf{I}-\mathbf{Y}\right)^{-1}\mathbf{z}\, \mathrm{d}u\label{eq:Eq6d52}\\
&=G_1
+\int_{a_u}^{b_u} g_1(u) \mathbf{x}\mathbf{Y}^{-1}\left(\mathbf{I}-\mathbf{Y}g_2(u)^{-1}\right)^{-1}\mathbf{z}\, \mathrm{d}u,
\end{align}
where $G_1\triangleq\int_{a_u}^{b_u} g_1(u)\, \mathrm{d}u$.
\end{theorem}

\begin{IEEEproof}
The expectation is
\begin{align}
   &  \mathbb{E}\{g(t)\}
=\int_0^\infty g(t) \mathbf{x}\mathrm{e}^{t\mathbf{Y}}\mathbf{z} \, \mathrm{d}t\notag\\
	&=\int_0^\infty \left( \int_{a_u}^{b_u} g_1(u) \mathrm{e}^{-t g_2(u)} \, \mathrm{d}u\right)\mathbf{x}\mathrm{e}^{t\mathbf{Y}}\mathbf{z} \, \mathrm{d}t \notag\\
&=\int_{a_u}^{b_u} g_1(u)\left(\int_0^\infty \mathbf{x}\mathrm{e}^{t(\mathbf{Y}-g_2(u)\mathbf{I})}\mathbf{z} \, \mathrm{d}t \right)\, \mathrm{d}u\notag\\
&=\int_{a_u}^{b_u} g_1(u) \mathbf{x}\left(g_2(u)\mathbf{I}-\mathbf{Y}\right)^{-1}\mathbf{z}\, \mathrm{d}u \notag\\
&=-\int_{a_u}^{b_u} g_1(u) \mathbf{x}\mathbf{Y}^{-1}\left(\mathbf{I}-g_2(u)\mathbf{Y}^{-1}\right)^{-1}\mathbf{z}\, \mathrm{d}u\notag\\
&=-\int_{a_u}^{b_u} g_1(u)\mathbf{x}\mathbf{Y}^{-1}\left(\mathbf{I} - \left(\mathbf{I}-\mathbf{Y}g_2(u)^{-1}\right)^{-1}\mathbf{z}\right)\, \mathrm{d}u \notag\\
&=\int_{a_u}^{b_u} g_1(u)\, \mathrm{d}u
+\int_{a_u}^{b_u} g_1(u) \mathbf{x}\mathbf{Y}^{-1}\left(\mathbf{I}-\mathbf{Y}g_2(u)^{-1}\right)^{-1}\mathbf{z}\, \mathrm{d}u \notag
\end{align}
\end{IEEEproof}
\begin{remark}
\label{rm:Rm6d4}
Integral representations for many different functions $g(t)$ with an integrand involving an exponential form are listed in standard mathematical tables, see e.g. \cite{gradshteyn07}. Some examples are $g(t)=n!(t+a)^{-(n+1)}=\mathrm{e}^{-ta}\int_0^\infty u^{n}\mathrm{e}^{-tu}\, \mathrm{d}u$,  $g(t)=\Gamma((n+1)/2)t^{-(n+1)/2}=\mathrm{e}^{-ta}\int_0^\infty u^{n}\mathrm{e}^{-tu^2}\, \mathrm{d}u$,  $g(t)=\ln(t/a)=\int_0^\infty (\mathrm{e}^{au}-\mathrm{e}^{tu})/u \, \mathrm{d}u$, $g(t)=2\pi I_0(\sqrt{t^2+c^2})=\int_0^{2\pi} \mathrm{e}^{t \cos{u}+c\sin{u}}\, \mathrm{d}u$.
\end{remark}

\begin{remark}
\label{rm:Rm6d5}
Note that \eqref{eq:Eq6d52} is a scalar integral with matrix parameters.
Hence, if  $\int_{a_u}^{b_u} g_1(u) \sum_{n=0}^\infty \left(y/g_2(u)\right)^n \mathrm{d}u=\int_{a_u}^{b_u} g_1(u) \left(1-y/g_2(u)\right)^{-1} \mathrm{d}u$ converges and have solution $f(y)$ for scalar $y$, then 
$\int_{a_u}^{b_u} g_1(u) \sum_{n=0}^\infty \left(\mathbf{Y}/g_2(u)\right)^n\, \mathrm{d}u=\int_{a_u}^{b_u} g_1(u) \left(\mathbf{I}-\mathbf{Y}/g_2(u)\right)^{-1}\, \mathrm{d}u$ converges too, if the spectral radius $\rho(\mathbf{Y})\leq1$, with solution $f(\mathbf{Y})$ where the functions in $f(\mathbf{Y})$ are the matrix counterpart of functions in $f(y)$. Mathematical softwares implement many standard matrix functions, e.g. $f(\mathbf{Y})=\mathbf{Y}^2,\mathbf{Y}^{-1}, \mathrm{e}^{\mathbf{Y}},\sqrt{\mathbf{Y}}$, while other, more exotic matrix functions, are often not defined, such as $\sin^{-1}(\mathbf{Y})$, matrix Bessel function $J_\nu(\mathbf{Y})$, matrix exponential integral $E_1(\mathbf{Y})$, etc. When $\mathbf{Y}$ is diagonalizable, i.e. $\mathbf{Y}$ having distinct eigenvalues, then $\mathbf{Y}=\mathbf{V}\mathbf{\Lambda}\mathbf{V}^{-1}$ and $f(\mathbf{Y})=\{\mathbf{V}f(\mathbf{\Lambda})\mathbf{V}^{-1}\}$, which allows direct use of existing scalar functions. In \cite[(64)]{LarssonRasmSkog16a}, we also considered an alternative approach when the matrix parameterized integrals dealt with resulted in a matrix expression that were not supported in mathematical softwares. We simply defined a new matrix function (more precisely the matrix incomplete gamma function), as a matrix extension of the corresponding scalar function and then expressed the performance in such matrix function.
\end{remark}

\begin{remark}
\label{rm:Rm6d6}
Let $y(s)$ be the characteristic polynomial of $\mathbf{Y}$. The LT of the ME-pdf is then on the form $F(s)=x(s)/y(s)$. Using \eqref{eq:Eq6d9}-\eqref{eq:Eq6d15}, we can also write \eqref{eq:Eq6d52} as
\begin{align}
\mathbb{E}\{g(t)\}
	&=\int_{a_u}^{b_u} \frac{g_1(u) x(g_2(u))}{y(g_2(u))}\, \mathrm{d}u.
\end{align}
\end{remark}

\section{Performance Analysis with ME-distributed Channel Fading} \label{sec:Sec6d5}
We now turn our attention to the performance analysis of various wireless communication systems with ME-distributed fading channels. Of course, ME-distributed fading may refer to slow fading, where the fading gain is assumed constant over a whole codeword/redundancy-block/data packet, or fast fading, where the fading gain is constant over a symbol, or any other time-scale of interest. For each time scale, suitable study cases and performance metrics may be chosen and studied. We will start with the slow fading case, and consider the fast fading case towards the end of the section. Nevertheless, we first address the effective channel SNR processing, which is generic for any coherence time. There, we generalize, e.g., the MRC and SDC cases in \cite{LarssonRasmSkog16a}, characterize the distribution of sum-interference, etc. Then, we examine the effective capacity of rate adaptive transmissions with known transmitter CSI. Subsequently, we treat ARQ, truncated-HARQ and persistent-HARQ wrt throughput, and related metrics, in detail. The throughput performance of NCBR with ARQ is characterized subsequently. Next, we introduce the bivariate ME-distribution and analyze ARQ where the signal and interfering channels are all ME-distributed. We also explore the outage probability, by means of the bivariate ME-distribution, of $2\times2$ SM-MIMO. Modulation and detection performance for common modulation formats are characterized for fast ME-distributed fading towards the end of this section. With those applications, we aim to illustrate the versatility and strength of the ME-distribution approach.

\subsection{Effective Channel SNR Processing} 
\label{sec:Sec6d5d1}

One important insight used in \cite[Sec. IV.F]{LarssonRasmSkog16a} was that the closure property of the convolution of ME-distributed r.v.s, Proposition \ref{pr:Pr6d1}, is a powerful tool for wireless system modeling and performance analysis. We now explore this, as well as other closure properties discussed in Section \ref{sec:Sec6d2d3}, \cite{RuizCastro13}, below. We start by generalizing the MRC case in \cite[Sec. IV.F]{LarssonRasmSkog16a} for iid to non-identical independent ME-distributed r.v.s. We illustrate the basic ideas below for two r.v.s, but the results are trivially extendable to more than two r.v.s.

\begin{example}
\label{ex:Ex6d8}
(MRC of two non-identical independent ME-distributed r.v.s)
Consider two ME-distributed r.v.s $z_u$, with pdfs $f_G^{(n)}(g)=\mathbf{p}_n\mathrm{e}^{g \mathbf{Q}_n}\mathbf{r}_n$, $n\in\{1,2\}$. Then, the effective SNR, $Z=G_1+G_2$, is also ME-distributed with pdf $f_Z(z)=\mathbf{\tilde p}\mathrm{e}^{z \mathbf{\tilde  Q}}\mathbf{\tilde  r}$, and parameters
\begin{align}
    \mathbf{\tilde Q}
    &=
    \begin{bmatrix}
        \mathbf{Q}_1 & \mathbf{P}_2 \\
        0 & \mathbf{Q}_2\\
    \end{bmatrix},\\
    \mathbf{\tilde p}
    &=
    \begin{bmatrix}
        \mathbf{p}_1 & \mathbf{0}
    \end{bmatrix},\\
    \mathbf{\tilde r}
    &=
    \begin{bmatrix}
        \mathbf{0} & \mathbf{r}_2^\textrm{T}
    \end{bmatrix}^\textrm{T},
\end{align}
where $\mathbf{Q}_1=\mathbf{S}-\mathbf{r}_1\mathbf{q}_1$, $\mathbf{Q}_2=\mathbf{S}-\mathbf{r}_2\mathbf{q}_2$, and $\mathbf{P}_2=\mathbf{r}_1\mathbf{p}_2$. The above follows from  Corollary~\ref{cr:Crl6d1}.
\end{example}

\begin{example}
\label{ex:Ex6d9}
(Sum-interference)
For two ME-distributed interfering signals, with non-identical independent SNR r.v.s. $G_1$ and $G_2$, the sum-interference is $Z_\textrm{I}=G_1+G_2$. Thus, due to the closure of the convolution for the ME-distribution class, $Z_\textrm{I}$ is also ME-distributed and has the same form as for the MRC case in Ex.~\ref{ex:Ex6d8}.
\end{example}

Moving on to SDC, where for the special case with iid exponentially distributed SNRs, a convolution view is applicable.
\begin{example}
\label{ex:Ex6d10}
(Effective channel of SDC and Rayleigh fading)
The pdf of the SDC effective channel with (unit-mean) exponentially distributed fading SNRs has, as given by Ex.~\ref{ex:Ex6d5}, LT $F(s)=N!/\prod_{n=1}^N(n+s)$, which gives
\begin{align}
    \mathbf{\tilde{Q}}_\textrm{um}
    &=
    \begin{bmatrix}
    -1 & 1 & \cdots & 0\\
    0 & -2 & \ddots& 0 \\
    \vdots & \ddots  & \ddots & \vdots  \\
    0 & 0  & \cdots  & -N \\
    \end{bmatrix},\\
    \mathbf{\tilde{p}}_\textrm{um}
    &=
    \begin{bmatrix}
    1 & 0 & \ldots & 0 \\
    \end{bmatrix}N!,\\
    \mathbf{\tilde{r}}_\textrm{um}
    &=
    \begin{bmatrix}
    0 & \ldots & 0 & 1 \\
    \end{bmatrix}^\textrm{T}.
\end{align}
\end{example}

\begin{example}
\label{ex:Ex6d11}
(SDC of 2 non-identical independent ME-distributed r.v.s)
For the more general case of selecting the maximum of non-identical independent ME-distributed r.v.s $Z=\max\{G_1,G_2\}$, Proposition~\ref{pr:Pr6d2}, or the more compactly formulated Theorem~\ref{thm:Thm6d2}, gives the cdf.
\end{example}

A different multi-antenna arrangement, compared to OSTBC+MRC which was considered in \cite{LarssonRasmSkog16a}, is ZF-MIMO.

\begin{example}
\label{ex:Ex6d12}
(Zero-forcing MIMO) 
For a zero-forcing $N_\textrm{rx}\times N_\textrm{tx}$-antenna MIMO system, $N_\textrm{rx}\geq N_\textrm{tx}$, with a complex Gaussian channel matrix with iid entries and mean SNR $S$, the per stream SNR is gamma-distributed, \cite{GoreHeathPaul02}, with $N_\textrm{rx}-N_\textrm{tx}+1$ degrees of freedom and the corresponding LT is $F(s)={1}/{(1+sS)^{N_\textrm{rx}-N_\textrm{tx}}}$. Thus, the ZF-MIMO per stream SNR r.v. is ME-distributed.
\end{example}

\begin{example}
\label{ex:Ex6d13}
(Mixed case and effective channel algebra)
Perhaps one of the more significant aspects of the proposed framework is that the SNR processing operations may also be mixed, e.g. $Z=G_1+\max(G_2,G_3)$ (SDC of branch 2 and 3, and then MRC with branch 1), or $Z=G_1+G_2+G_3-\min(G_1,G_2,G_3)$ (MRC of the two strongest branches), etc. As long as the r.v.s are ME-distributed, and the operations are closed, the effective channel r.v. $Z$ will also be ME-distributed. Such operations, with closure properties, on ME-distributed r.v.s can be seen as an \textit{Effective channel algebra}. Note that this framework naturally handles non-identical ME-distributed r.v.s.
\end{example}

\begin{example}
\label{ex:Ex6d14}
With random channel $\mathbf{H}\in\mathbb{C}^{N_\textrm{rx}\times N_\textrm{tx}}$, the MIMO channel capacity \cite{Telatar99}, $C=\ln{\det \left( \mathbf{I}+{S}{N_\textrm{tx}^{-1}}\mathbf{H}^H\mathbf{H}\right)}$,
corresponds to a scalar r.v. Since the ME-distribution is dense on $(0,\infty)$, we conjecture that the pdf of the MIMO channel capacity can (in principle) be approximated with a ME-distributed r.v., $Z$, with pdf $f_Z(z)=\mathbf{\tilde p}\mathrm{e}^{z\mathbf{\tilde Q}}\mathbf{\tilde r}$.
\end{example}

\subsection{Effective Capacity Analysis of Rate Adaptive Transmission} 
\label{sec:Sec6d5d2}

The effective capacity performance metric was introduced in \cite{WuNegi03}. The objective of the effective capacity is to quantify the maximum sustainable throughput under stochastic QoS guarantees with varying server rate. In \cite{LarssonRasmSkog16b}, we analyzed the effective capacity of ARQ, truncated- and persistent-HARQ, with respect to a ME-distributed effective channel $f_Z(z)=\mathbf{\tilde p}\mathrm{e}^{z\mathbf{\tilde Q}}\mathbf{\tilde q}$, with $F(s)=\tilde p(s)/\tilde q(s)$, and CSI known at the receiver.
Here, we determine the effective capacity for two cases, when the service rate is ME-distributed, and when the service rate equals the AWGN Shannon capacity (i.e. the CSI is also known at the transmitter) and the effective SNR is ME-distributed. Thus, we assume ideal rate adaptive (RA) transmissions, a.k.a adaptive modulation and coding (AMC). Those cases are addressed in turn in the theorems below.

\begin{theorem}
\label{thm:Thm6d4}
(Effective capacity with ME-distributed service rate)
Let the service rate $\zeta$ be iid, and have pdf $f_\zeta(\zeta)=\mathbf{\tilde p}\mathrm{e}^{\zeta \mathbf{\tilde Q}}\mathbf{\tilde r}$, $\mathbf{\tilde p}\in\mathbb{R}^{1 \times \tilde d}$, $\mathbf{\tilde Q}\in\mathbb{R}^{\tilde d\times \tilde d}$, $\mathbf{\tilde r}=[0 \ldots 0 \ 1]^\textrm{T}\in\mathbb{R}^{\tilde d \times 1}$. Then, the effective capacity for rate-adaptive transmission  is
\begin{align}
C_\textrm{eff}^\textrm{RA}
&=-\frac{1}{\theta}\ln \left(\mathbf{\tilde p}(\theta \mathbf{I}-\mathbf{\tilde Q})^{-1}\mathbf{\tilde r}\right)\notag\\
&=-\frac{1}{\theta}\ln \left(\frac{\tilde p(\theta)}{\tilde q(\theta)}\right).
\end{align}
where $\theta$ is the effective capacity quality-of-service exponent.
\end{theorem}
\begin{IEEEproof}
When $\zeta$ is iid, the effective capacity is $C_\textrm{eff}^\textrm{RA}\triangleq -\frac{1}{\theta}\ln \left( \mathbb{E}\left\{\mathrm{e}^{-\theta \zeta}\right\}\right)=-\frac{1}{\theta}\ln \left(\int_0 ^\infty \mathrm{e}^{-\zeta\theta}\mathbf{\tilde p}\mathrm{e}^{\zeta\mathbf{\tilde Q}}\mathbf{\tilde r} \, \mathrm{d}\zeta \right)$, and we have $\int_0 ^\infty \mathrm{e}^{-\zeta\theta}\mathbf{\tilde p}\mathrm{e}^{\zeta\mathbf{\tilde Q}}\mathbf{\tilde r} \, \mathrm{d}\zeta=\mathbf{\tilde p}(\theta \mathbf{I}-\mathbf{\tilde Q})^{-1}\mathbf{\tilde r}=\tilde p(\theta)/\tilde q(\theta)$.
\end{IEEEproof}

\begin{example}
\label{ex:Ex6d15}
Theorem~\ref{thm:Thm6d4} can, e.g., be used to compute the effective capacity for the (ME-distribution approximated) MIMO channel capacity in Ex.~\ref{ex:Ex6d14}.
\end{example}

\begin{theorem}
\label{thm:Thm6d5}
(Effective capacity with the effective channel ME-distributed and the service rate equals the AWGN Shannon-capacity)
Let the effective channel pdf be  $f_Z(z)=\mathbf{\tilde p}\mathrm{e}^{z \mathbf{\tilde Q}}\mathbf{\tilde r}$, $\mathbf{\tilde p}\in\mathbb{R}^{1 \times \tilde d}$, $\mathbf{\tilde Q}\in\mathbb{R}^{\tilde d\times \tilde d}$, $\mathbf{\tilde r}=[0 \ldots 0 \ 1]^\textrm{T}\in\mathbb{R}^{\tilde d \times 1}$.  Then, the effective capacity, with CSI known at the transmitter and perfect rate adaptation, is
\begin{align}
C_\textrm{eff}^\textrm{RA}
&\triangleq-\frac{1}{\theta}\ln \left(\int_0^\infty \mathrm{e}^{-\theta\ln(1+z)} \mathbf{\tilde p}\mathrm{e}^{z\mathbf{\tilde Q}}\mathbf{\tilde r} \, \mathrm{d}z \right)\notag\\
&=-\frac{1}{\theta}\ln \left(\int_0^\infty  \frac{u^{\theta-1}\mathrm{e}^{-u} }{\Gamma(\theta)}\mathbf{\tilde p}\left(u\mathbf{I}-\mathbf{\tilde Q}\right)^{-1}\mathbf{\tilde r} \, \mathrm{d}u\right)\\
&=-\frac{1}{\theta}\ln \left(\int_0^\infty  \frac{u^{\theta-1}\mathrm{e}^{-u} }{\Gamma(\theta)}\frac{\tilde p(u)}{\tilde q(u)} \, \mathrm{d}u\right).
\end{align}
\end{theorem}
\begin{IEEEproof}
We have $C_\textrm{eff}^\textrm{RA}
= -\frac{1}{\theta}\ln \left( \mathbb{E}\left\{\mathrm{e}^{-\theta \ln(1+z)}\right\}\right)$, where the expectation is
\begin{align}
&\mathbb{E}\left\{\mathrm{e}^{-\theta \ln(1+z)}\right\}\notag\\
&=\int_0^\infty \mathrm{e}^{-\theta\ln(1+z)} \mathbf{\tilde p}\mathrm{e}^{z\mathbf{\tilde Q}}\mathbf{\tilde r} \, \mathrm{d}z \notag\\
&\overset{(a)}{=}\int_0^\infty (1+z)^{-\theta} \mathbf{\tilde p}\mathrm{e}^{z\mathbf{\tilde Q}}\mathbf{\tilde r} \, \mathrm{d}z \notag\\
&=\int_0^\infty \left(\int_0^\infty \frac{u^{\theta-1}}{\Gamma(\theta)} \mathrm{e}^{-(1+z)u} \, \mathrm{d}u \right) \mathbf{\tilde p}\mathrm{e}^{z\mathbf{\tilde Q}}\mathbf{\tilde r} \, \mathrm{d}z \notag\\
&\overset{(b)}{=}\int_0^\infty  \frac{u^{\theta-1}\mathrm{e}^{-u} }{\Gamma(\theta)}\mathbf{\tilde p}\left(u\mathbf{I}-\mathbf{\tilde Q}\right)^{-1}\mathbf{\tilde r} \, \mathrm{d}u. \notag
\end{align}
In step (a), we exploit a trick where $(1+z)^{-\theta}$ is replaced with an integral representation, and where the gamma function is defined as $\Gamma(\theta)=\int_0^\infty x^{\theta-1}\mathrm{e}^{-x} \, \mathrm{d}x$. This also motives the generalization leading to Theorem~\ref{thm:Thm6d3}.
As in Theorem~\ref{thm:Thm6d3}, step (b) may also be expressed as $\overset{(b)}{=}1+\int_0^\infty  \frac{u^{\theta-1}\mathrm{e}^{-u} }{\Gamma(\theta)}\mathbf{\tilde p}\mathbf{\tilde Q}^{-1}\left(\mathbf{I}-\mathbf{\tilde Q}u^{-1}\right)^{-1}\mathbf{\tilde r} \, \mathrm{d}u$. Using the characteristic polynomial $\tilde q(\cdot)$ of $\mathbf{\tilde Q}$, gives $\mathbf{\tilde p}\left(u\mathbf{I}-\mathbf{\tilde Q}\right)^{-1}\mathbf{\tilde r}={\tilde p(u)}/{\tilde q(u)}$.
\end{IEEEproof}

An alternative effective capacity formulation, giving closed-form solutions when $\mathbf{Q}$ is diagonalizable, is discussed next. Moler and Van Loan gave various methods for computing the ME in \cite{MolerLoan03}. We explore matrix decomposition methods (based on similarity transformations), specifically their 14th and 16th method, the eigenvector decomposition and the Jordan canonical form. Those techniques are, of course, generally applicable to other related problems too. 

\begin{theorem}
\label{thm:Thm6d6}
(Effective capacity with the effective channel ME-distributed and the service rate equals the AWGN Shannon-capacity)
Let the assumptions be as in Theorem~\ref{thm:Thm6d5}. Then, the effective capacity is
\begin{align}
C_\textrm{eff}^\textrm{RA}
&=-\frac{1}{\theta} \ln \left(\mathbf{\tilde p}\mathbf{\tilde T}\mathbf{\tilde \Xi} \mathbf{\tilde T}^{-1}\mathbf{\tilde r} \right),
\label{eq:Eq6d64}
\end{align}
where
\begin{align}
	\mathbf{\tilde \Xi}
	&\triangleq \int_0 ^\infty (1+z)^{-\theta} 	\mathrm{e}^{z\mathbf{J}}\, \mathrm{d}z,
\label{eq:Eq6d65}\\
	\mathbf{\tilde J}
	&\triangleq \mathbf{T}^{-1}\mathbf{\tilde Q}\mathbf{T},
\end{align}
$\mathbf{\tilde J}$ is a Jordan-form matrix, and $\mathbf{\tilde T}$ is a non-singular matrix.
\end{theorem}
\begin{IEEEproof}
From the proof of Theorem~\ref{thm:Thm6d5}, we have
\begin{align}
C_\textrm{eff}^\textrm{RA}
&=-\frac{1}{\theta}\ln \left(\int_0^\infty (1+z)^{-\theta} \mathbf{\tilde p}\mathrm{e}^{z\mathbf{\tilde Q}}\mathbf{\tilde r} \, \mathrm{d}z \right)\notag\\
&=-\frac{1}{\theta}\ln \left(\mathbf{\tilde p}\mathbf{\tilde T}\left(\int_0 ^\infty (1+z)^{-\theta} \mathrm{e}^{z\mathbf{\tilde J}}\, \mathrm{d}z \right)\mathbf{\tilde T}^{-1}\mathbf{r}  \right),\notag
\end{align}
which yields \eqref{eq:Eq6d64}.
\end{IEEEproof}

\begin{corollary}
\label{cr:Crld2}
When $\mathbf{\tilde Q}$ is diagonalizable, $\mathbf{\tilde V}\mathbf{\tilde \Lambda}\mathbf{\tilde V}^{-1}=\mathbf{\tilde Q}$, $\mathbf{\tilde V}$ is a non-singular Vandermonde matrix, and all eigenvalues $\lambda_{j},\ j\in\{1,2,\ldots J\}$, are real negative,  then 
\begin{align}
\mathbf{\tilde \Xi}
&=\mathrm{diag}\{\xi_1,\xi_2,\ldots \xi_J\},\\
\xi_j
&=(-\lambda_j)^{\theta-1}\mathrm{e}^{-\lambda_j}\Gamma(1-\theta,-\lambda_j).
\label{eq:Eq6d68}
\end{align}
\end{corollary}
\begin{IEEEproof}
Solving $d_j
= \int_0 ^\infty\frac{\mathrm{e}^{z\lambda_j}}{ (1+z)^{\theta}} \, \mathrm{d}z$ gives \eqref{eq:Eq6d68}.
\end{IEEEproof}

\begin{remark}
\label{rm:Rm6d7}
When $\mathbf{Q}$ is not diagonalizable, solving the integral \eqref{eq:Eq6d65} analytically is more tedious but can (in principle) be handled with Jordan-form identities $\mathbf{\tilde Q}=\mathbf{\tilde T}\mathrm{diag}(\mathbf{\tilde J})\mathbf{\tilde T}^{-1}=\mathbf{\tilde T}\mathrm{diag}(\mathbf{\tilde J}_1,\mathbf{\tilde J}_2,\ldots \mathbf{\tilde J}_J)\mathbf{\tilde T}^{-1}$, and $f(z\mathbf{\tilde Q})=\mathbf{\tilde T}\mathrm{diag}(f(z\mathbf{\tilde J}_1),f(z\mathbf{\tilde J}_2),\ldots f(z\mathbf{\tilde J}_J))\mathbf{\tilde T}^{-1}$ \cite{Horn86}. Specifically, the matrices have the forms 
\begin{align}
\mathbf{\tilde J}=
\begin{bmatrix}
\mathbf{\tilde J}_1 & \mathbf{0}  & \cdots & \mathbf{0} \\
\mathbf{0}  & \mathbf{\tilde J}_2 & \ddots & \ddots  \\
 \vdots & \ddots  & \ddots  &\ddots \\
 \mathbf{0} & \ddots  & \ddots  &\mathbf{\tilde J}_j 
\end{bmatrix},
\end{align}
\begin{align}
\mathbf{\tilde J}_j=
\begin{bmatrix}
\lambda_j & 1 & 0 & \cdots & 0\\
0	& \lambda_j & 1 & \ddots & \ddots \\
\vdots	& \ddots & \ddots & \ddots & \ddots \\
\vdots	& \ddots & \ddots & \lambda_j & 1 \\
0	& \ddots & \ddots& 0 & \lambda_j 
\end{bmatrix} \in \mathbb{R}^{k_j\times k_j},
\end{align}
\begin{align}
\mathrm{e}^{z\mathbf{\tilde J}_{j}}=
\begin{bmatrix}
\mathrm{e}^{z\lambda_j} & z\mathrm{e}^{z\lambda_j} & \frac{z^2}{2!}\mathrm{e}^{z\lambda_j} & \cdots & \frac{z^{k_j-1}}{(k_j-1)!}\mathrm{e}^{z\lambda_j}\\
0	& \mathrm{e}^{z\lambda_j} & z\mathrm{e}^{z\lambda_j}  & \ddots & \ddots \\
\vdots	& \ddots & \ddots & \ddots & \ddots \\
\vdots	& \ddots & \ddots & \mathrm{e}^{z\lambda_j} & z\mathrm{e}^{z\lambda_j} \\
0	& \ddots & \ddots& 0 & \mathrm{e}^{z\lambda_j}
\end{bmatrix}.
\end{align}
\end{remark}

 \begin{remark}
\label{rm:Rm6d8}
Note that the ergodic capacity can be computed as $C_\textrm{erg}= \underset{\theta \rightarrow 0}{\lim}C_\textrm{eff}^\textrm{RA}$.
\end{remark}

To date, RA transmission, with transmitter CSI, has not been considered with the general ME-distributed fading channel. More specifically, the effective capacity has not been studied for RA transmissions and the ME-distributed fading channel. From the treatment here, we note that the ME-distribution, and the analysis framework, could be useful in this context.

\subsection{Outage Probability Analysis} 
\label{sec:Sec6d5d3}
After the ME-distributed effective channel SNR (or effective channel MI) has been characterized, the outage probability is a performance metric of interest to consider.
\begin{theorem}
\label{thm:Thm6d7}
(Outage probability for the ME-distributed effective channel)
Let the effective channel pdf be $f_Z(z)=\mathbf{\tilde p}\mathrm{e}^{z\mathbf{\tilde Q}}\mathbf{\tilde r}$, $\mathbf{\tilde p}\in\mathbb{R}^{1 \times \tilde d}$, $\mathbf{\tilde Q}\in\mathbb{R}^{\tilde d\times \tilde d}$, $\mathbf{\tilde r}=[0 \ldots 0 \ 1]^\textrm{T}\in\mathbb{R}^{\tilde d \times 1}$. Then, the outage probability, with decoding threshold $\Theta$, is
\begin{align}
    Q_\textrm{out}
    &=\mathbf{E}_{1,d^\textrm{I}},
\end{align}
where
\begin{align}
    \mathbf{E}
    &=\mathrm{e}^{\Theta\mathbf{Q}^\textrm{I}},\\
    d^\textrm{I}
    &= \tilde d+1,\\
    \mathbf{Q}^\textrm{I}
    &=
    \begin{bmatrix}
    0 & \mathbf{\tilde p}\\
    \mathbf{0} & \mathbf{\tilde Q}
    \end{bmatrix}.
\end{align}
\end{theorem}
\begin{IEEEproof}
From Theorem~\ref{thm:Thm6d1}, the outage probability can be directly computed as
$Q_\textrm{out}
=\mathbb{P}\{Z \leq  \Theta \}
=\int_0^{ \Theta} \mathbf{\tilde p}\mathrm{e}^{z\mathbf{\tilde Q}}\mathbf{\tilde r} \, \mathrm{d}z
=\mathbf{e}_1^\textrm{T}\mathrm{e}^{ \Theta\mathbf{Q}^\textrm{I}}\mathbf{e}_{d^\textrm{I}}
=\mathbf{E}_{1,d^\textrm{I}}$.
\end{IEEEproof}

\begin{example}
\label{ex:Ex6d16}
(OSTBC-MRC with exponential fading SNR order $\tilde N$ -- Product form)
The effective channel has $F(s)=1/(1+s)^{\tilde {N}}$, with threshold $\tilde \Theta=(\mathrm{e}^{\tilde R}-1)/{\tilde S}$. This is handled with Theorem~\ref{thm:Thm6d7} and with parameters as in Ex.~\ref{ex:Ex6d8}.
\end{example}

\begin{example}
\label{ex:Ex6d17}
(Effective OSTBC-channel of order $\tilde N$ -- Polynomial form)
The Effective OSTBC-channel is as in Ex.~\ref{ex:Ex6d12}. Since, $\tilde q(s)=(1+s)^{\tilde {N}}=\sum_{n=0}^{\tilde {N}} \binom{\tilde {N}} {n} s^n$, and $\tilde p(s)=1$, we use the companion form \eqref{eq:Eq6d9} with $\mathbf{\tilde Q}=\mathbf{S}-\mathbf{\tilde r}\mathbf{\tilde q}$, $\mathbf{\tilde p}=[1 \ 0 \ldots 0]$, $\mathbf{\tilde q}=[\binom{\tilde {N}} {0} \ \binom{\tilde {N}} {1}  \ldots  \binom{\tilde {N}} {N-1}]$, and $\mathbf{\tilde r}=[0 \ldots 0 \ 1]^\textrm{T}$.
\end{example}
As this form is more complicated than the form in Ex.~\ref{ex:Ex6d12}, it illustrates the benefit of judiciously choosing the best representation for the problem studied.

\begin{example}
\label{ex:Ex6d18}
(SDC of order $N$ with exponentially-distributed wireless channel - Product form)
Use Theorem~\ref{thm:Thm6d8} with $\mathbf{p}$, $\mathbf{Q}$, and $\mathbf{r}$, as in Ex.~\ref{ex:Ex6d10}, with $\Theta=(\mathrm{e}^R-1)/S$.
\end{example}

A related performance measure to the outage probability, is the outage capacity, $C_\textrm{out}$, defined by $\mathbb{P}\{\ln(1+Z)<C_\textrm{out}\}=Q_\textrm{out}$, where $Q_\textrm{out}$ is the desired outage probability. For the ME-distributed effective SNR, we get
\begin{align}
\mathbf{e}_1^\textrm{T}\mathrm{e}^{ (\mathrm{e}^{C_\textrm{out}}-1)\mathbf{Q}^\textrm{I}}\mathbf{e}_{d^\textrm{I}}=Q_\textrm{out},
\end{align}
which is implicit in $C_\textrm{out}$, but can be solved numerically.

As discussed in Section \ref{sec:Sec6d3}, the ME-distribution can be used to approximate non-ME-distributed r.v.s. To exemplify this, the MIMO outage probability at high SNR is considered next.
\begin{example}
\label{ex:Ex6d19}
It has been shown in \cite[Thm. 1]{WangGian04} that the LT of the AWGN $N_\textrm{rx}\times N_\textrm{tx}$-MIMO channel capacity (with iid equal power zero-mean complex Gaussian signals) can be written $F(s)=B^{-1}\det {(\mathbf{G}(s))}$, with matrix $\mathbf{G}(s)$ entries $g_{ij}(s)=\int_0^\infty \frac{\lambda^{i+j+d}\mathrm{e}^{-\lambda}}{(1+t\lambda )^s} \, \mathrm{d}\lambda \ i,j=\{0,1,\ldots, k-1\}$, and where $B=\prod_{i=1}^k\Gamma(d+i)=\prod_{i=0}^{k-1} (d+i)!$, $t=SN_\textrm{tx}^{-1}$, $k=\min{(N_\textrm{rx},N_\textrm{tx})}$, $d=\max{(N_\textrm{rx},N_\textrm{tx})}-N_\textrm{min}$.
We show in Appendix that for high SNR,  asymptotically $g_{ij}(s)\sim\frac{(i+j+d)!}{t^{i+j+d+1}}\prod_{n=1}^{i+j+d+1}\frac{1}{s-n}$. This is a rational expression in $s$, and so is the asymptotic $F(s)$. For example, for $2\times 2$-MIMO, $F(s)\sim 1/(s-1)(s-2)^2(s-3)t^4$. Thus, the high SNR asymptotic outage probability for $2\times 2$-MIMO can be expressed as $Q_\textrm{out} \sim t^{-4} \mathbf{e}_1^\textrm{T}\mathrm{e}^{R \mathbf{ Q^\textrm{I}}}\mathbf{e}_5$, with parameter
\begin{align}
    \mathbf{Q}^\textrm{I}
    &=
    \begin{bmatrix}
    0 & 1 & 0 & 0 & 0 \\
    0 & 1 & 1 & 0 & 0 \\
    0 & 0 & 2 & 1 & 0 \\
    0 & 0 & 0  & 2 & 1 \\
    0 & 0 & 0  & 0 & 3 \\
    \end{bmatrix}.
\end{align}
The above procedure can be extended to higher-order MIMO systems. Determining $F(s)$ with the rational approximation to $g_{ij}(s)$, when $N=N_\textrm{rx}=N_\textrm{tx}$, we observe that $F(s)\sim t^{-N^2}\prod_{n=0}^{N-1}n! \prod_{n=1}^{N}\frac{1}{(s-n)^n}\prod_{n=N+1}^{2N-1}\frac{1}{(s-n)^{2N-n}}$. The first order rational approximation of $g_{ij}(s)$ used here, could be further refined with a higher order approximation, albeit at the cost of increased complexity.
\end{example}

\subsection{ARQ Throughput Analysis} 
\label{sec:Sec6d5d4}
In our analysis of retransmission schemes, we consider ARQ first. Here, ARQ operates under the assumption that a data packet can be decoded if the mutual information, for a transmission, exceeds the information rate of the data packet. ARQ, under such assumptions, has been studied in many works, e.g., \cite{CaireTuni01, LarssonRasmSkog14b, BetteshSham06, ShenLiuFitz08}.
ARQ is, as seen below, quite straightforward to analyze for the ME-distributed channel, but, due to its fundamental nature, it is important to include.
\begin{theorem}
\label{thm:Thm6d8}
(Outage probability and ARQ throughput for the ME-distributed effective channel)
Let the effective channel pdf be $f_Z(z)=\mathbf{\tilde p}\mathrm{e}^{z\mathbf{\tilde Q}}\mathbf{\tilde r}$. Then, the throughput of ARQ is
\begin{align}
    T^\textrm{ARQ}
    &=R(1-\mathbf{E}_{1,d^\textrm{I}}),
    \label{eq:Eq6d78}
\end{align}
with $\mathbf{E}_{1,d^\textrm{I}}$ given by Theorem~\ref{thm:Thm6d7}.
\end{theorem}
\begin{IEEEproof}
The throughput is simply $T^\textrm{ARQ}=R(1-Q_\textrm{out})$, and we then use the outage probability in Theorem~\ref{thm:Thm6d7}.
\end{IEEEproof}

\begin{remark}
\label{rm:Rm6d9}
Note that \eqref{eq:Eq6d78}, in contrast  to \cite[(32)]{LarssonRasmSkog16a}, uses the integration form in Theorem~\ref{thm:Thm6d1}. It also aligns better with the throughput expression for truncated-HARQ in  \eqref{eq:Eq6d80} (for which we provide a refined expression of), and is more clearly formulated in the effective channel SNR parameters.
\end{remark}
\begin{remark}
\label{rm:Rm6d10}
Alternatively, the throughput of ARQ can be expressed as $T^\textrm{ARQ}=RP^\textrm{ARQ}$, where, due to \eqref{eq:Eq6d4}, the probability of a successful transmission is $P^\textrm{ARQ}=\int_{ \Theta}^\infty \mathbf{\tilde p}\mathrm{e}^{z\mathbf{\tilde Q}}\mathbf{\tilde r} \, \mathrm{d}z=-\mathbf{\tilde p}\mathrm{e}^{ \Theta\mathbf{\tilde Q}}\mathbf{\tilde Q}^{-1}\mathbf{\tilde r}$. However, the form in \eqref{eq:Eq6d78} aligns better with the throughput expression for truncated-HARQ.
\end{remark}

\begin{corollary}
\label{cr:Crl6d3}
(Optimal ARQ throughput for the ME-distributed effective channel)
Let the ARQ throughput be defined as in Theorem~\ref{thm:Thm6d8}. Then, the optimal throughput can be determined with the function
\begin{align}
g_\Theta(\Theta)=\frac{1-\mathbf{E}_{1,d^\textrm{I}}}{\Theta \mathbf{\tilde p}\mathrm{e}^{\Theta \mathbf{\tilde Q}}\mathbf{\tilde r}},
\end{align}
together with the auxiliary parametric optimization method in \cite{LarssonRasmSkog14b}, as also reviewed in Appendix \ref{sec:Sec7}.
\end{corollary}
\begin{IEEEproof}
In \cite{LarssonRasmSkog14b}, we define  $g_\Theta(\Theta)\triangleq f_\Theta(\Theta)/\Theta f'_\Theta(\Theta)$. We have $f_\Theta(\Theta)=1/(1-\mathbf{E}_{1,d^\textrm{I}})$. It is also noted that $\frac{\mathrm{d}}{\mathrm{d}\Theta} \mathbf{E}_{1,d^\textrm{I}}=\frac{\mathrm{d}}{\mathrm{d}\Theta} \int_0^\Theta \mathbf{\tilde p}\mathrm{e}^{z\mathbf{\tilde Q}}\mathbf{\tilde r} \, \mathrm{d}z=\mathbf{\tilde p}\mathrm{e}^{\Theta\mathbf{\tilde Q}}\mathbf{\tilde r}$.
\end{IEEEproof}

\subsection{Truncated-HARQ Throughput Analysis}
\label{sec:Sec6d5d5}
In this section, we consider truncated-HARQ where the number of transmission is limited to $K$ transmissions. Equipped with the powerful idea of a ME-distributed effective channel, and using Theorem~\ref{thm:Thm6d1}, the throughput expression is possible to give on a particularly simple form, as shown next.

\begin{theorem}
\label{thm:Thm6d9}
(Truncated-HARQ throughput for the ME-distributed effective channel)
Let the effective channel density be $f_Z(z)=\mathbf{\tilde p}\mathrm{e}^{z\mathbf{\tilde Q}}\mathbf{\tilde r}$. Then, the throughput of truncated-HARQ, with a maximum of $K$ transmissions and decoding threshold $\Theta$, is
\begin{align}
    T^\textrm{HARQ}_{K}
    &=\frac{R(1-\mathbf{E}_{1,(dK+1)})}{1+\sum_{k=1}^{K-1} \mathbf{E}_{1,(dk+1)}},
    \label{eq:Eq6d80}
\end{align}
where
\begin{align}
    \mathbf{E}
    &=\mathrm{e}^{\Theta\mathbf{Q}^\textrm{I}_{K\circledast}},
\end{align}
and the ME-parameters are
\begin{align}
    \mathbf{p}_{K\circledast}^\textrm{I}
    &=[\mathbf{\tilde p} \ \mathbf{0}] \in \mathbb{R}^{1 \times d^\textrm{I}},\\
    \mathbf{Q}^\textrm{I}_{K\circledast}
    &=
    \begin{bmatrix}
    0 & \mathbf{p}_{K\circledast} \\
    \mathbf{0} & \mathbf{Q}_{K\circledast} \\
    \end{bmatrix}
    \in \mathbb{R}^{ d^\textrm{I} \times d^\textrm{I}},\\
    \mathbf{r}_{K\circledast}^\textrm{I}
    &=\mathbf{e}_{d^\textrm{I}} \in \mathbb{R}^{ d^\textrm{I} \times 1},\\
    d^\textrm{I}
    &= \tilde dK+1,\\
    \mathbf{p}_{K\circledast}
    &=[\mathbf{\tilde p} \ \mathbf{0}] \in \mathbb{R}^{1 \times \tilde dK},\\
    \mathbf{Q}_{K\circledast}
    &=
    \begin{bmatrix}
    \mathbf{\tilde Q} & \mathbf{\tilde P} & \mathbf{0} & \cdots & \mathbf{0} & \mathbf{0} \\
    \mathbf{0} & \mathbf{\tilde Q} & \mathbf{\tilde P} & \cdots & \mathbf{0} & \mathbf{0} \\
    \mathbf{0} & \mathbf{0} & \mathbf{\tilde Q} & \ddots & \mathbf{0} & \mathbf{0} \\
    \vdots & \vdots & \vdots & \ddots & \vdots & \vdots \\
    \mathbf{0} & \mathbf{0} & \mathbf{0} & \cdots & \mathbf{\tilde Q} & \mathbf{\tilde P} \\
    \mathbf{0} & \mathbf{0} & \mathbf{0}  & \cdots & \mathbf{0} & \mathbf{\tilde Q} \\
    \end{bmatrix} 
    \in \mathbb{R}^{\tilde dK \times \tilde dK},
    \label{eq:Eq6d87}\\
    \mathbf{r}_{K\circledast}
    &=\mathbf{e}_{\tilde dK} \in \mathbb{R}^{ \tilde dK \times 1},\\
    \mathbf{\tilde Q}
    &=\mathbf{S}-\mathbf{\tilde r}\mathbf{\tilde q} \in \mathbb{R}^{ \tilde d \times \tilde d},\\
    \mathbf{\tilde P}
    &=\mathbf{\tilde r}\mathbf{\tilde p} \in \mathbb{R}^{ \tilde d \times \tilde d}.
\end{align}
\end{theorem}
\begin{IEEEproof}
The throughput for truncated-HARQ, expressed on a LT-form rather than outage-probability form, is $T_K^\textrm{HARQ}=R \left(1-\mathcal{L}_\Theta^{-1}\{s^{-1}F(s)^K\}\right) /\left(1+\mathcal{L}_\Theta^{-1} \{\sum_{k=1}^{K-1}s^{-1}F(s)^k\}\right)$. For the numerator, the K-fold convolution, is needed, whereas for the denominator, the $k$-fold convolutions, $k\in\{1,2,\ldots K-1\}$, are needed. Thanks to the upper triangular block-structure of \eqref{eq:Eq6d87}, we propose to  compute all required convolutions at the same time when the $K$-fold convolution is computed. To see this, consider first $\mathrm{e}^{\mathbf{X}_1}=\mathbf{Y}_1$, where $\mathbf{X}_1=\mathbf{X}_{11}$, and $\mathbf{Y}_1=\mathbf{Y}_{11}$. Then, extend the $\mathbf{X}$-matrix to $\mathbf{X}_2=\begin{bmatrix}\mathbf{X}_{11} & \mathbf{X}_{12}\\ \mathbf{0} & \mathbf{X}_{22} \end{bmatrix}$. Then, the $\mathbf{Y}$-matrix is $\mathbf{Y}=\begin{bmatrix}\mathbf{Y}_{11} & \mathbf{Y}_{12}\\ \mathbf{0} & \mathbf{Y}_{22} \end{bmatrix}$. Hence, by  computing $\mathrm{e}^{\mathbf{X}_2}$, one gets $\mathrm{e}^{\mathbf{X}_1}=\mathbf{Y}_{1}=\mathbf{Y}_{11}$ at the same time. By induction, this procedure can be extended to $K$-fold convolution.
\end{IEEEproof}

\begin{remark}
\label{rm:Rm6d11}
Observe that \eqref{eq:Eq6d80} is, in contrast to \cite[(30)-(31)]{LarssonRasmSkog16a} which uses vector convolution \cite[(28)]{LarssonRasmSkog16a}, expressed on a more structured, simpler  and more intuitive form thanks to the convolution formulation of \eqref{eq:Eq6d87}. Another difference is that \eqref{eq:Eq6d80} only requires one ME computation, whereas \cite[(30)-(31)]{LarssonRasmSkog16a} requires two, i.e. in addition to the vector convolution.
\end{remark}

\begin{remark}
\label{rm:Rm6d12}
Similar to ARQ, the optimal throughput for truncated-HARQ can be determined with the auxiliary parametric optimization method reviewed in Appendix \ref{sec:Sec7d1}. Then, $g_\Theta(\Theta)\triangleq f_\Theta(\Theta)/\Theta f'_\Theta(\Theta)$ with $f_\Theta(\Theta)=
(1+\sum_{k=1}^{K-1} \mathbf{E}_{1,(dk+1)})/
(1-\mathbf{E}_{1,(dK+1)})$. While the expression is easy to determine, it is not tidy enough to be given here.
\end{remark}

\subsection{Persistent-HARQ Throughput Analysis}
\label{sec:Sec6d5d6}
We now consider persistent-HARQ with no upper retransmission limit, i.e. $K=\infty$. In HARQ, if a data packet can not be decoded, all previous  transmissions for the same packets are combined. We assume that if the accumulated mutual information exceeds the initial information rate $R$, the packet can be correctly decoded. This operation is also the foundation in many modern works on HARQ, e.g., \cite{CaireTuni01, LarssonRasmSkog14b, WuJind10, SzczecinskiKDR13}. For truncated-HARQ, the dimension of the ME-vectors and matrices grows linearly with $K$. Clearly, this is a problem when $K\rightarrow \infty$, as for persistent-HARQ. Despite this apparent issue, the ME-distribution framework can, as shown below, handle the persistent-HARQ case too. In contrast to the ARQ and truncated-HARQ analysis, the effective channel on a polynomial rational LT form, $F(s)=\tilde{p}(s)/\tilde{q}(s)$, is preferred in the following analysis.

\begin{theorem}
\label{thm:Thm6d10}
(Persistent-HARQ throughput for the ME-distributed effective channel)
Let the effective channel pdf $f_Z(z)=\mathbf{\tilde p}\mathrm{e}^{z\mathbf{\tilde Q}}\mathbf{\tilde r}$ have LT $F(s)=\tilde p(s)/\tilde q(s)$. Then, the throughput, with decoding threshold $\Theta$, is
\begin{align}
    T^\textrm{HARQ}_{\infty}
    &=\frac{R}{1+\mathbf{E}_{1,d^\textrm{I}}},
\label{eq:Eq6d91}
\end{align}
where
\begin{align}
    \mathbf{E}
    &=\mathrm{e}^{\Theta\mathbf{Q}^\textrm{I}},\\
    \mathbf{Q}^\textrm{I}
    &=\begin{bmatrix}
    0 & \mathbf{\tilde p}\\
    \mathbf{0} & \mathbf{S}-\mathbf{\tilde r}(\mathbf{\tilde q}-\mathbf{\tilde p})
    \end{bmatrix}.
\end{align}
\end{theorem}
\begin{IEEEproof}
The mean number of transmissions is
\begin{align} 
&\mathfrak{L}^{-1}_\Theta\left\{\frac{1}{s}\frac{1}{1-F(s)}\right\}\notag
    =\mathfrak{L}^{-1}_\Theta\left\{\frac{1}{s} \frac{1}{1-\tilde p(s)/ \tilde q(s)}\right\}\notag\\
   & =1+\mathfrak{L}^{-1}_\Theta\left\{\frac{1}{s} \frac{\tilde p(s)}{\tilde q(s)-\tilde p(s)}\right\}\notag\\
    &=1+\mathbf{e}_1\textrm{T}\mathrm{e}^{\Theta\mathbf{Q}^\textrm{I}}\mathbf{e}_{d^\textrm{I}}
    =1+\mathbf{E}_{1,d^\textrm{I}}, \ \mathbf{E}\triangleq \mathrm{e}^{\Theta\mathbf{Q}^\textrm{I}}.\notag
\end{align}
\end{IEEEproof}

\begin{remark}
\label{rm:Rm6d13}
Observe that \eqref{eq:Eq6d91} differ wrt \cite[(21)]{LarssonRasmSkog16a}, in a similar way as discussed in Remark~\ref{rm:Rm6d9}. In addition, we have now generalized \eqref{eq:Eq6d91} to include arbitrary diversity order $N$ in Corollary~\ref{cr:Crl6d4}. In \cite[Sec. IV.F.4]{LarssonRasmSkog16a}, only $N=2$ for the ME-distributed channel $F(s)=p(s)/q(s)$ was handled.
\end{remark}

For persistent-HARQ, it is preferred using the polynomial LT-form for the effective channel. This cater for that the mean number of transmission metric is easy to express on a ME-form. However, an exception that can also be handled is for diversity, where the LT of the effective channel SNR takes the form $F(s)=(p(s)/q(s))^N$.
\begin{corollary}
\label{cr:Crl6d4}
(Persistent-HARQ throughput for the ME-distributed wireless channel and $N$-fold diversity)
Let the effective channel pdf $f_Z(z)=\mathbf{\tilde p}\mathrm{e}^{z\mathbf{\tilde Q}}\mathbf{\tilde r}$  have LT $F(s)=\tilde p(s)/\tilde q(s)= ( p(s)/ q(s))^N, N\in \mathbb{N}^+$. Then, the throughput, with decoding threshold $\Theta$, is
\begin{align}
    T^\textrm{HARQ}_{\infty}
    &=\frac{R}{1+\mathbf{E}_{1,d^\textrm{I}}},
\end{align}
where
\begin{align}
    \mathbf{E}
    &=\mathrm{e}^{\Theta\mathbf{Q}^\textrm{I}_{N\circledast}},
\end{align}
and the ME-parameters are
\begin{align}
    \mathbf{p}_{N\circledast}^\textrm{I}
    &=[\mathbf{\tilde p} \ \mathbf{0}] \in \mathbb{R}^{1 \times d^\textrm{I}}, \\
    \mathbf{Q}^\textrm{I}_{N\circledast}
    &=
    \begin{bmatrix}
    0 & \mathbf{p}_{N\circledast} \\
    \mathbf{0} & \mathbf{Q}_{N\circledast} \\
    \end{bmatrix}
    \in \mathbb{R}^{ d^\textrm{I} \times d^\textrm{I}},\\
    \mathbf{r}_{N\circledast}^\textrm{I}
    &=\mathbf{e}_{d^\textrm{I}} \in \mathbb{R}^{ d^\textrm{I} \times 1},\\
    d^\textrm{I}&= 1+\tilde d,\\
    \mathbf{p}_{N\circledast}
    &=[\mathbf{p} \ \mathbf{0}] \in \mathbb{R}^{1 \times \tilde d},\\
    \mathbf{Q}_{N\circledast}
    &\triangleq
    \begin{bmatrix}
    \mathbf{\breve{Q}}_1 & \mathbf{P} & \mathbf{0} & \cdots & \mathbf{0} & \mathbf{0} \\
    \mathbf{0} & \mathbf{\breve{Q}}_2 & \mathbf{P} & \cdots & \mathbf{0} & \mathbf{0} \\
    \mathbf{0} & \mathbf{0} & \mathbf{\breve{Q}}_3 & \ddots & \mathbf{0} & \mathbf{0} \\
    \vdots & \vdots & \vdots & \ddots & \vdots & \vdots \\
    \mathbf{0} & \mathbf{0} & \mathbf{0} & \cdots & \mathbf{\breve{Q}}_{N-1} & \mathbf{P}  \\
    \mathbf{0} & \mathbf{0} & \mathbf{0}  & \cdots & \mathbf{0} & \mathbf{\breve{Q}}_N \\
    \end{bmatrix}
    \in \mathbb{R}^{\tilde d \times \tilde d},\\
    \mathbf{r}_{N\circledast}
    &=\mathbf{e}_{\tilde d} \in \mathbb{R}^{\tilde d \times 1},\\
    \tilde d
    &=dN,\\
    \mathbf{P}
    &=\mathbf{r}\mathbf{p} \in \mathbb{R}^{d \times d},\\
    \mathbf{\breve{Q}}_n
    &=\mathbf{S}-\mathbf{r}\left(\mathbf{q}- \mathbf{p}\mathrm{e}^{\frac{2\pi \mathrm{i} n}{N}} \right) \in \mathbb{R}^{d \times d}, n\in \{0,1,\ldots N-1\}.
\end{align}
\end{corollary}

\begin{IEEEproof}
The mean number of transmissions is
\begin{align} 
&\mathfrak{L}^{-1}_\Theta\left\{\frac{1}{s}\frac{1}{1-F(s)}\right\}\notag
=1+ \mathfrak{L}^{-1}_\Theta\left\{\frac{1}{s}\frac{p(s)^N}{q(s)^N-p(s)^N}\right\}\notag\\    
&\overset{(a)}{=}1+\mathfrak{L}^{-1}_\Theta\left\{\frac{1}{s} \prod_{n=0}^{N-1}\frac{p(s)}{q(s)-p(s) \mathrm{e}^{2\pi\mathrm{i}n/N}}\right\}\notag\\
    &=1+\mathbf{e}_1^\textrm{T}\mathrm{e}^{\Theta\mathbf{Q}^\textrm{I}_{N\circledast}}\mathbf{e}_{d^\textrm{I}}\notag\\
    &=1+\mathbf{E}_{1,d^\textrm{I}}, \ \mathbf{E}\triangleq \mathrm{e}^{\Theta\mathbf{Q}^\textrm{I}_{N\circledast}},\notag
\end{align}
Note that the expression at step (a) is expressed on a product form, which suggests the use of Proposition \ref{pr:Pr6d1} (Convolution).
\end{IEEEproof}

\begin{example}
\label{ex:Ex6d20}
(Persistent-HARQ with diversity order 2)
Consider Theorem~\ref{thm:Thm6d10} with $N=2$. Then,
\begin{align}
    \mathbf{Q}^\textrm{I}_{2\circledast}
    =\begin{bmatrix}
    0 & \mathbf{p} & \mathbf{0}\\
    \mathbf{0} & \mathbf{S}-\mathbf{r}(\mathbf{q}-\mathbf{p}) & \mathbf{r}\mathbf{p}\\
    \mathbf{0} & \mathbf{0} & \mathbf{S}-\mathbf{r}(\mathbf{q}+\mathbf{p})
    \end{bmatrix},
\end{align}
since
\begin{align}
    &\mathfrak{L}^{-1}_\Theta\left\{\frac{1}{s}\frac{1}{1-(p(s)/q(s))^2}\right\}\notag\\
   & =1+\mathfrak{L}^{-1}_\Theta\left\{\frac{1}{s}\frac{p(s)}{q(s)-p(s)} \frac{p(s)}{q(s)+p(s)}\right\}.\notag
\end{align}
\end{example}

The following example formulate an alternative expression for an OSTBC-MRC Nakagami-$m$ channel on a simple form.
\begin{example}
\label{ex:Ex6d21}
(Diversity order $\tilde{N}$ - Alternative form)
Consider an effective channel with $F(s)=1/(1+s)^{\tilde N},\tilde N\in \mathbb{N}^+$, and threshold $\tilde \Theta$. Then, the throughput can be compactly expressed as
\begin{align}
    T^\textrm{HARQ}_{\infty}
    &=\frac{R}{\mathbf{E}_{d^\textrm{I},d^\textrm{I}}},
\end{align}
where
\begin{align}
    \mathbf{E}
    &=\mathrm{e}^{\tilde \Theta\mathbf{Q}^\textrm{I}},
\end{align}
and the ME-parameters are
\begin{align}
    \mathbf{p}^\textrm{I}
    &=[0  \ldots  0 \ 1] \in \mathbb{R}^{1 \times d^\textrm{I}}, \\
    \mathbf{Q}^\textrm{I}
    &=\mathbf{S}-\mathbf{r}^\textrm{I} \mathbf{q}^\textrm{I}- \mathbf{I} \in \mathbb{R}^{ d^\textrm{I} \times d^\textrm{I}},\\
    \mathbf{q}^\textrm{I}
    &=[-1 \ 1 \ 0  \ldots  0 \ 1] \in \mathbb{R}^{1 \times d^\textrm{I}},\\
    \mathbf{r}^\textrm{I}
    &=\mathbf{e}_{d^\textrm{I}} \in \mathbb{R}^{ d^\textrm{I} \times 1},\\
    d^\textrm{I}
    &= \tilde{N}.
\end{align}
\begin{IEEEproof}
The mean number of transmissions is
\begin{align}
&\mathcal{L}_{\tilde \Theta}^{-1}
\left\{ \frac{1}{s}\frac{1}{1-F(s)}\right\}
=\mathcal{L}_{\tilde \Theta}^{-1} \left\{ \frac{(1+s)^{\tilde N}}{s((1+s)^{\tilde N}-1)}\right\} \notag\\
&\overset{(a)}{=} \mathrm{e}^{-\tilde\Theta}\mathcal{L}_{\tilde \Theta}^{-1} \left
\{\frac{s^{\tilde N}}{s^{{\tilde N}+1}-s^{{\tilde N}}-s+1}\right \},\notag
\end{align}
where we simplified the numerator and denominator by using the
frequency shift property of the Laplace transform,
$\mathcal{L}^{-1} \left \{ G(s+x) \right
\}=\mathrm{e}^{-x}\mathcal{L}^{-1} \left \{ G(s) \right \}$ in step (a). We then express the rational LT on the ME-distribution form.
\end{IEEEproof}
\end{example}

\begin{corollary}
\label{cr:Crl6d5}
(Optimal persistent-HARQ throughput for the ME-distributed effective channel)
Let the persistent-HARQ throughput be defined as in Theorem~\ref{thm:Thm6d10}. Then, the optimal throughput, using the auxiliary parametric optimization method in Appendix \ref{sec:Sec7}, can be determined with
\begin{align}
g_\Theta(\Theta)=\frac{1+\mathbf{E}_{1,d^\textrm{I}}} {\Theta \mathbf{p}_{N\circledast}\mathrm{e}^{\Theta \mathbf{Q}_{N\circledast}}\mathbf{r}_{N\circledast}}.
\end{align}
\end{corollary}
\begin{IEEEproof}
We have $f_\Theta(\Theta)=1+\mathbf{E}_{1,d^\textrm{I}}$. In \cite{LarssonRasmSkog14b} we defined $g_\Theta(\Theta)\triangleq f_\Theta(\Theta)/\Theta f'_\Theta(\Theta)$. We also note that $f'_\Theta(\Theta)=\frac{\mathrm{d}}{\mathrm{d}\Theta} \int_0^\Theta \mathbf{p}_{N\circledast} \mathrm{e}^{z\mathbf{Q}_{N\circledast}} \mathbf{r}_{N\circledast} \, \mathrm{d}z
=\mathbf{p}_{N\circledast}\mathrm{e}^{\Theta\mathbf{Q}_{N\circledast}} \mathbf{r}_{N\circledast}$.
\end{IEEEproof}
From Section \ref{sec:Sec6d5d4}-\ref{sec:Sec6d5d6}, we conclude that the analysis framework handles the truncated/persistent (H)ARQ schemes for any ME-distributed effective channel.

\subsection{3-phase Network Coded Bidirectional Relaying (NCBR)} 
\label{sec:Sec6d5d7} 
\begin{figure}[t]
 \centering
 \vspace{+.1 cm}
 \includegraphics[width=11.5cm]{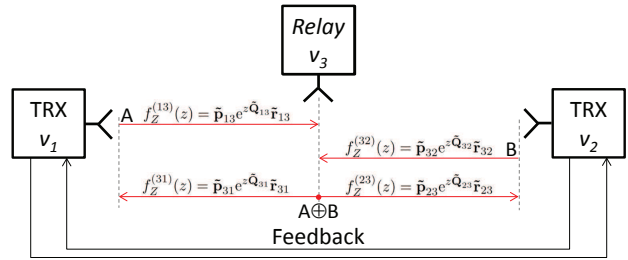}
 \vspace{-5.2cm}
 \caption{3-phase network coded bidirectional relaying (NCBR).}
 \label{fig:NCBR}
 \vspace{-0.3cm}
\end{figure}

To exemplify another use case of the ME-distribution approach, we consider 3-phase network coded bidirectional relaying (NCBR)  \cite{Larsson04P, LarssonJohaSune05, LarssonJohaSune06}, with end-to-end (ETE) ARQ, as schematically depicted in Fig.~\ref{fig:NCBR}. The system has 3 nodes, $\nu_1,\nu_2,\nu_3$, where $\nu_1$ ($\nu_2$) wants to communicate data $A$ ($B$) to $\nu_2$ ($\nu_1$), via relay node $\nu_0$.  The relay node  $\nu_0$ perform network coding, schematically illustrated as an XOR-operation of $A$ and $B$, and the receiving nodes, $\nu_1$ and $\nu_2$, perform corresponding decoding given knowledge of  $A$ and $B$, respectively. Data $A$ and $B$ are exchanged in only three phases thanks to the network coding operation. The network coding approach is assumed to allow for the data rates  $R_{12}$ and  $R_{21}$ to differ. We further assume that each links effective SNR (after any potential processing that may differ between the links) are iid ME-distributed.
\begin{theorem}
\label{thm:Thm6d11}
Consider the 3-phase NCBR model with nodes $\nu_1,\nu_2,\nu_3$, in Fig.~\ref{fig:NCBR}. Let each link between a node pair $\{\nu_i, \nu_j\}$, $\{ij\}=\{13,32,23,31\}$, be characterized by the effective channel SNR r.v. $Z_{ij}$ with pdf $f_Z^{(ij)}(z)=\mathbf{\tilde p}_{ij}\mathrm{e}^{z\mathbf{\tilde Q}_{ij}}\mathbf{\tilde r}_{ij}$, $\mathbf{\tilde p}_{ij}\in\mathbb{R}^{1 \times \tilde d_{ij}}$, $\mathbf{\tilde Q}_{ij}\in\mathbb{R}^{\tilde d_{ij}\times \tilde d_{ij}}$, $\mathbf{\tilde r}_{ij}=[0 \ldots 0 \ 1]^\textrm{T}\in\mathbb{R}^{\tilde d_{ij} \times 1}$, and the decoding threshold $\Theta_{12}=\mathrm{e}^{R_{12}}-1$,  $\Theta_{21}=\mathrm{e}^{R_{21}}-1$. Then, the ETE sum-throughput is
\begin{align}
    T^\textrm{NCBR}
	&=\frac{R_{12}(1-Q_{12})+R_{21}(1-Q_{21})}{3},
    \label{eq:Eq6d115}
\end{align}
where
\begin{align}
Q_{ij}
&=1-
(1-\mathbf{E}^{(i3)}_{1,d^\textrm{I}_{i3}})
(1-\mathbf{E}^{(3j)}_{1,d^\textrm{I}_{3j}}), \label{eq:Eq6d116}\\
\mathbf{E}^{(ij)}
    &\triangleq \mathrm{e}^{\Theta_{ij} \mathbf{Q}_{ij}^\textrm{I}},
    \label{eq:Eq6d117}\\
    \mathbf{Q}_{ij}^\textrm{I}
    &=
    \begin{bmatrix}
    0 & \mathbf{\tilde p}_{ij}\\
    \mathbf{0} & \mathbf{\tilde Q}_{ij}
    \end{bmatrix}.
    \label{eq:Eq6d118}
\end{align}
\end{theorem}
\begin{IEEEproof}
The ETE outage probability is
\begin{align}
Q_{ij}
&=1-\mathbb{P}\left\{\ln(1+\min(Z_{i3},Z_{3j}))>R_{ij}\right\}\notag\\
&=1-\mathbb{P}\left\{\min(Z_{i3},Z_{3j})>\Theta_{ij}\right\}\notag\\
&=1-\mathbb{P}\left\{Z_{i3}>\Theta_{ij}\}\mathbb{P}\{Z_{3j}>\Theta_{ij}\right\}\notag\\
&=1-(1-\mathbb{P}\left\{Z_{i3}<\Theta_{ij}\right\})(1-\mathbb{P}\left\{Z_{3j}<\Theta_{ij}\right\})\notag\\
&=1-
(1-\mathbf{E}^{(i3)}_{1,d^\textrm{I}_{i3}})
(1-\mathbf{E}^{(3j)}_{1,d^\textrm{I}_{3j}}),\notag
\end{align}
with $ \mathbf{E}^{(ij)}$ given by \eqref{eq:Eq6d117}.
\end{IEEEproof}
Note that  \eqref{eq:Eq6d116} corresponds to the new form of the minimum operator given in Remark~\ref{rm:Rm6d3}.
With symmetry, $f^{(13)}_Z(z)=f^{(23)}_Z(z)$,   $f^{(31)}_Z(z)=f^{(32)}_Z(z)$, and $R_{12}=R_{21}$, the throughput simplifies to
$T^\textrm{NCBR}=2R_{13}(1-\mathbf{E}^{(13)}_{1,d^\textrm{I}_{13}})(1-\mathbf{E}^{(32)}_{1,d^\textrm{I}_{32}})/3$.
From the above, we observe that the ME-distribution approach can also be useful for studies of more complicated, cooperating, multi-node, systems.

\subsection{Throughput Analysis of ARQ with ME-distributed Signal and Interferers}
\label{sec:Sec6d5d8} 
In many wireless systems, such as cellular systems, the interference case is the normal state of operation, rather than an interference-free state. This problem, ARQ performance in presence of interference, has not been studied in many works. The authors studied this problem in \cite{LarssonRasmSkog14a}, where the analytical framework only gave outage probabilities for the case when the desired signal was exponentially distributed, and the interference signals where Nakagami-$m$ distributed, or vice versa. In this section, we generalize the problem in \cite{LarssonRasmSkog14a}, incorporate the ME-distribution framework, and consider the case where the signal of interest and the interferes are all ME-distributed.

\subsubsection{System Model of ME-distributed Signal and Interferers} 
\label{sec:Sec6d5d8d1}
To handle interference, we need to revise the system model somewhat. The probability of successful decoding, with the signal of interest affected by interference, is $P_\textrm{Int}^\textrm{ARQ} =\mathbb{P}\left\{\ln\left(1+ Z/(1+Z_\textrm{I})\right)> R\right\}=\mathbb{P}\{Z\leq \Theta(1+Z_\textrm{I})\}$, where $Z$ is the SNR r.v. of the signal of interest, $Z_\textrm{I}$ is the sum-interference SNR, and $\Theta=\mathrm{e}^R-1$. The sum-interference SNR is $ Z_\textrm{I}= \sum_{u=1}^U Z_u$,
where the interfering users $u$ have SNR pdf $f_{Z}^{(u)}(z)\sim \mathbf{p}_u\mathrm{e}^{z \mathbf{Q}_u}\mathbf{r}_u$, each with a rational LT $F^{(u)}(s)=p_u(s)/q_u(s)$. This scenario was discussed in Ex.~\ref{ex:Ex6d9} where the ME-parameters were given for two interfering users. Thus, we have in total, $f_{Z_\textrm{I}}(z_\textrm{I})= \mathbf{p}_\textrm{I}\mathrm{e}^{z \mathbf{Q}_\textrm{I}}\mathbf{r}_\textrm{I}$ with rational LT $F_\textrm{I}(s)=\prod_{u=1}^U p_u(s)/ q_u(s)$. The signal of interest is also ME-distributed with density $f_Z(z)=\mathbf{p}\mathrm{e}^{z \mathbf{Q}}\mathbf{r}$. Later in this section, we generalize this model with independent signal and sum-interferer fading, to a joint density $f_{Z_\textrm{I},Z}(z_\textrm{I},z)= \mathbf{p}_\textrm{I}\mathrm{e}^{z_\textrm{I} \mathbf{Q}_\textrm{I}}\mathbf{P}_{12} \mathrm{e}^{z \mathbf{Q}}\mathbf{r}$, where $\mathbf{P}_{12}$ is a matrix, allowing for dependencies between the signal and sum-interference SNRs.

\subsubsection{Throughput Analysis} \label{sec:Sec6d5d8d2}
For the analysis, the following the lemma is helpful.
\begin{lemma} 
(Integral of product of independent ME-densities) \label{lm:Lm6d1}
\begin{align}
    &\int_0^\infty \mathbf{x}_1\mathrm{e}^{t \mathbf{Y}_1}\mathbf{z}_1 \mathbf{x}_2\mathrm{e}^{t \mathbf{Y}_2}\mathbf{z}_2 \, \mathrm{d}t\notag\\
    &=-\left(\mathbf{x}_1\otimes \mathbf{x}_2\right)(\mathbf{Y}_1\oplus \mathbf{Y}_2)^{-1}\left(\mathbf{z}_1 \otimes\mathbf{z}_2\right).
\end{align}
\end{lemma}
\begin{IEEEproof}
Using the same properties as in Theorem~\ref{thm:Thm6d2}, the integral is computed as
\begin{align}
    &\int_0^\infty \mathbf{x}_1\mathrm{e}^{t \mathbf{Y}_1}\mathbf{z}_1 \mathbf{x}_2\mathrm{e}^{t \mathbf{Y}_2}\mathbf{z}_2 \, \mathrm{d}t \notag\\
    &=\int_0^\infty \mathbf{x}_1\mathrm{e}^{t \mathbf{Y}_1}\mathbf{z}_1 \otimes \mathbf{x}_2\mathrm{e}^{t \mathbf{Y}_2}\mathbf{z}_2 \, \mathrm{d}t\notag\\
    &=\int_0^\infty  \left(\mathbf{x}_1\mathrm{e}^{t \mathbf{Y}_1} \otimes \mathbf{x}_2\mathrm{e}^{t \mathbf{Y}_2}\right)\left(\mathbf{z}_1 \otimes \mathbf{z}_2\right) \, \mathrm{d}t\notag\\
    &=\int_0^\infty  \left(\mathbf{x}_1\otimes \mathbf{x}_2\right)\left(\mathrm{e}^{t \mathbf{Y}_1} \otimes \mathrm{e}^{t \mathbf{Y}_2}\right)\left(\mathbf{z}_1 \otimes\mathbf{z}_2\right) \, \mathrm{d}t\notag\\
    &=\int_0^\infty  \left(\mathbf{x}_1\otimes \mathbf{x}_2\right)\left(\mathrm{e}^{t  (\mathbf{Y}_1\oplus \mathbf{Y}_2)}\right)\left(\mathbf{z}_1 \otimes\mathbf{z}_2\right) \, \mathrm{d}t \notag\\
    &=-\left(\mathbf{x}_1\otimes \mathbf{x}_2\right)(\mathbf{Y}_1\oplus \mathbf{Y}_2)^{-1}\left(\mathbf{z}_1 \otimes\mathbf{z}_2\right). \notag
\end{align}
\end{IEEEproof}

Using Lemma~\ref{lm:Lm6d1}, the throughput of ARQ with ME-distributed signal and sum-interference is now given.
\begin{theorem}
\label{thm:Thm6d12}
(ARQ throughput for iid ME-distributed signal and interferers)
Let the signal and sum-interferer be given by the system model. Then, the throughput is
\begin{align}
    T^\textrm{ARQ}_\textrm{Int}
    &=R\left(\mathbf{p}_\textrm{I}\otimes \mathbf{p}\right)
    \left((\mathbf{Q}_\textrm{I}\oplus \Theta \mathbf{Q})(\mathbf{I}\otimes \mathbf{Q}\mathrm{e}^{-\Theta \mathbf{Q}})\right)^{-1} \left(\mathbf{r}_\textrm{I}\otimes\mathbf{r}\right).
\end{align}
\end{theorem}
\begin{IEEEproof}
The throughput is $T^\textrm{ARQ}_\textrm{Int}=RP^\textrm{ARQ}_\textrm{Int}$, where $P_\textrm{Int}^\textrm{ARQ}=\mathbb{P}\{Z> \Theta(1+Z_\textrm{I})\}$ is determined via the integral
\begin{align}
    P^\textrm{ARQ}_\textrm{Int}
    &= \int_0^\infty \int_{\Theta(1+z_\textrm{I})}^\infty \mathbf{p}_\textrm{I}\mathrm{e}^{z_\textrm{I} \mathbf{Q}_\textrm{I}}\mathbf{r}_\textrm{I}
    \mathbf{p}\mathrm{e}^{z \mathbf{Q}}\mathbf{r} \, \mathrm{d}z_\textrm{I} \, \mathrm{d}z\notag\\
    &= -\int_0^\infty \mathbf{p}_\textrm{I}\mathrm{e}^{z_\textrm{I} \mathbf{Q}_\textrm{I}}\mathbf{r}_\textrm{I}
    \mathbf{p}\mathbf{Q}^{-1}\mathrm{e}^{\Theta \mathbf{Q}}\mathrm{e}^{z_\textrm{I} \Theta \mathbf{Q}}\mathbf{r} \, \mathrm{d}z_\textrm{I}    \label{eq:Eq6d105}\\
    &\overset{(a)}{=}\left(\mathbf{p}_\textrm{I}\otimes \left(\mathbf{p}\mathbf{Q}^{-1}\mathrm{e}^{\Theta \mathbf{Q}}\right)\right)(\mathbf{Q}_\textrm{I}\oplus \Theta \mathbf{Q})^{-1}\left(\mathbf{r}_\textrm{I}\otimes\mathbf{r}\right)\notag\\
    &\overset{(b)}{=}\left(\mathbf{p}_\textrm{I}\otimes \mathbf{p}\right)
    \left(\mathbf{I}\otimes \mathbf{Q}^{-1}\mathrm{e}^{\Theta \mathbf{Q}}\right) (\mathbf{Q}_\textrm{I}\oplus \Theta \mathbf{Q})^{-1}\left(\mathbf{r}_\textrm{I}\otimes\mathbf{r}\right)\notag\\
    &\overset{(c)}{=}\left(\mathbf{p}_\textrm{I}\otimes \mathbf{p}\right)
    \left(\mathbf{I}\otimes \mathbf{Q}\mathrm{e}^{-\Theta \mathbf{Q}}\right)^{-1} (\mathbf{Q}_\textrm{I}\oplus \Theta \mathbf{Q})^{-1}\left(\mathbf{r}_\textrm{I}\otimes\mathbf{r}\right)\notag\\
    &\overset{(d)}{=}\left(\mathbf{p}_\textrm{I}\otimes \mathbf{p}\right)
    \left((\mathbf{Q}_\textrm{I}\oplus \Theta \mathbf{Q})(\mathbf{I}\otimes \mathbf{Q}\mathrm{e}^{-\Theta \mathbf{Q}})\right)^{-1}\left(\mathbf{r}_\textrm{I}\otimes\mathbf{r}\right),\notag
\end{align}
where Lemma~\ref{lm:Lm6d1} is used in step (a), and the identities $(\mathbf{X}_1 \otimes \mathbf{Y}_1)(\mathbf{X}_2 \otimes \mathbf{Y}_2)=(\mathbf{X}_1 \mathbf{X}_2)\otimes(\mathbf{Y}_1 \mathbf{Y}_2)$, $(\mathbf{X} \otimes \mathbf{Y})^{-1}=(\mathbf{X}^{-1} \otimes \mathbf{Y}^{-1})$, and $\mathbf{X}^{-1} \mathbf{Y}^{-1}=(\mathbf{Y} \mathbf{X})^{-1}$, are used in step (b)-(d), respectively.
\end{IEEEproof}
\begin{remark}
\label{rm:Rm6d14}
Using the integration approach in Theorem~\ref{thm:Thm6d1}, we could alternatively express the throughput as $T^\textrm{ARQ}_\textrm{Int}=R(1-Q^\textrm{ARQ}_\textrm{Int})$, where we now get the integral
$Q^\textrm{ARQ}_\textrm{Int}
= \int_0^\infty \mathbf{p}_\textrm{I}\mathrm{e}^{z_\textrm{I} \mathbf{Q}_\textrm{I}}\mathbf{r}_\textrm{I}
\mathbf{e}_1^\textrm{T}\mathrm{e}^{\Theta \mathbf{Q}^\textrm{I}}\mathrm{e}^{z_\textrm{I} \Theta \mathbf{Q}^\textrm{I}}\mathbf{e}_d \, \mathrm{d}z_\textrm{I}$, $\mathbf{Q}^\textrm{I}=[0 \ \mathbf{p};\mathbf{0} \ \mathbf{Q}]$, which can then be determined by means of Lemma~\ref{lm:Lm6d1}.
\end{remark}

Let us consider some basic examples.
\begin{example}
\label{ex:Ex6d22}
(Exponentially distributed signal SNR and ME-distributed sum-interference)
Consider an ARQ system as defined in the system model, where we now assume that the SNR of the signal of interest is exponentially distributed, implying $\mathbf{p}_\textrm{um}=1$, $\mathbf{Q}_\textrm{um}=-1$, $\mathbf{r}_\textrm{um}=1$, and $\Theta=(\mathrm{e}^R-1)/S$. Then, \eqref{eq:Eq6d105} reduces to
\begin{align}
    P^\textrm{ARQ}_\textrm{Int}
    &=\mathrm{e}^{-\Theta}\int_0^\infty \mathbf{p}_\textrm{I}\mathrm{e}^{z_\textrm{I} (\mathbf{Q}_\textrm{I}-\Theta \mathbf{I})}\mathbf{r}_\textrm{I}\, \mathrm{d}z_\textrm{I}\notag\\
    &=\mathrm{e}^{-\Theta} \mathbf{p}_\textrm{I}(\Theta \mathbf{I}-\mathbf{Q}_\textrm{I})^{-1}\mathbf{r}_\textrm{I}    \label{eq:Eq6d106}\\
&=\mathrm{e}^{-\Theta} {p_\textrm{I}(\Theta)}/{q_\textrm{I}(\Theta)}.
\end{align}
This results generalizes the probability for successful decoding expression, $P^\textrm{ARQ}_\textrm{Int}
=\mathrm{e}^{-\Theta}/\prod_{u=1}^U (1+\Theta S_u/m^\textrm{N}_u)^{m^\textrm{N}_u}$, in \cite[(3)]{LarssonRasmSkog14a}, from $U$ iid Nakagami-$m$ interferers only, to ME-distributed interfering signals. A special case of \eqref{eq:Eq6d106} is for the exponentially distributed interferer with mean SNR $S_\textrm{I}$, and $\mathbf{p}_\textrm{I}=S_\textrm{I}^{-1}$, $\mathbf{Q}_\textrm{I}=-S_\textrm{I}^{-1}$, $\mathbf{r}_\textrm{I}=1$, which gives $P^\textrm{ARQ}_\textrm{Int}=\mathrm{e}^{-\Theta}/(1+S_\textrm{I}\Theta)$.
\end{example}

\begin{example}
\label{ex:Ex6d23}
(ME-distributed signal SNR and exponentially-distributed interference)
Consider an ARQ system as defined in the system model, where we now assume that the SNR of the signal of interest is ME-distributed, whereas the sum-interference is exponentially distributed with mean SNR $S_\textrm{I}$, implying $\mathbf{p}_\textrm{I}=S_\textrm{I}^{-1}$, $\mathbf{Q}_\textrm{I}=-S_\textrm{I}^{-1}$, $\mathbf{r}_\textrm{I}=1$, and $ \Theta=\mathrm{e}^{R}-1$. The integral \eqref{eq:Eq6d105} then reduces to
\begin{align}
    P^\textrm{ARQ}_\textrm{Int}
    &=-S_\textrm{I}^{-1}\int_0^\infty \mathbf{p}\mathbf{Q}^{-1}\mathrm{e}^{\Theta \mathbf{Q}}\mathrm{e}^{z_\textrm{I} (\Theta \mathbf{Q}-\mathbf{I} S_\textrm{I}^{-1})}\mathbf{r} \, \mathrm{d}z_\textrm{I}\notag\\
    &=\mathbf{p}\mathbf{Q}^{-1}\mathrm{e}^{\Theta \mathbf{Q}}(\mathbf{I}-\Theta S_\textrm{I} \mathbf{Q})^{-1} \mathbf{r}.
\end{align}
\end{example}

The throughput expression in Theorem~\ref{thm:Thm6d12} can be optimized with respect to the rate, but the resulting expression becomes somewhat complicated. Instead, we treat the throughput optimization problem, for a more general case, in Corollary~\ref{cr:Crl6d8}.

We note that when the upper integration interval in Lemma~\ref{lm:Lm6d1} is finite, the following corollary gives a new, more compact, expression. This integral property may be useful in different situations, e.g. if the integrand below is a pdf and its cdf is sought after.
\begin{corollary} 
(New integral of independent bivariate ME-form)
\label{cr:Crl6d6}
The integral of $f(t)=\mathbf{x}_1\mathrm{e}^{t\mathbf{Y}_1} \mathbf{z}_1\mathbf{x}_2\mathrm{e}^{t\mathbf{Y}_2} \mathbf{z}_2$, where $\mathbf{x}_j\in\mathbb{R}^{1 \times d_j}$, $\mathbf{Y}_j\in\mathbb{R}^{d_j \times d_j}$, $\mathbf{z}_j=[0\ldots 0 \ 1]^\textrm{T}\in\mathbb{R}^{d_j \times 1}$, $j=\{1,2\}$, and with interval $(0,b)$, is
\begin{align}
    \int_0^b \mathbf{x}_1\mathrm{e}^{t \mathbf{Y}_1}\mathbf{z}_1 \mathbf{x}_2\mathrm{e}^{t \mathbf{Y}_2}\mathbf{z}_2 \, \mathrm{d}t
    &=\mathbf{E}_{1,d_1d_2+1},
\end{align}
where
\begin{align}
    \mathbf{E}
    &\triangleq\mathrm{e}^{b\mathbf{Q}^\textrm{I}},\\
    \mathbf{Q}^\textrm{I}
    &=
    \begin{bmatrix}
    0 & \mathbf{x}_1\otimes \mathbf{x}_2\\
    \mathbf{0} & \mathbf{Y}_1\oplus \mathbf{Y}_2
    \end{bmatrix}.
\end{align}
\end{corollary}
\begin{IEEEproof}
The integral is
\begin{align}
    &\int_0^b \mathbf{x}_1\mathrm{e}^{t \mathbf{Y}_1}\mathbf{z}_1 \mathbf{x}_2\mathrm{e}^{t \mathbf{Y}_2}\mathbf{z}_2 \, \mathrm{d}t\notag\\
    &=\int_0^b  \left(\mathbf{x}_1\otimes \mathbf{x}_2\right)\left(\mathrm{e}^{t  (\mathbf{Y}_1\oplus \mathbf{Y}_2)}\right) \left(\mathbf{z}_1\otimes\mathbf{z}_2\right) \, \mathrm{d}t\notag \\
    &=\mathbf{e}_1^\textrm{T}
    \mathrm{e}^{b \mathbf{Q^\textrm{I}}}\mathbf{e}_{d_1d_2+1}=\mathbf{E}_{1,d_1d_2+1},\notag
\end{align}
where the rearrangements in Lemma~\ref{lm:Lm6d1} is combined with Theorem~\ref{thm:Thm6d1}.
\end{IEEEproof}

Next, we explore an alternative, perhaps less straightforward, but more general, solution approach to the ARQ problem with ME-distributed signal and interferers. This section is also motivated by that the more generally formulated bivariate ME-density, and associated analysis, may find applications beyond the studied case.

\subsubsection{Throughput Analysis - Dependent Signal and Interference} 
\label{sec:Sec6d5d8d3}
In the previous section, we analyzed the throughput of ARQ when the signal and interference SNR r.v.s where independent. This section gives an alternative analysis framework for this case, but, by introducing a more general joint pdf, also enables analysis of ARQ throughput when the signal and interference SNR r.v.s may be dependent. The dependent case may be less common, but could occur if, e.g., both the signal and interferer(s) experience a common varying channel attenuation, or a common diffractive object whilst the receiver is moving. Regardless of this, for performance optimization, this framework offer an attractive solution. Yet another motivation for this section is that this, more  general, joint pdf, and associated analytical framework, should be amenable for analyzing other wireless communication problems involving two dependent r.v.s. We start with the generalization of the joint pdf below, and then proceed with the analysis.

\begin{definition}
\label{def:Def6d1}
(Bivariate ME-distribution)
We define the joint ME-density of the wireless channel SNR r.v.s $(Z_1,Z_2)$, as $f_{Z_1,Z_2}(z_1,z_2) =\mathbf{p}_1\mathrm{e}^{z_1 \mathbf{Q}_1}\mathbf{P}_{12}\mathrm{e}^{z_2 \mathbf{Q}_2}\mathbf{r}_2$, $z_1\geq 0$, $z_2\geq 0$, where $\mathbf{p}_1\in \mathbb{R}^{1\times d_1}$, $\mathbf{Q}_1\in \mathbb{R}^{d_1\times d_1}$, $\mathbf{P}_{12}\in \mathbb{R}^{d_1\times d_2}$, $\mathbf{Q}_2\in \mathbb{R}^{d_2\times d_2}$, $\mathbf{r}_2\in \mathbb{R}^{d_2 \times 1}$. The parameters defining the joint density are, in a similar manner as for the univariate-ME-distribution, assumed selected to have a corresponding bivariate CDF fulfilling necessary characteristics, e.g. $0\leq F_{Z_1,Z_2}(z_1,z_2)\leq 1$, $F_{Z_1,Z_2}(z_1=0,z_2=0)=0$, $F_{Z_1,Z_2}(z_1\rightarrow \infty,z_2\rightarrow \infty)=1$, and $\frac{\mathrm{d}}{\mathrm{d}z_j}F_{Z_1,Z_2}(z_1,z_2)\geq 0$,  $j=\{1,2\}$.
\end{definition}

\begin{remark}
\label{rm:Rm6d15}
The bivariate-ME-distribution in Definition~\ref{def:Def6d1} is a generalization of the case with independent fading SNRs, with pdf $f_{Z_1,Z_2}(z_1,z_2) =\mathbf{p}_1\mathrm{e}^{z_1 \mathbf{Q}_1}\mathbf{r}_{1}\mathbf{p}_{2}\mathrm{e}^{z_2 \mathbf{Q}_2}\mathbf{r}_2$, considered in the previous Section. This is so since $\mathbf{P}_{12}$ may have full rank, but $\mathbf{r}_{1}\mathbf{p}_{2}$ has rank one. The joint density of this form has also been considered in the literature, e.g. in \cite{BladtNiel10, BodrogHorvTele08}.
\end{remark}

\begin{remark}
\label{rm:Rm6d16}
The LT of the bivariate-ME-density in Definition~\ref{def:Def6d1} is $F(s_1,s_2)=\int_0^\infty \int_0^\infty \mathrm{e}^{-s_1z_1-s_2z_2} f_{Z_1,Z_2}(z_1,z_2)\, \mathrm{d}z_1 \mathrm{d}z_2 =\mathbf{p}_1(\mathbf{Q}_1-s_1 \mathbf{I}_{d_1})^{-1}\mathbf{P}_{12}(\mathbf{Q}_2-s_2 \mathbf{I}_{d_2})^{-1}\mathbf{r}_2$.
Expanding the inverses, we get the rational form $F(s_1,s_2)=p(s_1,s_2)/ q_1(s_1) q_2(s_2)$. Due to the product of polynomials, $q_1(s_1)q_2(s_2)$, in the denominator, rather than a more general polynomial $q(s_1,s_2)$, it is clear that the considered bivariate ME-distribution is not on the most general form possible.
\end{remark}

To find the information-outage probability for ARQ with ME-distributed signal and interferers, the integral of the bivariate ME-density is of interest, and hence considered next. 

\begin{lemma}
\label{lm:Lm6d2}
(Integral of dependent bivariate ME-form - Sylvester's Equation, \cite[Lemma~3]{WahlstromAxelGust14})
Consider the function $f(t)=\mathbf{x}_1\mathrm{e}^{t \mathbf{Y}_1}\mathbf{Y}_{12}\mathrm{e}^{t \mathbf{Y}_2}\mathbf{z}_2$. Then, the  integral of $f(t)$, with intervals $(a,b)$ is
\begin{align}
    \int_a^b \mathbf{x}_1\mathrm{e}^{t \mathbf{Y}_1}\mathbf{X}_{12}\mathrm{e}^{t \mathbf{Y}_2}\mathbf{z}_2 \, \mathrm{d}t
    =\mathbf{x}_1\mathbf{X}\mathbf{z}_2,
    \label{eq:Eq6d128}
\end{align}
where $\mathbf{X}$ is given by solving Sylvester's equation
\begin{align}
    \mathbf{Y}_1\mathbf{X}+\mathbf{X}\mathbf{Y}_2
    &=\mathrm{e}^{b\mathbf{Y}_1}\mathbf{X}_{12}
    \mathrm{e}^{b\mathbf{Y}_2}-\mathrm{e}^{a\mathbf{Y}_1}
    \mathbf{X}_{12}\mathrm{e}^{a\mathbf{Y}_2},
    \label{eq:Eq6d129}
\end{align}
and $\mathbf{X}\triangleq \int_a^b\mathrm{e}^{t\mathbf{Y}_1} \mathbf{X}_{12}\mathrm{e}^{t\mathbf{Y}_2} \, \mathrm{d}t$.
\end{lemma}
\begin{IEEEproof}
It is straightforward to check that $\mathbf{X}
\triangleq \int_a^b \mathrm{e}^{t \mathbf{Y}_1}\mathbf{X}_{12}\mathrm{e}^{t \mathbf{Y}_2} \, \mathrm{d}t$ fulfills Sylvester's equation \eqref{eq:Eq6d129}, just by inserting the former expression in the latter.
\end{IEEEproof}

We note that the integral expression in Lemma~\ref{lm:Lm6d2} is well-known in control theory \cite[Lemma~3]{WahlstromAxelGust14}, but, to the best of our knowledge, it has not been used in the context of ME-distribution analysis, nor in the context of outage probability analysis for wireless communication systems.

\begin{remark}
\label{rm:Rm6d17}
When the interval is $(a,b)=(0,\infty)$, it is well-known that no eigenvalues of $\mathbf{Y}_1$, and $-\mathbf{Y}_2$ can be the same for a unique solution to Sylvester's equation. If all eigenvalues are located in the open left half-plane, and $(a,b)=(0,\infty)$, then the RHS of \eqref{eq:Eq6d129} is simply $-\mathbf{X}_{12}$.
\end{remark}

\begin{remark}
\label{rm:Rm6d18}
Sylvester's equation, $\mathbf{Y_1}\mathbf{X}+\mathbf{X}\mathbf{Y_2}=-\mathbf{X}_{12}$, can be numerically solved in \textsc{Matlab} by the command $X= \mathrm{sylvester}(Y_1,Y_2,-X_{12})$.
\end{remark}

\begin{theorem}
\label{thm:Thm6d13}
(ARQ throughput solution with Sylvester's equation)
Let the signal of interest, and the sum-interference, SNRs have joint density $f_{Z_\textrm{I},Z}(z_\textrm{I},z)=\mathbf{p}_\textrm{I}\mathrm{e}^{z_\textrm{I} \mathbf{Q}_\textrm{I}}\mathbf{P}_{12}\mathrm{e}^{z\mathbf{Q}}\mathbf{r} $. Then, the throughput is
\begin{align}
    T^\textrm{ARQ}_\textrm{Int}
    &=R\mathbf{p}_\textrm{I} \mathbf{X} \mathbf{r},
    \label{eq:Eq6d130}
\end{align}
where $\mathbf{X}$ is given by the solution $\mathbf{X}$ to the Sylvester equation
\begin{align}
    \mathbf{Q}_1\mathbf{X}+\mathbf{X}\Theta\mathbf{Q}_2
    &=-\mathbf{\breve{P}}_{12},
    \label{eq:Eq6d131}
\end{align}
with
\begin{align}
    \mathbf{\breve{P}}_{12}
    &\triangleq-\mathbf{P}_{12}\mathbf{Q}^{-1}\mathrm{e}^{\Theta \mathbf{Q}},
\end{align}
and $\Theta=\mathrm{e}^R-1$.
\end{theorem}

\begin{IEEEproof}
The throughput is $T^\textrm{ARQ}_\textrm{Int}=RP^\textrm{ARQ}_\textrm{Int} $, where, analogously to Theorem~\ref{thm:Thm6d12}, the decoding probability is
\begin{align}
    P^\textrm{ARQ}_\textrm{Int}
    &=\int_0^\infty \int_{\Theta(1+z_\textrm{I})}^\infty \mathbf{p}_\textrm{I}\mathrm{e}^{z_\textrm{I} \mathbf{Q}_\textrm{I}}\mathbf{P}_{12}\mathrm{e}^{z\mathbf{Q}}\mathbf{r} \, \mathrm{d}z_\textrm{I} \, \mathrm{d}z\notag\\
    &= -\int_0^\infty \mathbf{p}_\textrm{I}\mathrm{e}^{z_\textrm{I} \mathbf{Q}_\textrm{I}}\mathbf{P}_{12}\mathbf{Q}^{-1}\mathrm{e}^{\Theta(1+z_\textrm{I}) \mathbf{Q}}\mathbf{r} \, \mathrm{d}z_\textrm{I} \notag\\
    &= \int_0^\infty \mathbf{p}_\textrm{I}\mathrm{e}^{z_\textrm{I} \mathbf{Q}_\textrm{I}}\mathbf{\breve{P}}_{12}\mathrm{e}^{z_\textrm{I}  \Theta \mathbf{Q}}\mathbf{r} \, \mathrm{d}z_\textrm{I}, \ \mathbf{\breve{P}}_{12}
    \triangleq -\mathbf{P}_{12} \mathbf{Q}^{-1}\mathrm{e}^{\Theta \mathbf{Q}},  \notag\\
    &=\mathbf{p}_\textrm{I}\mathbf{X}\mathbf{r},\ \mathbf{X}
    \triangleq \int_0^\infty \mathrm{e}^{z_\textrm{I} \mathbf{Q}_\textrm{I}}\mathbf{\breve{P}}_{12}\mathrm{e}^{z_\textrm{I}  \Theta \mathbf{Q}} \, \mathrm{d}z_\textrm{I}.\notag
\end{align}
Based on Lemma~\ref{lm:Lm6d2}, we then solve for $\mathbf{X}$ in the Sylvester equation \eqref{eq:Eq6d131}.
\end{IEEEproof}
Eq. \eqref{eq:Eq6d130} can be expressed more explicitly, as done in Corollary~\ref{cr:Crl6d7} by means of the well-known vectorization solution in the following lemma.

\begin{lemma}
\label{lm:Lm6d3}
Sylvester's equation, $\mathbf{Y}_1\mathbf{X}+\mathbf{X}\mathbf{Y}_2
=-\mathbf{X}_{12}$, has solution
\begin{align}
    \mathrm{vec}(\mathbf{X})
    &=-\left(\mathbf{Y}_2^\textrm{T} \oplus \mathbf{Y}_1\right)^{-1}\mathrm{vec}\left(\mathbf{X}_{12}\right).
\end{align}
\begin{IEEEproof}
Vectorization of the LHS gives
$\mathbf{Y}_1\mathbf{X}+\mathbf{X}\mathbf{Y}_2
=\left(\mathbf{Y}_2^\textrm{T} \oplus \mathbf{Y}_1\right)\mathrm{vec}(\mathbf{X})$, and then $\mathrm{vec}(\mathbf{X})$ is solved for.
\end{IEEEproof}
\end{lemma}

\begin{corollary}
\label{cr:Crl6d7}
(Explicit ARQ throughput solution with Sylvester's equation)
Let the throughput be defined as in Theorem~\ref{thm:Thm6d13}. Then, the throughput can be written (more) explicitly as
\begin{align}
    T^\textrm{ARQ}_\textrm{Int}
    &=R(\mathbf{r}^\textrm{T} \otimes\mathbf{p}_\textrm{I}) \left(\Theta \mathbf{Q}^\textrm{T} \oplus \mathbf{Q}_\textrm{I}\right)^{-1}\mathrm{vec}\left(\mathbf{P}_{12}\mathbf{Q}^{-1}\mathrm{e}^{\Theta \mathbf{Q}}\right).
\end{align}
\end{corollary}

\begin{IEEEproof}
\begin{align}
    P^\textrm{ARQ}_\textrm{Int}
    &=\mathbf{p}_\textrm{I}\mathbf{X}\mathbf{r}\notag\\
    &\overset{(a)}{=}\mathrm{vec}(\mathbf{p}_\textrm{I}\mathbf{X}\mathbf{r})\notag\\
    &\overset{(b)}{=}(\mathbf{r}^\textrm{T} \otimes \mathbf{p}_\textrm{I})\mathrm{vec}(\mathbf{X})\notag\\
    &\overset{(c)}{=}-(\mathbf{r}^\textrm{T} \otimes \mathbf{p}_\textrm{I})
    \left(\Theta \mathbf{Q}^\textrm{T} \oplus \mathbf{Q}_\textrm{I}\right)^{-1}\mathrm{vec}\left(\mathbf{\breve{P}}_{12}\right)\notag.
\end{align}
The scalar is vectorized in step (a), the vectorization identity $\mathbf{x}\mathbf{Y}\mathbf{z}=(\mathbf{z}^\textrm{T} \otimes \mathbf{x})\mathrm{vec}(\mathbf{Y})$ is used in step (b), and Lemma~\ref{lm:Lm6d3} is invoked in step (c).  An alternative, more direct, proof is to vectorize the integral in the proof of Theorem~\ref{thm:Thm6d13} directly.
\begin{align}
    & \mathrm{vec} \left(\int_0^\infty \mathbf{p}_\textrm{I}\mathrm{e}^{z_\textrm{I} \mathbf{Q}_\textrm{I}}\mathbf{\breve{P}}_{12}\mathrm{e}^{z_\textrm{I}  \Theta \mathbf{Q}}\mathbf{r} \, \mathrm{d}z_\textrm{I}\right) \notag\\
    &= (\mathbf{r}^\textrm{T} \otimes \mathbf{p}_\textrm{I})\mathrm{vec} \left( \int_0^\infty \mathrm{e}^{z_\textrm{I} \mathbf{Q}_\textrm{I}}\mathbf{\breve{P}}_{12}\mathrm{e}^{z_\textrm{I}  \Theta \mathbf{Q}} \, \mathrm{d}z_\textrm{I}\right) \notag\\
    &= (\mathbf{r}^\textrm{T} \otimes \mathbf{p}_\textrm{I}) \left(\int_0^\infty \mathrm{e}^{z_\textrm{I}  \Theta \mathbf{Q}^\textrm{T}} \otimes \mathrm{e}^{z_\textrm{I} \mathbf{Q}_\textrm{I}} \, \mathrm{d}z_\textrm{I} \right)\mathrm{vec} \left( \mathbf{\breve{P}}_{12}\right) \notag\\
    &= (\mathbf{r}^\textrm{T} \otimes \mathbf{p}_\textrm{I}) \left(\int_0^\infty \mathrm{e}^{z_\textrm{I}  (\Theta \mathbf{Q}^\textrm{T} \oplus \mathbf{Q}_\textrm{I})}\, \mathrm{d}z_\textrm{I} \right)\mathrm{vec} \left( \mathbf{\breve{P}}_{12}\right) \notag\\
    &=-(\mathbf{r}^\textrm{T} \otimes \mathbf{p}_\textrm{I})
    \left(\Theta \mathbf{Q}^\textrm{T} \oplus \mathbf{Q}_\textrm{I}\right)^{-1}\mathrm{vec}\left(\mathbf{\breve{P}}_{12}\right)\notag.
\end{align}
\end{IEEEproof}

Inspired by the vectorization approach in Corollary~\ref{cr:Crl6d7}, an alternative (and potentially useful) integral expression to Lemma~\ref{lm:Lm6d2} with $a=0$ can also be given.
\begin{lemma}
\label{lm:Lm6d4}
The integral of $f(t)=\mathbf{x}_1\mathrm{e}^{t\mathbf{Y}_1} \mathbf{X}_{12}\mathrm{e}^{t\mathbf{Y}_2} \mathbf{z}_2$ with integration interval $(0,b)$, is
\begin{align}
    \int_0^b \mathbf{x}_1\mathrm{e}^{t \mathbf{Y}_1} \mathbf{X}_{12}\mathrm{e}^{t \mathbf{Y}_2}\mathbf{z}_2 \, \mathrm{d}t
    &=\mathbf{E}_{1,:}\mathrm{vec}(\mathbf{X}_{12}),
\end{align}
where
\begin{align}
    \mathbf{E}
    &\triangleq\mathrm{e}^{b\mathbf{Q}^\textrm{I}},\\
    \mathbf{Q}^\textrm{I}
    &=
    \begin{bmatrix}
    0 & \mathbf{z}_2^\textrm{T}\otimes \mathbf{x}_1\\
    \mathbf{0} & \mathbf{Y}_2^\textrm{T}\oplus \mathbf{Y}_1
    \end{bmatrix}.
\end{align}
\end{lemma}
\begin{IEEEproof}
From the proof of Corollary~\ref{cr:Crl6d7}, we see that the integral can be written
\begin{align}
    &\int_0^b  \left(\mathbf{z}_2^\textrm{T}\otimes \mathbf{x}_1\right)\left(\mathrm{e}^{t  (\mathbf{Y}_2^\textrm{T}\oplus \mathbf{Y}_1)}\right)\mathrm{vec}(\mathbf{X}_{12}) \, \mathrm{d}t\notag \\
    &=\mathbf{e}_1^\textrm{T}
    \mathrm{e}^{b \mathbf{Q^\textrm{I}}}\mathrm{vec}(\mathbf{X}_{12}) =\mathbf{E}_{1,:}\mathrm{vec}(\mathbf{X}_{12}),\notag
\end{align}
and in the last steps, Theorem~\ref{thm:Thm6d1} is used.
\end{IEEEproof}

Next, we consider throughput optimization when the throughput is based on a solution to a Sylvester equation. To be able to benefit from the auxiliary parameter method \cite{LarssonRasmSkog14b}, we assume $\Theta=(\mathrm{e}^R-1)/S$, and work with the unit-mean ME-parameters for the signal of interest.
\begin{corollary}
\label{cr:Crl6d8}
(ARQ Optimal throughput solution with Sylvester's equation)
Consider the ARQ throughput expression \eqref{eq:Eq6d130} with \eqref{eq:Eq6d131}. Then, the auxiliary parameter function $g_\Theta(\Theta)$ in \cite{LarssonRasmSkog14b}, for the optimal throughput, is
\begin{align}
g_\Theta(\Theta)=\frac{\mathbf{p}_\textrm{I}\mathbf{X}\mathbf{r}}{\Theta \mathbf{p}_\textrm{I}\mathbf{X}'_\Theta\mathbf{r}},
\end{align}
where $\mathbf{X}$ is the solution to the Sylvester equation
\begin{align}
    \mathbf{Q}_\textrm{I}\mathbf{X}+\mathbf{X}\Theta\mathbf{Q}_\textrm{um}
    &=-\mathbf{\breve{P}}_{12},
    \label{eq:Eq6d139} \\
 \mathbf{\breve{P}}_{12}
    &\triangleq-\mathbf{P}_{12}\mathbf{Q}_\textrm{um}^{-1}\mathrm{e}^{\Theta \mathbf{Q}_\textrm{um}},
\end{align}
and with $\mathbf{X}$ known, $\mathbf{X}'_\Theta$ is then the solution to the Sylvester equation
\begin{align}
    \mathbf{Q}_\textrm{I}\mathbf{X}'_\Theta+\mathbf{X}'_\Theta\Theta\mathbf{Q}_\textrm{um}
    &=-(\mathbf{\breve{P}}_{12}+\mathbf{X})\mathbf{Q}_\textrm{um}.
    \label{eq:Eq6d141}
\end{align}
\end{corollary}
\begin{IEEEproof}
Taking the implicit derivative of \eqref{eq:Eq6d139} wrt $\Theta$ gives $\mathbf{Q}_\textrm{I}\mathbf{X}'_\Theta+\mathbf{X}'_\Theta \Theta\mathbf{Q}_\textrm{um}+\mathbf{X}\mathbf{Q}_\textrm{um} =-\mathbf{\breve{P}}_{12}\mathbf{Q}_\textrm{um}$,
which rearranged gives \eqref{eq:Eq6d141}. We then have $P^\textrm{ARQ}_\textrm{Int}=\mathbf{p}_\textrm{I}\mathbf{X}\mathbf{r}$ and $\frac{\mathrm{d}}{\mathrm{d}\Theta}P^\textrm{ARQ}_\textrm{Int}=\mathbf{p}_\textrm{I}\mathbf{X}'_\Theta\mathbf{r}$ which is inserted in the definition for $g_\Theta(\Theta)$.
\end{IEEEproof}

\begin{remark}
\label{rm:Rm6d19}
(Integral of product of two MEs -- Van Loans method)
An alternative way of computing the integral in Lemma~\ref{lm:Lm6d2}, and then used for Corollary~\ref{cr:Crl6d8}, is via the method by Van Loan \cite{VanLoan78}. With interval $(0,b)$, the integral is
\begin{align}
    \int_0^b \mathbf{x}_1\mathrm{e}^{z \mathbf{Y}_1}\mathbf{X}_{12}\mathrm{e}^{t \mathbf{Y}_2}\mathbf{z}_2 \, \mathrm{d}t
    =\mathbf{x}_1\mathbf{Y}_{11}^{-1}\mathbf{Y}_{12}\mathbf{z}_2,
\end{align}
where
\begin{align}
    \mathbf{Y}^\textrm{I}
    &=\begin{bmatrix}
    -\mathbf{Y}_1 & \mathbf{X}_{12}\\
    \mathbf{0} & \mathbf{Y}_2
    \end{bmatrix},\\
    \begin{bmatrix}
    \mathbf{Y}_{11} & \mathbf{Y}_{12}\\
    \mathbf{0} & \mathbf{Y}_{22}
    \end{bmatrix}
    &=\mathrm{e}^{b \mathbf{Y}^\textrm{I}}.
\end{align}
This method has less computational complexity than the vectorization approaches. However, as $b$ is finite, $b$ has to be chosen "large enough" to give a good approximation for the case when $b\rightarrow \infty$. The proof is by Van Loan \cite{VanLoan78}.
\end{remark}

\begin{remark}
\label{rm:Rm6d20}
Pertinent to results on the bivariate integrals presented here, if $\mathbf{X}_{12}$ is square and invertible, we may also rewrite the integral as
\begin{align}
 &   \int \mathbf{x}_1\mathrm{e}^{z \mathbf{Y}_1}\mathbf{X}_{12}\mathrm{e}^{t \mathbf{Y}_2}\mathbf{z}_2 \, \mathrm{d}t\notag\\
&=  \int \mathbf{x}_1\mathbf{X}_{12}\mathrm{e}^{z \mathbf{X}_{12}^{-1}\mathbf{Y}_1\mathbf{X}_{12}}\mathrm{e}^{t \mathbf{Y}_2}\mathbf{z}_2 \, \mathrm{d}t .
\label{eq:Eq6d145}
\end{align}
If the matrix commutation $[\mathbf{Y}_2,\mathbf{X}_{12}^{-1}\mathbf{Y}_1\mathbf{X}_{12}]=0$ holds, we may further simplify \eqref{eq:Eq6d145} to
\begin{align}
       \int \mathbf{x}_1\mathbf{X}_{12}\mathrm{e}^{z \mathbf{X}_{12}^{-1}\mathbf{Y}_1\mathbf{X}_{12}+t \mathbf{Y}_2}\mathbf{z}_2 \, \mathrm{d}t. \notag
\end{align}
\end{remark}

\begin{remark}
\label{rm:Rm6d21}
We note that for persistent-HARQ with interference, the LT for the SNR pdf  is required which can be expressed as
\begin{align}
   F(s)
    &=\int_0^\infty\mathrm{e}^{-sz}
    \left(\int_0^\infty
    (1+z_\textrm{I})\mathbf{p}_\textrm{I}\mathrm{e}^{z_\textrm{I} \mathbf{Q}_\textrm{I}} \mathbf{r}_\textrm{I}\mathbf{p}\mathrm{e}^{z (1+z_\textrm{I})\mathbf{Q}} \mathbf{r} \, \mathrm{d}z_\textrm{I} \right)
    \, \mathrm{d}z\notag\\
    &=
    \int_0^\infty
    (1+z_\textrm{I})\mathbf{p}_\textrm{I}\mathrm{e}^{z_\textrm{I} \mathbf{Q}_\textrm{I}} \mathbf{r}_\textrm{I}\mathbf{p}(s\mathbf{I}-(1+z_\textrm{I})\mathbf{Q})^{-1} \mathbf{r} \, \mathrm{d}z_\textrm{I}\notag\\
    &=
    \int_1^\infty
    t\mathbf{p}_\textrm{I}\mathrm{e}^{- \mathbf{Q}_\textrm{I}}\mathrm{e}^{t \mathbf{Q}_\textrm{I}} \mathbf{r}_\textrm{I}\mathbf{p}(s\mathbf{I}-t\mathbf{Q})^{-1} \mathbf{r} \, \mathrm{d}t \notag\\
    &=
    -\int_1^\infty
    \mathbf{p}_\textrm{I}\mathrm{e}^{- \mathbf{Q}_\textrm{I}}\mathrm{e}^{t \mathbf{Q}_\textrm{I}} \mathbf{r}_\textrm{I}\mathbf{p}\mathbf{Q}^{-1}(\mathbf{I}-s\mathbf{Q}^{-1}t^{-1})^{-1} \mathbf{r} \, \mathrm{d}t \notag\\
    &=1+
    \int_1^\infty
    \mathbf{p}_\textrm{I}\mathrm{e}^{- \mathbf{Q}_\textrm{I}}\mathrm{e}^{t \mathbf{Q}_\textrm{I}} \mathbf{r}_\textrm{I}\mathbf{p}\mathbf{Q}^{-1}(\mathbf{I}-t\mathbf{Q}s^{-1})^{-1} \mathbf{r} \, \mathrm{d}t.
\end{align}
The reason why we do not analyze the throughput of persistent-HARQ with interference becomes apparent. This is because the corresponding LT is not on a rational form.
\end{remark}
From this section, it can be concluded that a ME-distributed signal in ME-distributed interference can be handled in an efficient structured manner, whereas a more traditional (non-ME-distribution-based) analysis would be untractable.

\subsection{$2\times 2$ SM-MIMO} 
\label{sec:Sec6d5d9}
Below, we consider $2\times 2$ SM-MIMO in AWGN with capacity $C=\ln \det \left( \mathbf{I}+SN_\textrm{tx}^{-1}\mathbf{H}^H\mathbf{H}\right)$, where $\mathbf{H}^H\mathbf{H}$ is Wishart unit-variance distributed, and $N_\textrm{tx}=2$. We merely aim to demonstrate the use of the bivariate ME-distribution in an outage probability context, not to fully solve the problem.
The following integral will be useful
\begin{lemma}
\label{lm:Lm6d5}
(ME-form integral related to the modified Bessel function of the Second kind) 
\begin{align}
    &K_1^\textrm{ME}(a,\mathbf{x}_1,\mathbf{Y}_1,\mathbf{X}_{12},\mathbf{Y}_2)\notag\\
&\triangleq \int_a^\infty \mathbf{x}_1\mathrm{e}^{t \mathbf{Y}_1}\mathbf{X}_{12}\mathrm{e}^{t^{-1} \mathbf{Y}_2}\mathbf{z}_2 \, \mathrm{d}t \notag\\
 &=((\mathbf{T}_2^{-1}\mathbf{z}_2)^\textrm{T} \otimes \mathbf{x}_1\mathbf{T}_1) \mathbf{\Xi} \mathrm{vec}(\mathbf{T}_1^{-1}\mathbf{X}_{12}\mathbf{T}_2),
\end{align}
where
\begin{align}
\mathbf{\Xi}
&\triangleq \int_a^\infty \mathrm{e}^{t^{-1}\mathbf{J}_2^\textrm{T} \oplus t \mathbf{J}_1} \, \mathrm{d}t,\\
\mathbf{T}_j\mathbf{J}_j\mathbf{T}_j^{-1}
&=\mathbf{Y}_j, \ j \in\{1,2\}.
\end{align}
\end{lemma}
\begin{IEEEproof}
Using earlier basic vectorization identities, we get
\begin{align}
K_1^\textrm{ME}(\cdot)
&=\int_a^\infty \mathbf{x}_1\mathbf{T}_1\mathrm{e}^{t \mathbf{J}_1}\mathbf{T}_1^{-1}\mathbf{X}_{12}\mathbf{T}_2 \mathrm{e}^{t^{-1}\mathbf{J}_2}\mathbf{T}_2^{-1}\mathbf{z}_2 \, \mathrm{d}t \notag\\
    &=((\mathbf{T}_2^{-1}\mathbf{z}_2)^\textrm{T} \otimes \mathbf{x}_1\mathbf{T}_1)\left(\int_a^\infty \mathrm{e}^{t^{-1}\mathbf{J}_2 \oplus t \mathbf{J}_1} \, \mathrm{d}t\right) \notag\\
&\times \mathrm{vec}(\mathbf{T}_1^{-1}\mathbf{X}_{12}\mathbf{T}_2)\notag.
\end{align}
\end{IEEEproof}

\begin{corollary}
\label{cr:Crl6d9}
When $\mathbf{Y}_1$ and  $\mathbf{Y}_2$ are both diagonalizable, and all eigenvalues are real negative,  then 
\begin{align}
\mathbf{\Xi}
&=\mathrm{diag}\{d_{11},d_{12},\ldots d_{1J_2},d_{21},\ldots d_{I_1J_2}\},\\
d_{ij}
&=2\sqrt{\frac{\lambda_{i}^{(1)}}{\lambda_{j}^{(2)}}}K_1\left(a\sqrt{\frac{\lambda_{i}^{(1)}}{\lambda_{j}^{(2)}}},2\sqrt{\lambda_{i}^{(1)}\lambda_{j}^{(2)}}\right).
\label{eq:Eq6d151}
\end{align}
where $K_1(\cdot)$ is the scalar modified Bessel function of the second kind.
\end{corollary}
\begin{IEEEproof}
The diagonal matrix have entries $d_{ij}
= \int_a^\infty
\mathrm{e}^{t\lambda_{j}^{(1)}+t^{-1}\lambda_{i}^{(2)}} \, \mathrm{d}z$, which is solved with
$K_1(a',b')\triangleq\int_{a'}^\infty \mathrm{e}^{-tb'-t^{-1}b'} \, \mathrm{d}t$ gives \eqref{eq:Eq6d151}.
\end{IEEEproof}

The next lemma illustrates that the Wishart eigenvalue density for $2\times 2$-matrix can be expressed on a bivariate ME-density form.
\begin{lemma}
\label{lm:Lm6d6}
Let $z_1, z_2$ denote the eigenvalues of the Wishart distribution with density 
$f(z_1,z_2)=\mathrm{e}^{-z_1-z_2}(z_1-z_2)^2, \ 0\leq z_1\leq z_2$. Then, a corresponding bivariate ME-density is $f(z_1,z_2)=\mathbf{p}_1\mathrm{e}^{z_1 \mathbf{Q}_1}\mathbf{P}_{12}\mathrm{e}^{z_2 \mathbf{Q}_2}\mathbf{r}_2, \ 0\leq z_1\leq z_2$, with parameters 
$\mathbf{p}_1=[1  \ 0 \ 0]$, $\mathbf{Q}_1=\mathbf{Q}_2=[-1  \ 1 \ 0;0 \ -1 \ 1;0 \ 0 \ -1]$, $\mathbf{P}_{12}=2[1  \ 0 \ 0;0 \ -1 \ 0;0 \ 0 \ 1]$, $\mathbf{r}_2=[0  \ 0 \ 1]^\textrm{T}$.
\end{lemma}

Below, we give the outage probability for $2\times 2$ SM-MIMO.
\begin{align}
    &Q_\textrm{out}^{2\times 2}
=\underset{ \underset{0\leq z_1 \leq z_2}{\tilde z\leq R}}{\iint} \mathbf{p}_1\mathrm{e}^{z_1 \mathbf{Q}_1}\mathbf{P}_{12}\mathrm{e}^{z_2 \mathbf{Q}_2}\mathbf{r}_2 \, \mathrm{d}z_1 \, \mathrm{d}z_2\notag\\
    &=\underset{\underset{t_1\geq 1, t_2\geq 1}{t_1t_2\leq \Theta}}{\iint} 2^{-1}\mathbf{p}_1\mathrm{e}^{t_1 \mathbf{Q}_1}\mathrm{e}^{ -\mathbf{Q}_1}\mathbf{P}_{12}\mathrm{e}^{- \mathbf{Q}_2}\mathrm{e}^{t_2 \mathbf{Q}_2}\mathbf{r}_2 \, \mathrm{d}t_1 \, \mathrm{d}t_2\notag\\
    &=\int_1^\Theta 2^{-1}\mathbf{p}_1\mathrm{e}^{t_1 \mathbf{Q}_1}\mathrm{e}^{ -\mathbf{Q}_1}\mathbf{P}_{12}\mathrm{e}^{- \mathbf{Q}_2}\mathbf{Q}_2^{-1}(\mathrm{e}^{t_2 \mathbf{Q}_2}-\mathbf{I})\mathbf{r}_2|_1^{\Theta/t_1} \, \mathrm{d}t_1\notag\\
    &=\int_1^\Theta 2^{-1}\mathbf{p}_1\mathrm{e}^{t_1 \mathbf{Q}_1}\mathrm{e}^{ -\mathbf{Q}_1}\mathbf{P}_{12}\mathrm{e}^{- \mathbf{Q}_2}\mathbf{Q}_2^{-1}(\mathrm{e}^{{\Theta/t_1} \mathbf{Q}_2}-\mathrm{e}^{ \mathbf{Q}_2})\mathbf{r}_2 \, \mathrm{d}t_1\notag\\
    &=\int_1^\Theta 2^{-1}\mathbf{p}_1\mathrm{e}^{t_1 \mathbf{Q}_1}\mathrm{e}^{ -\mathbf{Q}_1}\mathbf{P}_{12}\mathrm{e}^{- \mathbf{Q}_2}\mathbf{Q}_2^{-1}(\mathrm{e}^{{\Theta/t_1} \mathbf{Q}_2})\mathbf{r}_2 \, \mathrm{d}t_1\notag\\
&-\int_1^\Theta 2^{-1}\mathbf{p}_1\mathrm{e}^{t_1 \mathbf{Q}_1}\mathrm{e}^{ -\mathbf{Q}_1}\mathbf{P}_{12}\mathbf{Q}_2^{-1}\mathbf{r}_2 \, \mathrm{d}t_1\notag\\
&=\int_1^\Theta 2^{-1}\mathbf{p}_1\mathrm{e}^{t_1 \mathbf{Q}_1}\mathrm{e}^{ -\mathbf{Q}_1}\mathbf{P}_{12}\mathrm{e}^{- \mathbf{Q}_2}\mathbf{Q}_2^{-1}(\mathrm{e}^{{\Theta/t_1} \mathbf{Q}_2})\mathbf{r}_2 \, \mathrm{d}t_1\notag\\
& +2^{-1}(1+\mathbf{p}_1\mathrm{e}^{(\Theta-1) \mathbf{Q}_1}\mathbf{Q}_1^{-1}\mathbf{P}_{12}\mathbf{Q}_2^{-1}\mathbf{r}_2),
\end{align}
where $\tilde z \triangleq \ln(1+z_1)+\ln(1+z_2)$, $\Theta=\mathrm{e}^R$, and the last integral have the same form as in Lemma~\ref{lm:Lm6d5}.
While we can not solve the outage probability exactly, we demonstrate that the outage probability characterization can be formulated within the bivariate ME-distribution framework, and we observe that the derived outage probability expression also handle more intricate channels with bivariate joint densities going beyond the Wishart eigenvalues.

\subsection{Modulation and Detection} \label{sec:Sec6d5d10}
So far, we have considered slow fading, where the fading state spans the time of a redundancy block, or packet, transmission. Of course, the ME-distribution can be applied to fast fading too, with the fading state randomly varying from symbol-to-symbol.\footnote{The SER/BER can also evaluated when the fading state remains constant over a redundancy block, or packet. However, for this case, the packet error rate (PER) is of more relevance than the SER/BER.} Numerous standard textbooks on performance analysis of wireless systems, such as \cite{Wilson96, ProakisMano96, Shankar12, Rappaport01, TseVisw04, SimonAlouini05, Molisch05}, consider this symbol-scale scenario where a significant emphasis (of those and similar works) are (often) put on the analysis of modulation and detection schemes for fading channels. Performance on this level is commonly evaluated wrt symbol error rate (SER), or bit error rate (BER). In the following, we exemplify the ME-distribution method for non-coherent detection, binary DPSK and non-coherent FSK, as well as for coherent detection, binary PSK and FSK, in the AWGN channel.
\begin{theorem}
\label{thm:Thm6d14}
(Differential binary PSK (DBPSK) and FSK BER with non-coherent detection). 
Let the conditional error probability have the generic form $P(z)=\mathbf{e}^{-az}/2$, where $z$ is the instantaneous SNR, and $a$ is constant for the specific modulation and detection method  (DBPSK: $a=1$,  FSK: $a=1/2$), \cite{Wilson96, ProakisMano96, Molisch05}. Then, the BER can be written as
\begin{align} 
    P_b
	&=\int_0^\infty \frac{1}{2}\mathbf{e}^{-az} \mathbf{\tilde p}\mathrm{e}^{z\mathbf{\tilde Q}}\mathbf{\tilde r} \, \mathrm{d}z\notag\\
&=\frac{1}{2}\mathbf{\tilde p}(a\mathbf{I}-\mathbf{\tilde Q})^{-1}\mathbf{\tilde r}=\frac{1}{2}\frac{\tilde p(a)}{\tilde q(a)}.
\end{align}
\end{theorem}
Theorem~\ref{thm:Thm6d14} parallels \cite[(3.6.16),(3.6.17)]{Wilson96}, which deals with unprocessed exponentially distributed SNR fading, whereas the theorem handles fading when the effective channel SNR is ME-distributed.
The approach for the next Theorem, which deals with coherent detection, is similar to analyzing the effective capacity in Theorem~\ref{thm:Thm6d5}.
\begin{theorem}
\label{thm:Thm6d15}
(Binary PSK (BPSK) and FSK BER with coherent detection). 
Let the conditional error probability have the generic form $P(z)=Q(\sqrt{2az})$, where $z$ is the instantaneous SNR, and $a$ is constant for the specific modulation and detection method  (BPSK: $a=1$,  FSK: $a=1/2$) \cite{Wilson96, ProakisMano96, Molisch05}. Then, the BER is
\begin{align}
    P_\textrm{b}
	&=\int_0^\infty Q(\sqrt{2az}) \mathbf{\tilde p}\mathrm{e}^{z\mathbf{\tilde Q}}\mathbf{\tilde r} \, \mathrm{d}z\notag\\
&=\frac{1}{2}\left(1+\mathbf{\tilde p}\mathbf{\tilde Q}^{-1}(\mathbf{I}-\mathbf{\tilde Q}a^{-1})^{-1/2}\mathbf{\tilde r}\right).
\end{align}
\end{theorem}
\begin{IEEEproof}
The integral can be solved by integration by parts, as in \cite[(3.6.7)]{Wilson96}, or by using Craig's integral representation, \cite{Craig91}, as in \cite{ SimonAlouini05, Molisch05, SimonAlou98}, or by Theorem
\ref{thm:Thm6d3}. We have
\begin{align}
      P_\textrm{b}
	&=\int_0^\infty Q(\sqrt{2az}) \mathbf{\tilde p}\mathrm{e}^{z\mathbf{\tilde Q}}\mathbf{\tilde r} \, \mathrm{d}t\notag\\
	&\overset{(a)}{=}\int_0^\infty \left(\frac{1}{\pi}\int_0^{\pi/2} \mathrm{e}^{-\frac{az}{\sin^2(t)}}\, \mathrm{d}t\right) \mathbf{\tilde p}\mathrm{e}^{z\mathbf{\tilde Q}}\mathbf{\tilde r} \, \mathrm{d}z\notag\\
	&\overset{(b)}{=}\frac{1}{2}+\frac{1}{\pi}\int_0^{\pi/2} \mathbf{\tilde p}\mathbf{\tilde Q}^{-1}\left(\mathbf{ I}-\sin^2(t)\mathbf{\tilde Q}a^{-1}\right)^{-1}\mathbf{\tilde r} \, \mathrm{d}t \notag\\
&\overset{(c)}{=}\frac{1}{2}\left(1+\mathbf{\tilde p}\mathbf{\tilde Q}^{-1}(\mathbf{\mathbf{I}-\tilde Q}a^{-1})^{-1/2}\mathbf{\tilde r}\right),\notag
\end{align}
where Craig's integral representation, \cite{Craig91}, was used for the $Q$-function in step (a), Theorem
\ref{thm:Thm6d3} was used in step (b), and $\mathbf{\tilde p}\mathbf{\tilde Q}^{-1}\mathbf{\tilde r}=-1$, as well as the scalar integral $\int_0^{\pi/2}1/(1-\sin^2(t)c)/ \pi \, \mathrm{d}t=(c-1)^{-1/2}$, were used in step (c).
\end{IEEEproof}

Having expressed the BER in ME-distribution parameters, we now quantify the diversity gain, i.e. the BER slope at high SNR. For differential binary PSK and FSK with non-coherent detection, the diversity gain is
\begin{align}
d(R)
&\triangleq-\lim_{S\rightarrow \infty} \frac{\ln(P_\textrm{b}(R,S))}{\ln(S)}\notag\\
&=-\lim_{S\rightarrow \infty} \frac{\ln\left(\frac{1}{2}\mathbf{\tilde p}_\textrm{um}S^{-1}(a\mathbf{I}-\mathbf{\tilde Q}_\textrm{um}S^{-1})^{-1}\mathbf{\tilde r}\right)}{\ln(S)}\notag\\
&=-\lim_{S\rightarrow \infty} \frac{\ln\left(\frac{1}{2}{\tilde p_\textrm{um}(aS)}/{\tilde q_\textrm{um}(aS)}\right)}{\ln(S)} \notag\\
&=\deg{\left(\tilde q(\cdot)\right)}.
\end{align} 
Correspondingly, for binary PSK and FSK with coherent detection, the diversity gain is
\begin{align}
d(R)
&=-\lim_{S\rightarrow \infty} \frac{\ln \left(\frac{1}{2}\left(1+\mathbf{\tilde p}_\textrm{um}\mathbf{\tilde Q}_\textrm{um}^{-1}(\mathbf{I}-\mathbf{\tilde Q}_\textrm{um}a^{-1}S^{-1})^{-1/2}\mathbf{\tilde r}\right)\right)}{\ln(S)}\notag\\
&=\deg{\left(\tilde q(\cdot)\right)}.
\end{align} 

The ME-distribution can also be applied to more complicated forms of pairwise-error-probabilities, e.g. for  channel coding with interleaving and independent fading, \cite[(6.6.9)]{Wilson96}, or for space-time coding,  \cite[(3.84)]{TseVisw04}. A generic form of such PEP is given in the following theorem. 
\begin{theorem}
\label{thm:Thm6d16}
(Pairwise error probability). 
Let the conditional pairwise error probability have the generic form $P(\mathbf{c}\rightarrow \mathbf{e}| z_1,\dots z_N)=Q\left(\sqrt{2\sum_{n=1}^N {a_nz_n}}\right)$, see e.g. \cite[(6.6.9)]{Wilson96}, \cite[(3.84)]{TseVisw04},  or \cite[Chap. 13]{SimonAlouini05}, where $a_n, n\in\{1,2,\ldots N\}$, are constants, and $z_n$ are ME-distributed r.v.s. Then, the average PEP is
\begin{align}
    PEP(\mathbf{c}\rightarrow \mathbf{e})
	&=\frac{1}{\pi}\int_0^{\pi/2} \prod_{n=1}^N \mathbf{\tilde p}_n\left(\frac{a_n}{\sin^2(t)}\mathbf{I}-\mathbf{\tilde Q}_n\right)^{-1}\mathbf{\tilde r}_n \, \mathrm{d}t.
\label{eq:Eq6d157}
\end{align}
\end{theorem}
\begin{IEEEproof}
\begin{align}
&PEP(\mathbf{c}\rightarrow \mathbf{e})\notag\\
	&=\int_0^\infty \! \! \! \! \cdots \int_0^\infty \! \! Q\left(\sqrt{2\sum_{n=1}^N {a_nz_n}}\right) \prod_{n=1}^N\mathbf{\tilde p}_n\mathrm{e}^{z_n\mathbf{\tilde Q}_n}\mathbf{\tilde r}_n \, \mathrm{d}z_1 \dots \mathrm{d}z_N \notag\\
	&\overset{(a)}{=} \!  \int_0^\infty \! \! \! \! \cdots \int_0^\infty \! \! \left(\frac{1}{\pi}\int_0^{\pi/2} \mathrm{e}^{-\frac{\sum_{n=1}^N {a_nz_n}}{\sin^2(t)}}\, \mathrm{d}t\right)\notag\\
&\times \prod_{n=1}^N\mathbf{\tilde p}_n\mathrm{e}^{z_n\mathbf{\tilde Q}_n}\mathbf{\tilde r}_n \, \mathrm{d}z_1 \dots \mathrm{d}z_N\notag\\
	&=\frac{1}{\pi}
\int_0^{\pi/2} \prod_{n=1}^N \mathbf{\tilde p}_n\left(\frac{a_n}{\sin^2(t)}\mathbf{I}-\mathbf{\tilde Q}_n\right)^{-1}\mathbf{\tilde r}_n \, \mathrm{d}t,\notag
\end{align}
where Craig's integral representation \cite{Craig91}, similar to \cite[Chap. 13]{SimonAlouini05}, was (again) used in step (a).
\end{IEEEproof}

\begin{remark}
Note that, instead of an $N$-fold integral, the average PEP, \eqref{eq:Eq6d157}, is now expressed in a single variable that can be solved by numerical integration. A work that avoids such numerical integration altogether is \cite{MartinezFabrCair07}, which studies bit-interleaved BPSK in Nakagami-$m$ fading and gives an approximate PEP.
\end{remark}

In this section,we observe that the ME-distribution approach gives very simple closed-form BER-expressions, even for SNR pdfs that would have been considered untractable if expressed on regular, non-ME-distribution, forms. For example, the Nakagami-$m$ fading case, a sub-case of ME-distributed fading, is known to have a relatively complicated average error probability expression, see e.g. \cite[(3,37)]{TseVisw04}. In general, only special cases have been possible to handle, \cite{PatenaudeLodgChou97, WinWint99}. Naturally, the analysis approach given here can be extended to other, more advanced, modulation and detection schemes, such as $M$-ary PSK, QAM, etc.

\section{ME-distributed Discrete-time Signals} 
\label{sec:Sec6d6}
So far, we have considered a fading channel characterized by a ME-distribution. Here, we propose yet another use of the ME-distribution, namely to characterize the statistical properties of discrete-time signals, such as sampled information bearing signals, noise signals, or alike.

\subsection{Entropy, ME-distribution Channel, and Mutual Information} 
\label{sec:Sec6d6d1}

\subsubsection{Entropy of ME-distributed r.v.} 
\label{sec:Sec6d6d1d1}
One common key measure of r.v. signals is the entropy.
Let $T$ be a ME-distributed iid discrete signal with density $f_T(t)=\mathbf{x}\mathrm{e}^{t \mathbf{Y}}\mathbf{z}$. Then, the differential entropy is by definition 
\begin{align}
h&=-\int_0^\infty \mathbf{x}\mathrm{e}^{t \mathbf{Y}}\mathbf{z}\ln {\left(\mathbf{x}\mathrm{e}^{t \mathbf{Y}}\mathbf{z}\right)} \, \mathrm{d}t.
\label{eq:Eq6d158}
\end{align}
This integral is hard to determine in a closed-form. When $f_T(t)$ is gamma-distributed, with $m\in \mathbb{N}^+$, the entropy is known to be $h=m+\ln{(S(m-1)!/m)}+(1-m)\psi(m)$ \cite{MichalowiczNichBuch13}, where $\psi(x)=\Gamma'(x)/\Gamma(x)$ is the Digamma function. There are two challenges with \eqref{eq:Eq6d158}, the logarithm of the ME-density, and the product of the ME-density and the logarithm of the ME-density. Following the notion of approach of substituting (complicated) functions with integral representations, we get the alternative expression
\begin{align}
h
&=-\int_0^1 \int_0^\infty \frac{\mathbf{x}\mathrm{e}^{t \mathbf{Y}}\mathbf{z}\left(\mathbf{x}\mathrm{e}^{t \mathbf{Y}}\mathbf{z}-1\right)}{1+u\left(\mathbf{x}\mathrm{e}^{t \mathbf{Y}}\mathbf{z}-1\right)} \, \mathrm{d}t \, \mathrm{d}u,
\end{align}
where the integral representation $\ln(1+x)=\int_0^1 x/(1+ux)\, \mathrm{d}u$ has been used. Although, the logarithm is substituted with a rational function of ME-distributions, this integral is not easily solvable. It would be desirable with an integrand containing only a single ME-distribution. Hence, another representation, inspired by the effective capacity definition, see Theorem~\ref{thm:Thm6d4}, is given by the following lemma.
\begin{lemma}
The entropy of a density $f_T(t)$, where we assume  $f_T(t)=\mathbf{x}\mathrm{e}^{t \mathbf{Y}}\mathbf{z}$, can be written
\begin{align}
h
&=\lim_{\theta \rightarrow 0} \frac{1}{\theta}\ln\left( \int_0^\infty  f_T(t)^{1-\theta} \, \mathrm{d}t \right).
\label{eq:Eq6d160}
\end{align}
\end{lemma}
\begin{IEEEproof}
\begin{align}
&\lim_{\theta \rightarrow 0} \frac{1}{\theta}\ln\left( \int_0^\infty  f_T(t)^{1-\theta}\, \mathrm{d}t \right)\notag\\
&=\lim_{\theta \rightarrow 0} \frac{1}{\theta}\ln\left( \int_0^\infty  \mathrm{e}^{-\theta \ln(f_T(t))}f_T(t)\, \mathrm{d}t \right)\notag\\
&=
\lim_{\theta \rightarrow 0} \frac{1}{\theta}\ln\left( \int_0^\infty  \left(1-\theta \ln\left(f_T(t)\right)\right)f_T(t) \, \mathrm{d}t \right)\notag\\
&=
\lim_{\theta \rightarrow 0} \frac{1}{\theta}\ln\left( 1-\theta \int_0^\infty  f_T(t) \ln\left(f_T(t)\right) \, \mathrm{d}t \right)\notag\\
&=-\int_0^\infty  f_T(t) \ln\left(f_T(t)\right) \, \mathrm{d}t.\notag
\end{align}
\end{IEEEproof}
For $0<\theta<1$, we may now rewrite the integral in \eqref{eq:Eq6d160} as 
\begin{align}
&\int_0^\infty \left( \mathbf{x}\mathrm{e}^{t \mathbf{Y}}\mathbf{z}\right)^{1-\theta} \, \mathrm{d}t\notag\\
&\overset{(a)}{=}\frac{ \sin{\left(\pi(1-\theta)\right)}}{\pi(1-\theta)}\int_0^\infty \! \int_0^\infty \! \! \frac{\mathbf{x}\mathrm{e}^{t \mathbf{Y}}\mathbf{z}}{u^{\frac{1}{1-\theta}}+\mathbf{x}\mathrm{e}^{t \mathbf{Y}}\mathbf{z}} \, \mathrm{d}t \, \mathrm{d}u\notag\\
&\overset{(b)}{=}\frac{ \sin{\left(\pi(1-\theta)\right)}}{\pi(1-\theta)}\int_0^\infty \!  \int_0^\infty  \! \! \!
\mathbf{x}\left(u^{\frac{1}{1-\theta}}\mathrm{e}^{-t \mathbf{Y}}+\mathbf{z}\mathbf{x}\right)^{-1}\mathbf{z}
\, \mathrm{d}t \, \mathrm{d}u,
 \label{eq:Eq6d161} 
\end{align}
where we used the integral representation \cite[(11)]{HasanHasaSchar00} for the $n$th root, $x^{1/n}=\frac{ \sin{(\pi/n)}}{\pi/n}\int_0^\infty \frac{x}{u^n+x}\, \mathrm{d}u$ in step (a), and the Sherman-Morison identity, $\mathbf{a}(\mathbf{B}^{-1}+\mathbf{c}\mathbf{a})^{-1}\mathbf{c}=\mathbf{a}\mathbf{B}\mathbf{c}/(1+\mathbf{a}\mathbf{B}\mathbf{c})$ in step (b). Since the matrix $\mathbf{z}\mathbf{x}$ inside the inverse is singular, the integral is hard to solve. However, we believe the current integral forms can be helpful for further analysis.

\subsubsection{Communication Channel with ME-distributed r.v.s and Mutual Information} 
\label{sec:Sec6d6d1d2}
Like the AWGN channel with Gaussian input distribution, 
it is natural to ponder about a communication channel where the input distribution and noise are assumed ME-distributed. This could potentially represent, or approximate, an optical channel with information carried in the intensity. Consider the baseband model
\begin{align}
y=x+w, \, x\geq0, \, w\geq0,
\end{align}
where $y$ is the received signal, $x$ is the transmitted signal, and $w$ the noise. Both $x$ and $w$ are assumed ME-distributed, characterized by ME-parameters $(\mathbf{p}_x,\mathbf{Q}_x,\mathbf{r}_x)$, and $(\mathbf{p}_w,\mathbf{Q}_w, \mathbf{r}_w)$. Exploiting the closure of the convolution operation for ME-distributions, Proposition \ref{pr:Pr6d1}, it is recognized that $y$ is also ME-distributed with ME-parameters $(\mathbf{p}_y,\mathbf{Q}_y, \mathbf{r}_y)$, where $\mathbf{p}_y=[\mathbf{p}_x \ \mathbf{0}]$,  $\mathbf{Q}_y=[\mathbf{Q}_x \ \mathbf{r}_x\mathbf{p}_w; \mathbf{0} \ \mathbf{Q}_w]$, $\mathbf{r}_y=[\mathbf{0} \ \mathbf{r}_w]$. 

One key aspect of interest, of this channel, is the mutual information. The MI is $I\triangleq h(y)-h(y|x)=h(y)-h(w)$, where $h(\cdot)$ denotes the differential entropy. The MI can now be expressed, in unit mean parameters, as
\begin{align}
I
=&h(y)-h(y|x)=h(y)-h(w)\notag\\
=&-\int_0^\infty \mathbf{x}_\textrm{y}\mathrm{e}^{t \mathbf{Y}_\textrm{y}}\mathbf{z}_\textrm{y}\ln {\left(\mathbf{x}_\textrm{y}\mathrm{e}^{t \mathbf{Y}_\textrm{y}}\mathbf{z}_\textrm{y}\right)} \, \mathrm{d}t\notag\\
&+\int_0^\infty \mathbf{x}_\textrm{w}\mathrm{e}^{t \mathbf{Y}_\textrm{w}}\mathbf{z}_\textrm{w}\ln {\left(\mathbf{x}_\textrm{w}\mathrm{e}^{t \mathbf{Y}_\textrm{w}}\mathbf{z}_\textrm{w}\right)} \, \mathrm{d}t,\notag\\
&\left\{\mathbf{x}_\textrm{y}
=\mathbf{x}_\textrm{yun}
\mathbf{S}_\textrm{y}^{-1},\ \mathbf{x}_\textrm{yun}
=[\mathbf{x}_\textrm{xun} \ \mathbf{0}] ,\ \mathbf{S}_\textrm{y}
=[\mathbf{S}_\textrm{x} \ \mathbf{0}; \ \mathbf{0} \ \mathbf{S}_\textrm{w}]\right., \notag\\
&\mathbf{Y}_\textrm{y}
=\mathbf{Y}_\textrm{yun}\mathbf{S}_\textrm{y}^{-1}, \ \mathbf{Y}_\textrm{yun}=[\mathbf{Y}_\textrm{xun} \ \mathbf{z}_\textrm{yun}\mathbf{x}_\textrm{wun}; \ \mathbf{0} \ \mathbf{Y}_\textrm{wun}],\notag\\
&\left. \mathbf{z}_\textrm{y}
=\mathbf{z}_\textrm{yun},\ \mathbf{z}_\textrm{yun}
=[\mathbf{0}; \ \mathbf{z}_\textrm{wun} ] \right\},\notag\\
=&-\int_0^\infty \mathbf{x}_\textrm{yun}\mathbf{S}_\textrm{y}^{-1}\mathrm{e}^{t \mathbf{Y}_\textrm{yun}\mathbf{S}_\textrm{y}^{-1}}\mathbf{z}_\textrm{yun} \ln {\left(\mathbf{x}_\textrm{yun}\mathbf{S}_\textrm{y}^{-1}\mathrm{e}^{t \mathbf{Y}_\textrm{yun}\mathbf{S}_\textrm{y}^{-1}}\mathbf{z}_\textrm{yun}\right)} \, \mathrm{d}t\notag\\
&+\int_0^\infty S_\textrm{w}^{-1}\mathbf{x}_\textrm{wun}\mathrm{e}^{t S_\textrm{w}^{-1}\mathbf{Y}_\textrm{wun}}\mathbf{z}_\textrm{w}\ln {\left(S_\textrm{w}^{-1}\mathbf{x}_\textrm{wun}\mathrm{e}^{t S_\textrm{w}^{-1}\mathbf{Y}_\textrm{wun}}\mathbf{z}_\textrm{w}\right)} \, \mathrm{d}t\notag\\
=&-\int_0^\infty \mathbf{x}_\textrm{yun}\mathrm{e}^{u \mathbf{Y}_\textrm{yun}}\mathbf{z}_\textrm{yun} \ln {\left(\mathbf{x}_\textrm{yun}\mathbf{S}_\textrm{y}^{-1}\mathrm{e}^{u \mathbf{Y}_\textrm{yun}}\mathbf{z}_\textrm{yun}\right)} \, \mathrm{d}u\notag\\
&+\int_0^\infty \mathbf{x}_\textrm{wun}\mathrm{e}^{u \mathbf{Y}_\textrm{wun}}\mathbf{z}_\textrm{wun}\ln {\left(S_\textrm{w}^{-1}\mathbf{x}_\textrm{wun}\mathrm{e}^{u \mathbf{Y}_\textrm{wun}}\mathbf{z}_\textrm{wun}\right)} \, \mathrm{d}u\notag\\
=&+\ln(S_\textrm{x})-\int_0^\infty \mathbf{x}_\textrm{yun}\mathrm{e}^{u \mathbf{Y}_\textrm{yun}}\mathbf{z}_\textrm{yun} \ln {\left(\mathbf{x}_\textrm{yun}\mathrm{e}^{u \mathbf{Y}_\textrm{yun}}\mathbf{z}_\textrm{yun}\right)} \, \mathrm{d}u\notag\\
&-\ln(S_\textrm{w})+\int_0^\infty \mathbf{x}_\textrm{wun}\mathrm{e}^{u \mathbf{Y}_\textrm{wun}}\mathbf{z}_\textrm{wun}\ln {\left(\mathbf{x}_\textrm{wun}\mathrm{e}^{u \mathbf{Y}_\textrm{wun}}\mathbf{z}_\textrm{wun}\right)} \, \mathrm{d}u\notag\\
=&\ln\left({S_\textrm{x}}/{S_\textrm{w}}\right)-\left(h(y_\textrm{um})-h(w_\textrm{um})\right).
\end{align}
Hence, if the unit mean entropies can be determined in a closed-form for ME-distributed r.v.s, then so can the MI.
Since $h(y_\textrm{um})-h(w_\textrm{um})\geq 0$, we get $I\leq \ln\left({S_\textrm{x}}/{S_\textrm{w}}\right)$.

\subsubsection{Quantization} 
\label{sec:Sec6d6d1d3}
Facing a new signal distribution, the ME-distribution, quantization is yet another aspect of interest to explore. We will not delve deeply into this rich topic, but merely illustrate two possible applications.

For example, the Lloyd-Max quantization algorithm, see e.g. \cite{GrayNeuh98}, iteratively computes quantization limits and centroids, of a r.v. with a given PDF and $M$ levels that minimizes the mean square error (MSE).
For the ME-distribution, the decision thresholds and the centroids in closed-form, are
\begin{align}
l_q&=\frac{1}{2}\left(\hat u_{q-1}+\hat u_{q}\right), \, q=\{1,2,\ldots M-1 \},\notag\\
\hat u_{q}
&=\frac{\int_{l_q}^{l_{q+1}}t f_T(t) \, \mathrm{d}t}{\int_{l_q}^{l_{q+1}} f_T(t) \, \mathrm{d}t}, \, q=\{0,1,\ldots M-1\},
\end{align}
where the $q$th centroid for the ME-density is given in the closed-form
\begin{align}
\hat u_{q}
&=\frac{\int_{l_q}^{l_{q+1}}t\mathbf{x}\mathrm{e}^{t \mathbf{Y}}\mathbf{z} \, \mathrm{d}t}{\int_{l_q}^{l_{q+1}} \mathbf{x}\mathrm{e}^{t \mathbf{Y}}\mathbf{z} \, \mathrm{d}t}\notag\\
&=\frac{\mathbf{x}\mathrm{e}^{t \mathbf{Y}}\left(t\mathbf{Y}^{-1}-\mathbf{Y}^{-2} \right)\mathbf{z} |_{l_q}^{l_{q+1}}}{\mathbf{x}\mathrm{e}^{t \mathbf{Y}}\mathbf{Y}^{-1}\mathbf{z} |_{l_q}^{l_{q+1}}}.
\end{align}

The max-quantization problem, assuming infinite number of quantization levels, is another classical quantization problem worth touching on. The Panter-Dite formula, see e.g. \cite{GrayNeuh98}, expresses  the MSE for this case as
\begin{align}
MSE&
\approx \frac{1}{12M^2} \left(\int_0^\infty \left( \mathbf{x}\mathrm{e}^{t \mathbf{Y}}\mathbf{z}\right)^{1/3} \, \mathrm{d}t\right)^3.
\label{eq:Eq6d166}
\end{align}
A special case that can be solved is when the ME-pdf can be decomposed as $\mathbf{x}\mathbf{T}^{-1}=\mathbf{\breve x}\otimes \mathbf{\breve x}\otimes \mathbf{\breve x}$, $\mathbf{T}\mathbf{Y}\mathbf{T}^{-1}=\mathbf{\breve Y}\oplus \mathbf{\breve Y}\oplus \mathbf{\breve Y}$,  $\mathbf{T}\mathbf{z}=\mathbf{\breve z}\otimes \mathbf{\breve z}\otimes \mathbf{\breve z}$, where $\mathbf{T}$ is a transform matrix of choice. The integral in \eqref{eq:Eq6d166} is then
\begin{align}
&\int_0^\infty \left( \mathbf{x}\mathrm{e}^{t \mathbf{Y}}\mathbf{z}\right)^{1/3} \, \mathrm{d}t\notag\\
&=\int_0^\infty \left( \mathbf{x}\mathbf{T}^{-1}\mathrm{e}^{t \mathbf{T}\mathbf{Y}\mathbf{T}^{-1}}\mathbf{T}\mathbf{z}\right)^{1/3} \, \mathrm{d}t\notag\\
&=\int_0^\infty 
\left( \left( \mathbf{\breve x} \otimes \mathbf{\breve x} \otimes \mathbf{\breve x} \right)
\mathrm{e}^{t \left( \mathbf{\breve Y} \oplus \mathbf{\breve Y} \oplus \mathbf{\breve Y} \right)}
\left( \mathbf{\breve z}  \otimes  \mathbf{\breve z}  \otimes  \mathbf{\breve z} \right) \right)^{1/3} \, \mathrm{d}t\notag\\
&=\int_0^\infty \left( \left(\mathbf{\breve x}\mathrm{e}^{t \mathbf{\breve Y}}\mathbf{\breve z}\right)^3\right)^{1/3} \, \mathrm{d}t\notag\\
&=-\mathbf{\breve x} \mathbf{\breve Y}^{-1}\mathbf{\breve z}.
\end{align}
For the more general case, without such decomposition, we have not found a solution. Yet, one strategy, that may facilitate the solution of \eqref{eq:Eq6d166}, could be to exploit \eqref{eq:Eq6d161}, which implies $\int_0^\infty \left( \mathbf{x}\mathrm{e}^{t \mathbf{Y}}\mathbf{z}\right)^{1/3} \, \mathrm{d}t
=\frac{3\sqrt{3}}{2 \pi}\int_0^\infty \int_0^\infty \frac{\mathbf{x}\mathrm{e}^{t \mathbf{Y}}\mathbf{z}}{u^3+\mathbf{x}\mathrm{e}^{t \mathbf{Y}}\mathbf{z}} \, \mathrm{d}t \, \mathrm{d}u$.

\section{ME-distribution Generalizations} 
\label{sec:Sec6d7}
Already in Section \ref{sec:Sec6d3d1}, and Tab. \ref{tab:Tab6d1}, we generalized the ME-distribution to two new distributions, paralleling the Rayleigh distribution, and the bivariate Gaussian distribution. In this section, we look at those two distributions, but also introduce a third ME-distribution generalization, paralleling the univariate Gaussian distribution.
The proposed distributions are of interest to consider for several reasons. On one hand, it is interesting to see what the new mathematical forms offers in terms of closed-form expressions for some basic properties, and general manipulability. On the other hand, due to the general matrix parameter form, we expect the proposed pdfs to be able to approximate a great number of practically relevant pdfs. This is so since the reference distribution, the ME-distribution, is dense on its domain. We see applications to characterize, e.g., channel fading, discrete-time signals, etc. Some basic characteristics of interests, in addition to the pdfs, are the moments, the cdfs, various integrals and, for the bivariate distribution, also the marginal densities. Below, we consider pdf definitions and moments.

\subsection{Type I -- Univariate Matrix Gaussian-like Distribution} 
\label{sec:Sec6d7d1}
\begin{figure}[t]
 \centering
 \vspace{+.1 cm}
 \includegraphics[width=9cm]{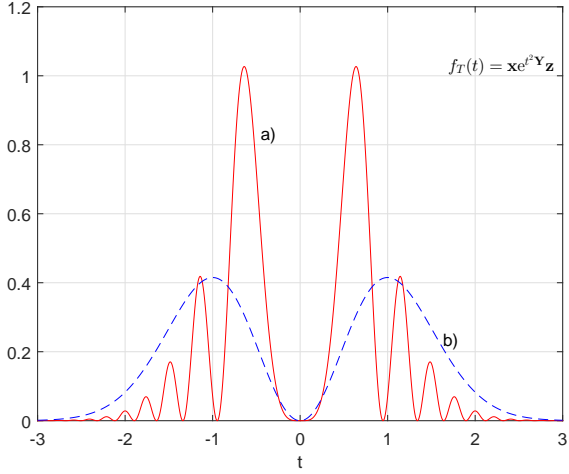}
 \vspace{-0.7cm}
 \caption{Example of type I distribution for a) $\mathbf{x}=[50 \ 0 \ 0]$, $\mathbf{y}=[50 \ 52 \ 3]$, $\mathbf{z}=[0 \ 0 \ 1]^\textrm{T}$, and b) $\mathbf{x}=[1  \ 0]$, $\mathbf{y}=[1 \ 2]$, $\mathbf{z}=[0 \ 1]^\textrm{T}$, with  $\mathbf{Q}=\mathbf{S}-\mathbf{r}\mathbf{p}$.}
 \label{fig:1DGME}
 \vspace{-0.3cm}
\end{figure}

\begin{definition} 
Let $f_T(t)=c\mathbf{x}\mathrm{e}^{t^2\mathbf{Y}}\mathbf{z}, \, t\in[-\infty,\infty],$ with $c=(\sqrt{\pi}\mathbf{x}\left(-\mathbf{Y}\right)^{-1/2}\mathbf{z})^{-1}$, denote the type I pdf.
\end{definition}
We exemplify two univariate type I pdfs in Fig. ~\ref{fig:1DGME}.
The odd moments are zeros since $f_T(t)$ is even. The even moments are
\begin{align}
\mathbb{E}\{T^n\}
&=c\int_{-\infty}^\infty
t^n\mathbf{x}\mathrm{e}^{t^2\mathbf{Y}}\mathbf{z} \, \mathrm{d}t, \, n=\{0,2,4,\ldots\}, \notag\\
&=2c\int_0^\infty
t^n\mathbf{x}\mathrm{e}^{t^2\mathbf{Y}}\mathbf{z} \, \mathrm{d}t \notag\\
&=c\int_0^\infty
y^{(n-1)/2}\mathbf{x}\mathrm{e}^{y\mathbf{Y}}\mathbf{z} \, \mathrm{d}y \notag\\
&=\frac{2c}{\sqrt{\pi}}\int_0^\infty \int_0^\infty
y^{n/2}\mathrm{e}^{-y x^2}\mathbf{x}\mathrm{e}^{y\mathbf{Y}}\mathbf{z} \, \mathrm{d}y \, \mathrm{d}x  \notag\\
&=c\Gamma\left(\frac{n+1}{2}\right)\mathbf{x}\left(-\mathbf{Y}\right)^{-(n+1)/2}\mathbf{z}.
\label{eq:Eq6d168}
\end{align}
For this distribution, if the matrix parameter, $(\mathbf{x},\mathbf{Y},\mathbf{z})$, are taken directly from a ME-distribution, normalization to unit probability is required. 
The normalization constant $c$ is determined by the condition $1=c\int_{-\infty}^\infty
\mathbf{x}\mathrm{e}^{t^2\mathbf{Y}}\mathbf{z} \, \mathrm{d}t $. Using \eqref{eq:Eq6d168} with $n=0$ yields
$c=(\sqrt{\pi}\mathbf{x}\left(-\mathbf{Y}\right)^{-1/2}\mathbf{z})^{-1}$.

\subsection{Type II -- Bivariate Matrix Gaussian-like Distribution} 
\label{sec:Sec6d7d2}
\begin{figure}[t]
 \centering
 \vspace{+.1 cm}
 \includegraphics[width=9cm]{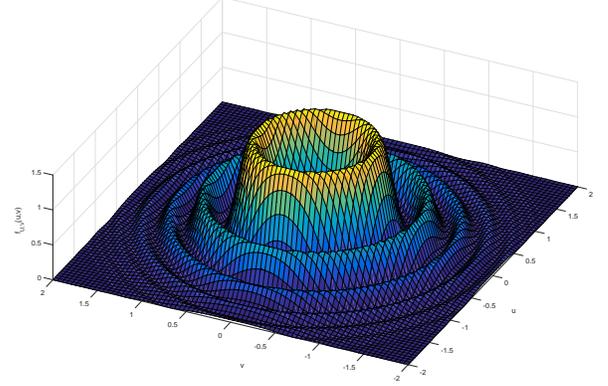}
 \vspace{-0.7cm}
 \caption{Example of type II distribution for $\mathbf{x}=[50 \ 0 \ 0]$, $\mathbf{y}=[50 \ 52 \ 3]$, $\mathbf{z}=[0 \ 0 \ 1]^\textrm{T}$, and $\mathbf{Y}=\mathbf{S}-\mathbf{z}\mathbf{x}$.}
 \label{fig:2DGME}
 \vspace{-0.3cm}
\end{figure}

\begin{definition}
Let $f_{U,V}(u,v)=\frac{1}{\pi}\mathbf{x}\mathrm{e}^{(u^2+v^2)\mathbf{Y}}\mathbf{z}, \, u\in[-\infty,\infty], \, v\in[-\infty,\infty],$ denote the type II pdf.
\end{definition}
We illustrate an example of a type II pdf in Fig. ~\ref{fig:2DGME}.
We note that due to the symmetry, any of the odd moments are zero. The even moments are
\begin{align}
\mathbb{E}\{U^nV^m\}
&=\frac{1}{\pi}\int_{-\infty}^\infty \int_{-\infty}^\infty
u^nv^m\mathbf{x}\mathrm{e}^{(u^2+v^2)\mathbf{Y}}\mathbf{z} \, \mathrm{d}u \, \mathrm{d}v,  \notag\\ 
&\, n=\{0,2,4,\ldots\}, \, m=\{0,2,4,\ldots\}, \notag\\
&=\frac{4}{\pi}\int_0^\infty
a^{(n-1)/2}b^{(m-1)/2}\mathbf{x}\mathrm{e}^{(a+b)\mathbf{Y}}\mathbf{z} \, \mathrm{d}a \, \mathrm{d}b \notag\\
&=\frac{4}{\pi}\Gamma\left(\frac{n+1}{2}\right)\Gamma\left(\frac{m+1}{2}\right)\notag\\
&\times\mathbf{x}\left(-\mathbf{Y}\right)^{-(n+1)/2}\left(-\mathbf{Y}\right)^{-(m+1)/2}\mathbf{z}.
\end{align}

The marginal density for $U$ is
\begin{align}
f_U(u)
&=  \frac{1}{\pi}\int_{-\infty}^\infty
\mathbf{x}\mathrm{e}^{(u^2+v^2)\mathbf{Y}}\mathbf{z}
\mathrm{d}v\notag\\
&= \frac{1}{\sqrt{\pi}}
\mathbf{x}\mathrm{e}^{u^2\mathbf{Y}}\left(-\mathbf{Y}\right)^{-1/2}\mathbf{z}.
\end{align}
The marginal density $f_V(v)$ is found analogously.

\subsection{Type III -- Matrix Rayleigh-like Distribution} 
\label{sec:Sec6d7d3}
\begin{figure}[t]
 \centering
 \vspace{+.1 cm}
 \includegraphics[width=9cm]{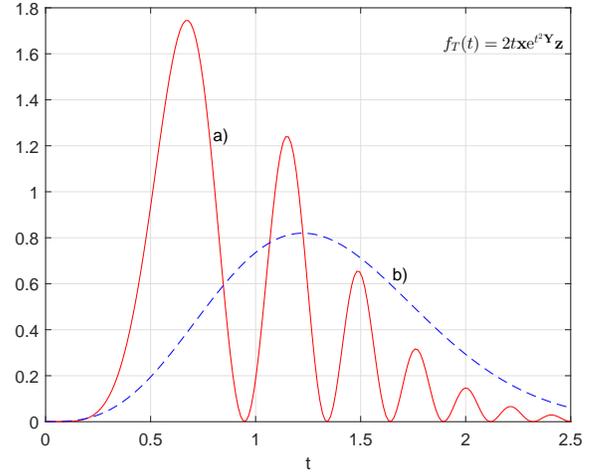}
 \vspace{-0.7cm}
 \caption{Example of type III distribution for a) $\mathbf{x}=[50 \ 0 \ 0]$, $\mathbf{y}=[50 \ 52 \ 3]$, $\mathbf{z}=[0 \ 0 \ 1]^\textrm{T}$, and b) $\mathbf{x}=[1  \ 0]$, $\mathbf{y}=[1 \ 2]$, $\mathbf{z}=[0 \ 1]^\textrm{T}$, with $\mathbf{Y}=\mathbf{S}-\mathbf{z}\mathbf{x}$.}
 \label{fig:1DRME}
 \vspace{-0.3cm}
\end{figure}

\begin{definition}
Let $f_{T}(t)=2t\mathbf{x}\mathrm{e}^{t^2\mathbf{Y}}\mathbf{z}, \, t\in(0,\infty],$ denote the type III pdf.
\end{definition}
We illustrate two examples of type III pdfs in Fig. ~\ref{fig:1DRME}.
For this case, the moments are
\begin{align}
\mathbb{E}\{T^n\}
&=\int_{0}^\infty 
2t^{n+1}\mathbf{x}\mathrm{e}^{t^2\mathbf{Y}}\mathbf{z} \, \mathrm{d}t  \, n=\{0,1,\ldots\} \notag\\
&=\int_0^\infty
u^{n/2}\mathbf{x}\mathrm{e}^{u\mathbf{Y}}\mathbf{z} \, \mathrm{d}u \notag\\
&=\Gamma \left(\frac{n+2}{2}\right)\mathbf{x}\left(-\mathbf{Y}\right)^{-(n+2)/2}\mathbf{z}.
\end{align}

We finally note that many other distributions, inspired by or extended from the ME-distribution form, as the above, can be formed.

\section{Summary and Conclusions}
\label{sec:Sec6d8}

In this work, we structured, refined and extended the ME-distribution approach for performance analysis of wireless communication systems with ME-distributed fading SNR. New tools were derived, new communication cases were analyzed, and new channel fading models were introduced.

It was demonstrated that the ME-distribution framework is useful to characterize the effective channel SNR (or MI) due to signal processing, communication schemes, and interference, as well as to analyze (H)ARQ systems wrt throughput, rate-adaptive systems wrt effective capacity, and modulation and detection schemes wrt SER/BER/PEP and diversity gain. We also exemplified its use to, e.g., 3-phase NCBR, and SM-MIMO. We extended the framework to a bivariate ME-distribution case, and showed its use for ARQ with non-identical (possibly dependent) ME-distributed signal and interferers. Towards the end, we let the ME-distribution represent, not fading, but discrete-time random variable signals, and looked at entropy and mutual information for the ME-distribution, but also generalized the ME-distribution to uni- and bivariate Gaussian like distributions.

To conclude, we believe that the ME-distribution approach can be helpful for wireless system performance analysis in communication theory, information theory, and related areas. It may be possible to extend this framework further, e.g. combining queuing and fading channel analysis under a unified framework. The ARQ-interference and NCBR analysis hints that the ME-distribution approach is also useful to analyze multinode systems.

\appendix
\label{sec:Sec7}

\subsubsection{Proofs } 
\label{sec:Sec7d2} Below, we give several of the proofs. \\
\begin{IEEEproof} (Ex.~\ref{ex:Ex6d5})
The Laplace transform is
\begin{align}
    &F(s)
    =\int_0^\infty \mathrm{e}^{-sz} \left(NS^{-1}\mathrm{e}^{-zS^{-1}} (1-\mathrm{e}^{-zS^{-1}})^{N-1}\right) \, \mathrm{d}z\notag\\
    &=\underbrace{\frac{\mathrm{e}^{-z(S^{-1}+s)}}{S^{-1}+s} NS^{-1}(1-\mathrm{e}^{-zS^{-1}})^{N-1}|_0^\infty}_{=0}+\frac{NS^{-1}}{S^{-1}+s}\notag\\
&\times\int_0^\infty \mathrm{e}^{-sz} \left((N-1)S^{-1}\mathrm{e}^{-2zS^{-1}}(1-\mathrm{e}^{-zS^{-1}})^{N-2}\right) \, \mathrm{d}z\notag\\
    &\overset{(a)}{=}\frac{N!}{\prod_{n=1}^N(n+sS)}
    =\frac{1}{\prod_{n=1}^N(1+sS/n)}\notag,
\end{align}
where partial integration is used repeatedly in step (a).
\end{IEEEproof}

\begin{IEEEproof} (Theorem~\ref{thm:Thm6d3})
The expectation is computed as
\begin{align}
     \mathbb{E}\{g(t)\}
&=\int_0^\infty g(t) \mathbf{x}\mathrm{e}^{t\mathbf{Y}}\mathbf{z} \, \mathrm{d}t\notag\\
	&=\int_0^\infty \left( \int_{a_u}^{b_u} g_1(u) \mathrm{e}^{-t g_2(u)} \, \mathrm{d}u\right)\mathbf{x}\mathrm{e}^{t\mathbf{Y}}\mathbf{z} \, \mathrm{d}t \notag\\
&=\int_{a_u}^{b_u} g_1(u)\left(\int_0^\infty \mathbf{x}\mathrm{e}^{t(\mathbf{Y}-g_2(u)\mathbf{I})}\mathbf{z} \, \mathrm{d}t \right)\, \mathrm{d}u\notag\\
&=\mathbf{x}\left(\int_{a_u}^{b_u} g_1(u) \left(g_2(u)\mathbf{I}-\mathbf{Y}\right)^{-1}\, \mathrm{d}u\right) \mathbf{z}\notag\\
&=-\mathbf{x}\mathbf{Y}^{-1}\left(\int_{a_u}^{b_u} g_1(u) \left(\mathbf{I}-g_2(u)\mathbf{Y}^{-1}\right)^{-1}\, \mathrm{d}u\right) \mathbf{z},\notag
\end{align}
where the last expression is a somewhat more convenient form to determine the scalar integral.
\end{IEEEproof}

\begin{IEEEproof} (Example~\ref{ex:Ex6d19}) The entries are expressed as
\begin{align}
g_{ij}(s)
&=\int_0^\infty \frac{\lambda^{i+j+d}\mathrm{e}^{-\lambda}}{(1+t\lambda )^s} \, \mathrm{d}\lambda \notag\\
& \overset{(a)}{=}\frac{1}{\Gamma(s)}\int_0^\infty \int_0^\infty u^{s-1}\mathrm{e}^{-u(1+t\lambda )} \lambda^{i+j+d}\mathrm{e}^{-\lambda} \, \mathrm{d}\lambda \  \mathrm{d}u \notag\\
&=\frac{1}{\Gamma(s)}\int_0^\infty \int_0^\infty u^{s-1}\mathrm{e}^{-u} \lambda^{i+j+d}\mathrm{e}^{-\lambda(1+tu)} \, \mathrm{d}\lambda \ \mathrm{d}u \notag\\
&\overset{(b)}{=}\frac{1}{\Gamma(s)}\int_0^\infty \int_0^\infty \frac{u^{s-1}\mathrm{e}^{-u}}{(1+tu)^{i+j+d+1}} v^{i+j+d}\mathrm{e}^{-v} \, \mathrm{d}v \ \mathrm{d}u \notag\\
&=\frac{(i+j+d)!}{\Gamma(s)} \int_0^\infty \frac{u^{s-1}\mathrm{e}^{-u}}{(1+tu)^{i+j+d+1}} \mathrm{d}u\notag\\
&\sim\frac{(i+j+d)!}{t^{i+j+d+1}\Gamma(s)} \int_0^\infty u^{s-i-j-d-2}\mathrm{e}^{-u} \mathrm{d}u\notag\\
&=\frac{(i+j+d)!\Gamma(s-i-j-d-1)}{t^{i+j+d+1}\Gamma(s)}\notag\\
&\overset{(c)}{=}\frac{(i+j+d)!}{t^{i+j+d+1}}\prod_{n=1}^{i+j+d+1}\frac{1}{s-n},\notag
\end{align}
where an integral representation, a variable substitution, and  $\Gamma(1+x)=x\Gamma(x)$ were used in step (a), (b), and (c).
\end{IEEEproof}

\subsubsection{Auxiliary Parametric Optimization} \label{sec:Sec7d1}
The Auxiliary parametric optimization approach in \cite{LarssonRasmSkog14b} is briefly reviewed here. We assume that the throughput has the form $T=R/f_\Theta(\Theta)$, where $\Theta=(\mathrm{e}^R-1)/S$, and give the following Corollary.
\begin{corollary} 
\label{cr:Crl6d10}
The optimal rate point, the optimal throughput, and the SNR are parametrically given (in the auxiliary parameter $\Theta$) by
\begin{align}
T^*( \Theta)&=\frac{R^*}{f_\Theta},\label{eq:Eq6d172}\\
S( \Theta)&=\frac{\mathrm{e}^{R^*}-1}{ \Theta},\label{eq:Eq6d173}\\
R^*( \Theta)&=g_\Theta+W_0(-g_\Theta\mathrm{e}^{-g_\Theta}) ,\label{eq:Eq6d174}\\
&\text{where}\notag\\
g_\Theta( \Theta)&\triangleq\frac{f_\Theta}{ \Theta f'_\Theta},\label{eq:Eq6d175}\\
 \Theta &\in [0,\infty).\notag
\end{align}
\end{corollary}
\begin{IEEEproof}
Taking the derivate of $T=R/f_\Theta(\Theta)$ wrt $R$, equating to zero, this can be expressed as $R\mathrm{e}^R/(\mathrm{e}^R-1)=f_\Theta(\Theta)/\Theta f'_\Theta(\Theta)\triangleq g_\Theta(\Theta)$, which is then solved for $R$.
\end{IEEEproof}


\begin{thebibliography}{10}
\providecommand{\url}[1]{#1}
\csname url@samestyle\endcsname
\providecommand{\newblock}{\relax}
\providecommand{\bibinfo}[2]{#2}
\providecommand{\BIBentrySTDinterwordspacing}{\spaceskip=0pt\relax}
\providecommand{\BIBentryALTinterwordstretchfactor}{4}
\providecommand{\BIBentryALTinterwordspacing}{\spaceskip=\fontdimen2\font plus
\BIBentryALTinterwordstretchfactor\fontdimen3\font minus
  \fontdimen4\font\relax}
\providecommand{\BIBforeignlanguage}[2]{{%
\expandafter\ifx\csname l@#1\endcsname\relax
\typeout{** WARNING: IEEEtran.bst: No hyphenation pattern has been}%
\typeout{** loaded for the language `#1'. Using the pattern for}%
\typeout{** the default language instead.}%
\else
\language=\csname l@#1\endcsname
\fi
#2}}
\providecommand{\BIBdecl}{\relax}
\BIBdecl

\bibitem{LarssonRasmSkog16a}
P.~Larsson, L.~K. Rasmussen, and M.~Skoglund, ``Throughput analysis of
  hybrid-{ARQ} -- {A} matrix exponential distribution approach,'' \emph{IEEE
  Transactions on Communications}, vol.~64, no.~1, pp. 416--428, Jan 2016.

\bibitem{LarssonRasmSkog16b}
P.~Larsson, J.~Gross, H.~Al-Zubaidy, L.~K. Rasmussen, and M.~Skoglund,
  ``Effective capacity of retransmission schemes: {A} recurrence relation
  approach,'' \emph{IEEE Transactions on Communications}, vol.~64, no.~11, pp.
  4817--4835, Nov 2016.

\bibitem{Wilson96}
S.~G. Wilson, \emph{Digital Modulation and Coding}, 1st~ed.\hskip 1em plus
  0.5em minus 0.4em\relax Delhi: Pearson Education, 1996.

\bibitem{ProakisMano96}
J.~G. Proakis and D.~G. Manolakis, \emph{Digital Signal Processing (3rd Ed.):
  {P}rinciples, Algorithms, and Applications}.\hskip 1em plus 0.5em minus
  0.4em\relax Upper Saddle River, NJ, USA: Prentice-Hall, Inc., 1996.

\bibitem{CaireTuni01}
G.~Caire and D.~Tuninetti, ``The throughput of hybrid-{ARQ} protocols for the
  {G}aussian collision channel,'' \emph{IEEE Transactions on Information
  Theory}, vol.~47, no.~5, pp. 1971--1988, Jul 2001.

\bibitem{LanemanTseWorn04}
\BIBentryALTinterwordspacing
J.~N. Laneman, D.~N. Tse, and G.~W. Wornell, ``Cooperative diversity in
  wireless networks: {E}fficient protocols and outage behavior,'' \emph{IEEE
  Transactions on Information Theory}, vol.~50, no.~12, pp. 3062--3080, Dec
  2004. [Online]. Available: \url{http://dx.doi.org/10.1109/TIT.2004.838089}
\BIBentrySTDinterwordspacing

\bibitem{HunterNosra06}
T.~E. Hunter and A.~Nosratinia, ``Diversity through coded cooperation,''
  \emph{IEEE Transactions on Wireless Communications}, vol.~5, no.~2, pp.
  283--289, Feb 2006.

\bibitem{Rappaport01}
T.~Rappaport, \emph{Wireless Communications: {P}rinciples and Practice},
  2nd~ed.\hskip 1em plus 0.5em minus 0.4em\relax Upper Saddle River, NJ, USA:
  Prentice Hall PTR, 2001.

\bibitem{Shankar12}
P.~M. Shankar, \emph{Fading and Shadowing in Wireless Systems}.\hskip 1em plus
  0.5em minus 0.4em\relax Springer-Verlag New York, 2012.

\bibitem{LarssonRasmSkog14b}
P.~Larsson, L.~K. Rasmussen, and M.~Skoglund, ``Throughput analysis of {ARQ}
  schemes in {G}aussian block fading channels,'' \emph{IEEE Transactions on
  Communications}, vol.~62, no.~7, pp. 2569--2588, July 2014.

\bibitem{BetteshSham06}
I.~Bettesh and S.~Shamai, ``Optimal power and rate control for minimal average
  delay: The single-user case,'' \emph{IEEE Transactions on Information
  Theory}, vol.~52, no.~9, pp. 4115--4141, Sept 2006.

\bibitem{ShenLiuFitz08}
C.~Shen, T.~Liu, and M.~Fitz, ``Aggressive transmission with {ARQ} in
  quasi-static fading channels,'' in \emph{Proc. {IEEE} International
  Conference on Communications ({ICC}�08)}, May 2008, pp. 1092--1097.

\bibitem{Mandelbaum74}
D.~Mandelbaum, ``An adaptive-feedback coding scheme using incremental
  redundancy (corresp.),'' \emph{IEEE Transactions on Information Theory},
  vol.~20, no.~3, pp. 388--389, May 1974.

\bibitem{Sindhu77}
P.~Sindhu, ``Retransmission error control with memory,'' \emph{IEEE
  Transactions on Communications}, vol.~25, no.~5, pp. 473--479, May 1977.

\bibitem{Chase85}
D.~Chase, ``Code combining -- a maximum-likelihood decoding approach for
  combining an arbitrary number of noisy packets,'' \emph{IEEE Transactions on
  Communications}, vol.~33, no.~5, pp. 385--393, May 1985.

\bibitem{Benelli85}
G.~Benelli, ``An {ARQ} scheme with memory and soft error detectors,''
  \emph{IEEE Transactions on Communications}, vol.~33, no.~3, pp. 285--288, Mar
  1985.

\bibitem{WuJind10}
P.~Wu and N.~Jindal, ``Performance of hybrid-{ARQ} in block-fading channels:
  {A} fixed outage probability analysis,'' \emph{IEEE Transactions on
  Communications}, vol.~58, no.~4, pp. 1129--1141, April 2010.

\bibitem{SzczecinskiKDR13}
L.~Szczecinski, S.~Khosravirad, P.~Duhamel, and M.~Rahman, ``Rate allocation
  and adaptation for incremental redundancy truncated {HARQ},'' \emph{IEEE
  Transactions on Communications}, vol.~61, no.~6, pp. 2580--2590, June 2013.

\bibitem{TseVisw04}
D.~Tse and P.~Viswanath, \emph{Fundamentals of Wireless Communications}.\hskip
  1em plus 0.5em minus 0.4em\relax New York, NY, USA: Cambridge Univ. Press,
  2004.

\bibitem{TarokhSeshCald98}
V.~Tarokh, N.~Seshadri, and A.~Calderbank, ``Space-time codes for high data
  rate wireless communication: {P}erformance criterion and code construction,''
  \emph{IEEE Transactions on Information Theory}, vol.~44, no.~2, pp. 744--765,
  Mar 1998.

\bibitem{AlamoutiTaro97P}
S.~Alamouti and V.~Tarokh, ``Transmitter diversity technique for wireless
  communications,'' U.S. Patent 6\,185\,258 B1, Sep 16, 1997.

\bibitem{Alamouti98}
S.~Alamouti, ``A simple transmit diversity technique for wireless
  communications,'' \emph{IEEE Journal on Selected Areas in Communications},
  vol.~16, no.~8, pp. 1451--1458, Oct 1998.

\bibitem{SimonAlouini05}
\BIBentryALTinterwordspacing
M.~K. Simon and M.-S. Alouini, \emph{Digital communication over fading
  channels}, ser. Wiley series in telecommunications and signal
  processing.\hskip 1em plus 0.5em minus 0.4em\relax Hoboken, N.J.
  Wiley-Interscience, 2005. [Online]. Available:
  \url{http://opac.inria.fr/record=b1102756}
\BIBentrySTDinterwordspacing

\bibitem{Molisch05}
A.~Molisch, \emph{Wireless Communications}.\hskip 1em plus 0.5em minus
  0.4em\relax Wiley-IEEE Press, 2005.

\bibitem{BiglieriProaSham98}
E.~Biglieri, J.~Proakis, and S.~Shamai, ``Fading channels:
  {I}nformation-theoretic and communications aspects,'' \emph{IEEE Transactions
  on Information Theory}, vol.~44, no.~6, pp. 2619--2692, Oct 1998.

\bibitem{WuNegi03}
D.~Wu and R.~Negi, ``Effective capacity: {A} wireless link model for support of
  quality of service,'' \emph{IEEE Transactions on Wireless Communications},
  vol.~2, no.~4, pp. 630--643, July 2003.

\bibitem{KoAlouSimo00}
Y.-C. Ko, M.~S. Alouini, and M.~K. Simon, ``Outage probability of diversity
  systems over generalized fading channels,'' \emph{IEEE Transactions on
  Communications}, vol.~48, no.~11, pp. 1783--1787, Nov 2000.

\bibitem{JabiSzczBenj12}
M.~Jabi, L.~Szczecinski, and M.~Benjillali, ``Accurate outage approximation of
  {MRC} receivers in arbitrarily fading channels,'' \emph{IEEE Communications
  Letters}, vol.~16, no.~6, pp. 789--792, June 2012.

\bibitem{BeanFackTayl08}
\BIBentryALTinterwordspacing
N.~G. Bean, M.~Fackrell, and P.~Taylor, ``Characterization of
  matrix-exponential distributions,'' \emph{Stochastic Models}, vol.~24, no.~3,
  pp. 339--363, 2008. [Online]. Available:
  \url{http://dx.doi.org/10.1080/15326340802232186}
\BIBentrySTDinterwordspacing

\bibitem{AsmussenBlad96}
S.~Asmussen and M.~Bladt, \emph{Renewal theory and queueing algorithms for
  matrix-exponential distributions}.\hskip 1em plus 0.5em minus 0.4em\relax
  Marcel Dekker Incorporated, 1996, pp. 313--341.

\bibitem{MehdiLiefReec95}
D.~Medhi, A.~van~de Liefvoort, and C.~S. Reece, ``Performance analysis of a
  digital link with heterogeneous multislot traffic,'' \emph{IEEE Transactions
  on Communications}, vol.~43, no. 2/3/4, pp. 968--976, Feb 1995.

\bibitem{McMillan95}
D.~McMillan, ``Delay analysis of a cellular mobile priority queueing system,''
  \emph{IEEE/ACM Transactions on Networking}, vol.~3, no.~3, pp. 310--319, Jun
  1995.

\bibitem{MolerLoan03}
C.~Moler and C.~V. Loan, ``Nineteen dubious ways to compute the exponential of
  a matrix, twenty-five years later,'' \emph{SIAM Review}, vol.~45, no.~1, pp.
  3--49, 2003.

\bibitem{Higham08}
N.~J. Higham, \emph{Functions of matrices -- {T}heory and computation}.\hskip
  1em plus 0.5em minus 0.4em\relax SIAM, 2008.

\bibitem{Fackrell03}
M.~W. Fackrell, ``Characterization of matrix-exponential distributions,'' Ph.D.
  dissertation, The University of Adelaide, Faculty of Engineering, Computer
  and Mathematical Sciences, 2003.

\bibitem{AsmussenOcin04}
\BIBentryALTinterwordspacing
S.~Asmussen and C.~A. O�cinneide, \emph{Matrix-Exponential
  Distributions}.\hskip 1em plus 0.5em minus 0.4em\relax John Wiley \& Sons,
  Inc., 2004. [Online]. Available:
  \url{http://dx.doi.org/10.1002/0471667196.ess1092.pub2}
\BIBentrySTDinterwordspacing

\bibitem{RuizCastro13}
J.~E. Ruiz-Castro, ``Matrix-exponential distributions: {C}losure properties,''
  \emph{International Journal of Advanced Statistics and Probability}, vol.~1,
  no.~2, pp. 44--52, 2013.

\bibitem{Neuts81}
M.~F. Neuts, \emph{Matrix-geometric solutions in stochastic models : {A}n
  algorithmic approach}, ser. Johns Hopkins series in the mathematical
  sciences.\hskip 1em plus 0.5em minus 0.4em\relax Baltimore: Johns Hopkins
  University Press, 1981.

\bibitem{gradshteyn07}
\BIBentryALTinterwordspacing
I.~S. Gradshteyn and I.~M. Ryzhik, \emph{Table of Integrals, Series, and
  Products}, 7th~ed., D.~Zwillinger and V.~H. Moll, Eds.\hskip 1em plus 0.5em
  minus 0.4em\relax Academic Press, 2007. [Online]. Available:
  \url{http://www.mathtable.com/gr/}
\BIBentrySTDinterwordspacing

\bibitem{GoreHeathPaul02}
D.~Gore, R.~W. Heath, and A.~Paulraj, ``On performance of the zero forcing
  receiver in presence of transmit correlation,'' in \emph{Proc. {IEEE}
  International Symposium on Information Theory ({ISIT}�02)}, 2002.

\bibitem{Telatar99}
I.~E. Telatar, ``Capacity of multi-antenna {G}aussian channels,''
  \emph{European Transactions on Telecommunications}, vol.~10, pp. 585--595,
  1999.

\bibitem{Horn86}
R.~A. Horn, \emph{Topics in Matrix Analysis}.\hskip 1em plus 0.5em minus
  0.4em\relax New York, NY, USA: Cambridge University Press, 1986.

\bibitem{WangGian04}
Z.~Wang and G.~B. Giannakis, ``Outage mutual information of space-time mimo
  channels,'' \emph{IEEE Transactions on Information Theory}, vol.~50, no.~4,
  pp. 657--662, April 2004.

\bibitem{Larsson04P}
P.~Larsson, N.~Johansson, and K.~Sunell, ``Method and arrangement for
  bidirectional relaying in wireless communication systems,'' U.S. Patent
  7\,920\,501 B1, Dec 30, 2004.

\bibitem{LarssonJohaSune05}
P.~Larsson, N.~Johansson, and K.~E. Sunell, ``Coded bidirectional relaying,''
  in \emph{Proc. Scandinavian Wireless Adhoc Workshop ({ADHOC}�05)}, May 2005.

\bibitem{LarssonJohaSune06}
------, ``Coded bidirectional relaying,'' in \emph{Proc. {IEEE} 63rd Vehicular
  Technology Conference ({VTC}�06)}, vol.~2, May 2006, pp. 851--855.

\bibitem{LarssonRasmSkog14a}
P.~Larsson, L.~K. Rasmussen, and M.~Skoglund, ``Analysis of rate optimized
  throughput for {ARQ} in fading interference channels,'' in \emph{Proc. {IEEE}
  International Conference on Communications ({ICC}�14)}, June 2014, pp.
  5926--5931.

\bibitem{BladtNiel10}
\BIBentryALTinterwordspacing
M.~Bladt and B.~F. Nielsen, ``On the construction of bivariate exponential
  distributions with an arbitrary correlation coefficient,'' \emph{Stochastic
  Models}, vol.~26, no.~2, pp. 295--308, 2010. [Online]. Available:
  \url{http://dx.doi.org/10.1080/15326341003756486}
\BIBentrySTDinterwordspacing

\bibitem{BodrogHorvTele08}
\BIBentryALTinterwordspacing
L.~Bodrog, A.~Horv{\'a}th, and M.~Telek, ``{M}oment characterization of matrix
  exponential and {M}arkovian arrival processes,'' \emph{Annals of Operations
  Research}, vol. 160, no.~1, pp. 51--68, 4 2008. [Online]. Available:
  \url{https://mycite.omikk.bme.hu/doc/28632.pdf}
\BIBentrySTDinterwordspacing

\bibitem{WahlstromAxelGust14}
N.~Wahlstr{\"o}m, P.~Axelsson, and F.~Gustafsson, ``Discretizing stochastic
  dynamical systems using {L}yapunov equations,'' in \emph{Proc. 19th World
  Congress of the International Federation of Automatic Control}, ser. World
  Congress, vol. Volume 19, Part 1, 2014, pp. 3726--3731.

\bibitem{VanLoan78}
C.~V. Loan, ``Computing integrals involving the matrix exponential,''
  \emph{IEEE Transactions on Automatic Control}, vol.~23, no.~3, pp. 395--404,
  Jun 1978.

\bibitem{Craig91}
J.~W. Craig, ``A new, simple and exact result for calculating the probability
  of error for two-dimensional signal constellations,'' in \emph{Proc. {IEEE}
  Military Communications Conference ({MILCOM}�91)}, Nov 1991, pp. 571--575
  vol.2.

\bibitem{SimonAlou98}
M.~K. Simon and M.~Alouini, ``A unified approach to the performance analysis of
  digital communication over generalized fading channels,'' \emph{Proceedings
  of the IEEE}, vol.~86, no.~9, pp. 1860--1877, Sep 1998.

\bibitem{MartinezFabrCair07}
A.~Martinez, A.~G. i~Fabregas, and G.~Caire, ``A closed-form approximation for
  the error probability of {BPSK} fading channels,'' \emph{IEEE Transactions on
  Wireless Communications}, vol.~6, no.~6, pp. 2051--2054, June 2007.

\bibitem{PatenaudeLodgChou97}
F.~Patenaude, J.~H. Lodge, and J.~Y. Chouinard, ``Error probability expressions
  for non-coherent diversity in {N}akagami fading channels,'' in \emph{Proc.
  {IEEE} 47th Vehicular Technology Conference ({VTC}�97)}, vol.~3, May 1997,
  pp. 1484--1487 vol.3.

\bibitem{WinWint99}
M.~Z. Win and J.~H. Winters, ``On maximal ratio combining in correlated
  {N}akagami channels with unequal fading parameters and {SNRs} among branches:
  {A}n analytical framework,'' in \emph{Proc. {IEEE} Wireless Communications
  and Networking Conference ({WCNC}�99)}, 1999, pp. 1058--1064 vol.3.

\bibitem{MichalowiczNichBuch13}
J.~V. Michalowicz, J.~M. Nichols, and F.~Bucholtz, \emph{Handbook of
  Differential Entropy}.\hskip 1em plus 0.5em minus 0.4em\relax Chapman \&
  Hall/CRC, 2013.

\bibitem{HasanHasaSchar00}
M.~A. Hasan, J.~A.~K. Hasan, and L.~Scharenbroich, ``New integral
  representations and algorithms for computing nth roots and the matrix sector
  function of nonsingular complex matrices,'' in \emph{Proc. 39th IEEE
  Conference on Decision and Control}, vol.~5, 2000, pp. 4247--4252 vol.5.

\bibitem{GrayNeuh98}
R.~M. Gray and D.~L. Neuhoff, ``Quantization,'' \emph{IEEE Transactions on
  Information Theory}, vol.~44, no.~6, pp. 2325--2383, Oct 1998.

\end{thebibliography}


\balance

\end{document}